\documentclass[aps,prx,twocolumn,showpacs,citeautoscript,superscriptaddress,longbibliography]{revtex4-1}

\usepackage{graphicx}  
\usepackage{epstopdf}
\usepackage{dcolumn}   
\usepackage{bm}        
\usepackage{amsmath}
\usepackage{amssymb}   
\usepackage{wrapfig}
\usepackage{array}
\usepackage[caption=false]{subfig}
\usepackage[bookmarks=false,linkcolor=blue,urlcolor=blue,colorlinks,citecolor=blue]{hyperref}
\usepackage{exscale,relsize}
\usepackage{color}
\usepackage{times, verbatim}
\usepackage{pstricks}

\usepackage{pst-all}

\newcommand{\bra}[1]{\langle #1|}
\newcommand{\ket}[1]{|#1\rangle}

\makeatletter

\begin{document}

\title{Scalable Designs for Quasiparticle-Poisoning-Protected Topological Quantum Computation
with Majorana Zero Modes}

\author{Torsten Karzig}
\affiliation{Station Q, Microsoft Research, Santa Barbara, California 93106-6105 USA}

\author{Christina Knapp}
\affiliation{Department of Physics, University of California, Santa Barbara,
	California 93106 USA}

\author{Roman M. Lutchyn}
\affiliation{Station Q, Microsoft Research, Santa Barbara, California 93106-6105 USA}

\author{Parsa Bonderson}
\affiliation{Station Q, Microsoft Research, Santa Barbara, California 93106-6105 USA}

\author{Matthew B. Hastings}
\affiliation{Station Q, Microsoft Research, Santa Barbara, California 93106-6105 USA}

\author{Chetan Nayak}
\affiliation{Station Q, Microsoft Research, Santa Barbara, California 93106-6105 USA}
\affiliation{Department of Physics, University of California, Santa Barbara,
	California 93106 USA}

\author{Jason Alicea}
\affiliation{Walter Burke Institute for Theoretical Physics and Institute for Quantum Information and Matter, California Institute of Technology, Pasadena, California 91125 USA}
\affiliation{Department of Physics, California Institute of Technology, Pasadena, California 91125 USA}

\author{Karsten Flensberg}
\affiliation{Center for Quantum Devices and Station Q Copenhagen, Niels Bohr Institute, University of Copenhagen, DK-2100 Copenhagen, Denmark}

\author{Stephan Plugge}
\affiliation{Center for Quantum Devices and Station Q Copenhagen, Niels Bohr Institute, University of Copenhagen, DK-2100 Copenhagen, Denmark}
\affiliation{Institut f\"ur Theoretische Physik, Heinrich-Heine-Universit\"at, D-40225 D\"usseldorf}

\author{Yuval Oreg}
\affiliation{Department of Condensed Matter Physics, Weizmann Institute of Science, Rehovot 76100, Israel.}

\author{Charles M. Marcus}
\affiliation{Center for Quantum Devices and Station Q Copenhagen, Niels Bohr Institute,
University of Copenhagen, DK-2100 Copenhagen, Denmark}

\author{Michael H. Freedman}
\affiliation{Station Q, Microsoft Research, Santa Barbara, California 93106-6105 USA}
\affiliation{Department of Mathematics, University of California, Santa Barbara,
California 93106 USA}

\begin{abstract}
We present designs for scalable quantum computers composed of qubits
encoded in aggregates of four or more Majorana zero modes, realized at the ends of
topological superconducting wire segments that are assembled into superconducting islands
with significant charging energy. Quantum information can be manipulated according to
a measurement-only protocol, which is facilitated by tunable couplings between Majorana
zero modes and nearby semiconductor quantum dots. Our proposed architecture designs have the following
principal virtues: (1) the magnetic field can be aligned in the direction of all
of the topological superconducting wires since they are all parallel; (2) topological
T-junctions are not used, obviating possible difficulties in their fabrication and utilization;
(3) quasiparticle poisoning is abated by the charging energy;
(4) Clifford operations are executed by a relatively standard measurement:
detection of corrections to quantum dot energy, charge, or
differential capacitance induced by quantum fluctuations;
(5) it is compatible with strategies for producing good
approximate magic states.
\end{abstract}

\date{\today}

\maketitle

\section{Introduction}

Non-Abelian topological phases of matter provide an attractive platform, in principle, for fault-tolerant
quantum computation. However, there are a number of obstacles that must be surmounted
in order to make this a reality. (1) A non-Abelian topological phase must be found or
engineered. (2) Quasiparticles must be braided in order to manipulate the quantum information that
is encoded in them; moving individual quasiparticle excitations is a feat
that has never been accomplished before, and it would have to be done routinely
during the operation of a topological quantum computer. (3) The topological charge
of a pair of quasiparticles must be measured in order to
determine the result of a calculation. The conceptually simplest way to do this would be
with an anyonic interferometry measurement~\cite{Fradkin98,Stern06,Bonderson06a,Nayak08,Bonderson08d,Bishara09a}, but that requires coherent transport, potentially over
long scales; neither an interferometry nor any other measurement has unambiguously
measured the topological charge of a pair of quasiparticles.
In this paper, we present a scheme for topological quantum computation
that obviates these difficulties.

A path surmounting the first obstacle noted above was opened up
by the advent of semiconductor-superconductor
heterostructures that combine superconductivity, strong spin-orbit coupling, and
magnetic fields to create a topological superconducting state that supports Majorana
zero modes (MZMs)~\cite{Sau10a,Lutchyn10,Oreg10}.
While originally envisioned in two dimensions~\cite{Read00}, such topological superconducting phases can also be hosted in one-dimensional systems, e.g., nanowires~\cite{Kitaev01,Fu08,Lutchyn10,Oreg10}, and braiding operations can be implemented
in wire networks~\cite{Alicea11}.  There is strong experimental evidence that a topological superconductor
has been realized with semiconductor nanowires~\cite{Mourik12,Rokhinson12,Deng12,Churchill13,Das12,Finck12,Albrecht16}.

The price that is paid in such an approach is that a topological superconductor
is not quite a topological phase of matter but, rather, a ``fermion parity-protected topological
phase''~\cite{Bonderson13b} and, therefore, is vulnerable to ``quasiparticle poisoning'' (QPP),
i.e.,  to processes that change the number of electrons in the device.
However, one can prevent QPP of MZMs on a superconducting island by
incorporating relatively large charging energies that provide a Coulomb blockade for the island, as utilized in the proposals of
Refs.~\onlinecite{Landau16,Plugge16a,Starkpatent16,Vijay16b,Plugge16b}.
(Charging energy does not protect MZMs from quasiparticle excitations occurring within
the device. { However, such excitations and the errors they cause are exponentially suppressed by $\Delta/T$ for energy gap $\Delta$ and temperature $T$}.) We refer to a Coulomb-blockaded superconducting island hosting MZMs as a ``MZM island.''

A recent experiment, inspired by the theoretical prediction of Ref.~\onlinecite{Fu10}, reported the first systematic measurement of the ground-state degeneracy splitting for proximitized nanowires in a Coulomb blockade regime and observed that it is exponential in the nanowire length $L$~\cite{Albrecht16}.
The transport measurements of Ref.~\onlinecite{Albrecht16} are in qualitative agreement with theoretical calculations~\cite{Heck16}. The combination of material science progress~\cite{Krogstrup15,Shabani16}, device quality and controllability~\cite{Albrecht16, Zhang16}, and theoretical advances involving semiconductor-superconductor heterostructures~\cite{Beenakker13a, Alicea12a, Leijnse12, Stanescu13b, DasSarma15} provides a pathway for topological quantum computation with semiconductor nanowires.

A way to circumvent the second obstacle, i.e.,  the need to move quasiparticles, is to use
a ``measurement-only'' protocol~\cite{Bonderson08b,Bonderson08c}, wherein a sequence of measurements
has the same effect as a braiding operation. Such methods eliminate the need to move the computational quasiparticles
and, thus, eliminate the need for coherent topological ``T-junctions''~\cite{Alicea11}, which may present banal engineering issues such as those identified in Ref.~\onlinecite{Nijholt16}.

The remaining obstacle is the measurement of the topological charge of quasiparticle pairs.
One might worry that measurements could still involve moving probe
quasiparticles through an interferometry loop, thereby reintroducing the second obstacle. However, this concern can be surmounted by taking
advantage of the distinction between a fermion parity-protected topological phase and a true
topological phase (which is a mathematical abstraction that may not quite correspond to
any real physical system anyway~\cite{Bonderson13b}): topological charge can be manipulated by
the process of an electron tunneling into a MZM~\cite{Flensberg11}.
As shown in Ref.~\onlinecite{Fu10},
transport through a pair of MZMs can provide a measurement of their combined topological charge in the presence of a large charging energy.

Majorana-based qubits with four MZMs residing on a Coulomb blockaded island have been studied recently. In particular, Refs.~\onlinecite{Landau16,Plugge16a} have focused on surface code architectures where the MZM islands form a hexagonal lattice. The large charging energies invoked in these papers distinguishes them from
other Majorana surface code proposals in which the charging energies are small~\cite{Vijay15,Vijay16a}.
The former surface code approach has the advantage that conductance measurement via interference is naturally built in, with the interfering paths involving co-tunneling through MZM islands. While the surface code aims for fault-tolerant computation, one can also think about a minimal setup in which islands with four MZMs constitute logical qubits, denoted as ``Majorana box qubits'' in Ref.~\onlinecite{Plugge16b}, and measurements are performed by detecting frequency shifts of double dot systems. In that work, a minimal demonstration of the Clifford gates was proposed using four such qubits.

In this paper, we design a modular system for measurement-only MZM topological quantum computation in which the basic module contains a small network of (4 or 6) MZMs and quantum dots for measurement.~\footnote{The coupling to quantum dots plays a different role here than in Ref.~\onlinecite{Hoffman16}, where the dot is a spin qubit.} Related ideas have appeared in the independent work of Ref.~\onlinecite{Plugge16b},
but they are sharpened here by quantum information requirements that lead us
to a scalable arrangement with novel features.

We analyze five new scalable architectures~\cite{hexon_patent16,tetron_patent16} for Majorana-based quantum computing, each of which overcomes all of the obstacles listed above. Each architecture is centered around a qubit composed of parallel sets of topological superconducting wires. The wires are electrically connected by normal superconductors, so that no individual wire has a charging energy, but the entire qubit is Coulomb blockaded at all times.  This fact is an important distinction with respect to the previous Majorana-based quantum computing
proposals~\cite{Alicea11,Hassler10,Jiang11a,Bonderson11b,Sau11a,Hyart13, Clarke16, Aasen16}.
Quantum information is manipulated by joint fermion parity measurements on pairs and quartets of MZMs.  These measurements allow for intra-qubit braiding operations via the measurement-only protocols, as well as for two-qubit entangling operations. Of our five proposed architectures, three involve six MZMs per superconducting island, which we refer to as ``hexons,'' and two involve four MZMs per island, which we call ``tetrons.''  We evaluate each hexon and tetron design on four axes:  (1) QPP time $\sim$ charging energy $E_C$; (2) signal visibility $\sim E_C^{-1}$; (3) fabrication simplicity; and (4) computational efficiency.

{ Due to the exponential suppression of errors, our proposed qubit designs should have sufficiently long coherence times to solve low-depth problems. For long enough computations, the exponentially small errors will eventually become important and must be addressed through some form of error-correction. The computational universality of our proposed qubits allows flexibility in the choice of code, though it would be wise to use codes that take advantage of having high fidelity Clifford gates.  While the implementation of an error correcting code on the system is an interesting and important question, it is not the focus of this paper and will instead be addressed in a future work~\cite{Hastings16}.}

This paper is organized as follows.
In Section~\ref{sec:overview}, we describe one of our five designs, the ``one-sided hexon,'' as an illustrative example of the key concepts utilized in our proposals.  All of our designs rely on measurement-only topological quantum computation~\cite{Bonderson08b,Bonderson08c}, so in Section~\ref{sec:measurements}, we explain how the fermion parity of an even number of MZMs
can be measured through their coupling to nearby quantum dots.
In Section~\ref{sec:architectures}, we give a detailed description of all of our
topological qubit designs: the one-sided hexon introduced in Section~\ref{sec:overview}, as well as two-sided hexons, linear hexons, two-sided tetrons, and linear tetrons. We elucidate the quantum information-theoretic
basis for achieving all Clifford operations, i.e., a ``Clifford complete'' gate set, { in a topologically protected manner} with these designs. { In Section~\ref{sec:design-summary}, we compare and contrast the proposed qubit designs using the axes (1)-(4) mentioned above.}  In Section~\ref{sec:universal}, we
describe how our proposed architectures support universal quantum computation by using approximate
magic state production and distillation.
Finally, in Section~\ref{sec:Conclusions},
we outline the next experimental steps towards realizing our qubit designs.

\section{Overview and design example}
\label{sec:overview}

\begin{figure*}
	\includegraphics[width=2\columnwidth]{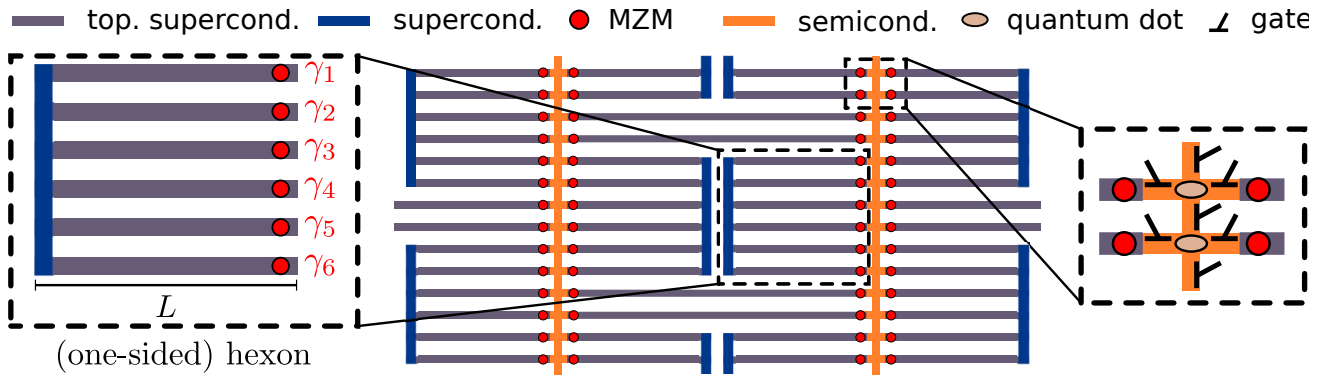}
	\caption{An example of a scalable hexon architecture. The minimal building block defining a qubit and an ancilla are one-sided hexons, which are topological Cooper pair boxes containing six MZMs (magnified in the left panel). Note: the illustration is not drawn to scale; in practice, the length $L$ of 1DTS wires is much larger than the coherence length $\xi$ and vertical separation distances between wires are much smaller than $\xi$. The measurement of joint parities of MZMs becomes possible by selective coupling to quantum dots. The latter are defined and controlled by gates as depicted in the magnification in the right panel. Two-MZM measurements within a hexon and four-MZM measurements involving two hexons (with two MZMs from a given hexon) enable Clifford complete operations on the array of qubits.}
	\label{fig:1}
\end{figure*}

In this section, we discuss the main principles of the scalable Majorana-based quantum computing architectures presented in this paper. For concreteness, we focus on a particular example of hexons consisting of six proximitized nanowires. In Section~\ref{sec:architectures}, we present additional architectures utilizing hexons and tetrons constructed from various numbers of proximitized nanowires.

The main building block of the presented design is a comb-like structure (see Fig.~\ref{fig:1}) consisting of six floating (i.e.,  not grounded) one-dimensional topological superconductors (1DTSs) of length $L$. These 1DTSs may be realized, for instance, using InAs wires coated by a superconducting half shell~\cite{Krogstrup15}. To form a single island hosting multiple MZMs, the 1DTSs are connected by a strip of ($s$-wave) superconductor at one side, which we refer to as the ``backbone.'' Since the superconducting backbone is a conventional (i.e.,  non-topological) superconductor, the 1DTSs and the magnetic field needed to bring them into the topological phase can all be aligned in the same direction. The vertical distance between neighboring 1DTS is chosen to be shorter than the superconducting coherence length, which will lead to a strong hybridization of the six MZMs located at the backbone side of the 1DTSs. Consequently, there remain only six MZMs in the structure, localized at the non-backbone side. We denote these MZMs by $\gamma_j $ for $j=1,\dots,6$, which we also use to represent the corresponding Majorana operators. We call this comb-like structure a ``one-sided hexon.''

{ The hexon acts as a topological Cooper pair box~\cite{Beri12,Altland13}}. If operated away from charge degeneracies, i.e.,  in Coulomb valleys, the overall parity of the hexon $-i\prod\limits_{j=1}^{6} \gamma_j$ will be fixed and the charging energy will protect the system from QPP. A QPP event would occur if an unpaired fermionic quasiparticle hopped onto or off of the hexon. However, due to the hexon charging energy $E_C$, such events
will be suppressed as $\exp(-E_C/T)$. As such, the hexon has a (nearly) degenerate ground state subspace that is four-dimensional, which we use to encode a logical qubit and an ancilla. A QPP event would be a ``leakage
error'' in which the system leaves the four-dimensional computation subspace.

{ One might additionally be concerned about thermally excited quasiparticles within the device. Provided the temperature is much smaller than the energy gap $\Delta,$ such excitations and the errors they cause are exponentially suppressed in $\Delta/T$.}

In order to avoid errors due to splitting the ground state degeneracies of the MZMs from accruing in the quantum information stored in a hexon, we require two crucial constraints for the one-sided hexons. First, the 1DTSs need to be long enough compared to the effective coherence length $\xi$ within the 1DTSs, i.e., $L\gg\xi$, to suppress the hybridization of the MZMs by a factor of $\exp(-2L/\xi)$. Secondly, we need to suppress the charging energy associated with the mutual capacitance between two 1DTSs within a hexon.  Both hybridization of the MZMs and relative charging energies between 1DTSs would result in splitting the degeneracy of the hexon ground states.  The relative charging energy decreases exponentially with the number of channels that connect the 1DTSs to the backbone~\cite{SCHON90}. In the limit of many weak channels (described by a Josephson energy $E_J$), the relative charging energy $E_{C0}$ is suppressed by a factor $\exp(-\sqrt{8E_J/E_{C0}})$. We assume that a direct connection of the backbone with the superconducting shell of a nanowire has a large area in units of the Fermi wavelength, i.e.,  the number of transverse channels in the junction exceeds thousands. Thus, the relative charging energy will be quenched with exponential accuracy so that one can characterize this system as a superconducting island with an overall charging energy $E_C$. In other words, it is a topological Cooper pair box.

{ As the superconducting island's charging energy $E_C$ is inversely related to its geometric capacitance, there is a trade-off between using long 1DTSs and maintaining a large charging energy.  When the wire length $L$ is much longer than the width $w$ of the island (i.e., the length of the superconducting backbone), the geometric capacitance of the island will approximately depend linearly in $L$; the dependence of the capacitance on $w$ will be more complicated, but can safely be estimated to be sub-linear.  Thus, the charging energy will roughly behave as $1/L$ and there will be an optimal value of $L$ that maximizes the combined protection, i.e., roughly when $E_C/T \approx 2L/\xi$ for the one-sided hexon. Based on estimates from experiments~\cite{Albrecht16}, it should not be difficult to reach a regime in which $E_C/T \sim L/\xi \gg 1$.}

With the above conditions, dynamical phases and QPP errors will be strongly suppressed by large exponentials. This opens the path to creating qubits with exceptionally long coherence times. In the next subsection, we discuss how these qubits can be manipulated and combined to a large scale quantum computer.

\subsection{Single qubit operations}

{ A universal gate set can be generated by the Clifford operations (which can be generated from the Hadamard gate, Phase gate, and CNOT gate) supplemented by an additional non-Clifford gate.  One benefit of Majorana-based quantum computing is that the Clifford operations may be implemented with topological protection, as we now explain for the hexon. We discuss how to implement the (non-Clifford) $T$ gate in  Section~\ref{sec:universal}.}

The hexon can be understood as a standard encoding of a topological qubit in four MZMs combined with an ancillary pair of MZMs. For concreteness, we let the topological qubit be encoded in MZMs $\gamma_1$, $\gamma_2$, $\gamma_5$, and $\gamma_6$, which are taken to have total fermion parity even. We can choose the basis states of the topological qubit to be $\left| 0 \right\rangle = \left| p_{12} = p_{56}= -1 \right\rangle $  and $\left| 1 \right\rangle = \left| p_{12} = p_{56} = +1 \right\rangle $, where $p_{jk}$ is the eigenvalue of $i\gamma_j\gamma_k$.

The ancillary pair of MZMs $\gamma_3$ and $\gamma_4$ is thus constrained to have $i \gamma_3 \gamma_4 = -1$ in this encoding. The presence of the ancillary pair of MZMs allows us to implement arbitrary braiding operations on the four MZMs of the topological qubit by appropriate measurements~\cite{Bonderson08b}. Moreover, we can use measurements to change which MZMs encode the computational qubit, shuttling around the ancillary MZMs via anyonic teleportation. As an example, performing a sequence of parity measurements of $i\gamma_3 \gamma_4$, $i\gamma_1\gamma_3$, $i\gamma_2\gamma_3$, and then $i\gamma_3\gamma_4$ generates the same operator obtained by exchanging $\gamma_1$ and $\gamma_2$ (see Section~\ref{sec:hexon_qi_basics} for details). In this way, intra-hexon measurements provide a precise way of generating all single-qubit Clifford gates (which can be generated by the Hadamard gate and Phase gate, for example) on the topological qubits.

These operations require us to have the ability to perform a sufficiently diverse set of parity measurements of MZM pairs. Our designs incorporate this via a quantum dot based measurement scheme. Quantum dots can be defined and selectively coupled to MZMs by tuning depletion gates in a nearby semiconducting wire that is connected to the hexon's MZM side (see Fig.~\ref{fig:1}). Measurements of the parity $i\gamma_j\gamma_k$ can then be done by connecting MZMs $\gamma_j$ and $\gamma_k$ to quantum dots in the semiconducting wire. In general, the eigenvalue $p_{jk}$ of $i\gamma_j\gamma_k$ will affect the ground-state energy as well as the average charge and differential capacitance of the quantum dots. This can be used in a variety of schemes to make the desired measurement, as is detailed in Section~\ref{sec:measurements}.

\subsection{Entangling operations and full quantum computation}

We must entangle different hexons in order to implement quantum operations corresponding to the full set of Clifford gates. Such entangling operations between hexons can be achieved by performing four-MZM measurements, involving two MZMs from each hexon. The latter can also be realized using quantum dots (see Section~\ref{sec:measurements} for details). The main idea is to use an interference effect~\cite{Landau16,Plugge16a} in the hybridization of two quantum dots arranged as in the magnified panel of Fig.~\ref{fig:1}. The pinch-off gates are tuned so that there is no direct connection between the two quantum dots. However, the two dots can hybridize via tunneling in and out of the MZM states of the nearby hexons. Coherently summing amplitudes along the paths through each nearby hexon leads to a detectable dependence of the hybridization energy on the overall parity of the four involved MZMs.

In order to achieve a fully-connected two dimensional graph for the entangling operations, some of the four-MZM measurements must involve MZMs that are separated by distances of approximately $2L$. Measurements involving these longer distances require additional structure to actualize. For this purpose, additional floating topological superconductors of length $2L$ can act as links to bridge these distances by MZM-mediated coherent electron tunneling~\cite{Fu10,Heck16}. Two such coherent links are placed above and below any superconducting backbone (see Fig.~\ref{fig:1}). The resulting (trivalent) connectivity graph of the hexon qubits is hexagonal.

Due to the freedom of arbitrary MZM exchanges within each hexon, a single entangling operation between adjacent hexon pairs is enough to realize CNOT operations between qubits and therefore make the hexons Clifford complete. The latter can be augmented to full quantum universality if we can also generate approximate magic states. The designs presented here naturally allow us to prepare very precise magic states, which lowers the overhead for magic state distillation (see Section~\ref{sec:universal}).

{ We further note that error correction may be implemented at the software level on the array of hexons, as Clifford complete physical qubits support all stabilizer codes~\cite{Terhal15}.}

\section{Majorana measurements}
\label{sec:measurements}

\begin{figure}
	\includegraphics[width=0.95\columnwidth]{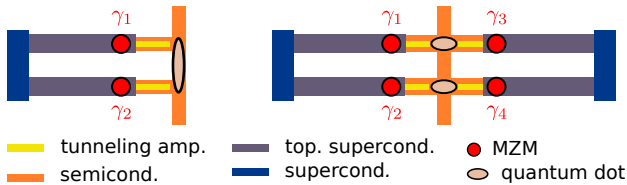}
	\caption{Appropriately tuning the gates shown in the magnification of the right of Fig.~\ref{fig:1} creates the scenarios depicted in the left and right panels here. {\it Left panel}: A device configuration for measuring the two-MZM parity { $p_{12}$ (eigenvalue of $i\gamma_1\gamma_2$)}. MZMs $\gamma_1$ and $\gamma_2$ are coupled to a single quantum dot with tunneling amplitudes $t_1$ and $t_2$ respectively.  {\it Right panel}: A device configuration for measuring the four-MZM parity $p=p_{12}p_{34}$, where { $p_{jk}$ is the eigenvalue of $i\gamma_j\gamma_k$}.  MZMs $\gamma_1, \gamma_3$ are tunnel coupled to the upper quantum dot, while MZMs $\gamma_2$ and $\gamma_4$ are tunnel coupled to the lower quantum dot.  Both geometries can be modified to measure non-adjacent pairs of MZMs, as demonstrated in Fig.~\ref{fig:operations}. }
	\label{fig:mst}
\end{figure}

A key feature of our approach to scalable topological quantum computing is the ability to perform projective measurements of the combined fermionic parity of multiple MZMs. Such measurements are initiated by appropriately tuning gates to couple MZMs to quantum dots, as seen in the magnified right panel of Fig.~\ref{fig:1}. This realizes the devices depicted in Fig.~\ref{fig:mst} with one quantum dot (left panel) or two quantum dots (right panel). The gates control the amplitudes $t_j$ for electrons to tunnel between the MZMs (red) and a quantum dot (light gray). At low temperature $T\ll E_C$, the probability of an excited state with an electron on the island is exponentially small, as it is proportional to $\text{exp}\left( -E_C/T\right)$. The virtual transitions of electrons to the island are state dependent and, therefore, shift the energy levels in a parity-dependent manner.
Suitable spectroscopy on the quantum dot system allows measurements of the two-MZM parity (left panel) or of the four-MZM parity (right panel) parity~\cite{Starkpatent16,Plugge16b}.

{ The amplitude $t_j$ is exponentially suppressed in the tunnel barrier separating $\gamma_j$ from the quantum dot, and as such may be accurately tuned to zero.  Before and after the measurement, all couplings are turned off, leaving the MZM island and the quantum dot with fixed charge.  In this decoupled state, environmental noise, which couples to charge, can cause decoherence of states with different occupancy on the quantum dot(s), but has no effect on the MZM island.  Thus, unless we are actively performing a measurement, noise cannot measure and collapse the qubit state.  }

There is a small probability that the final occupancy of the quantum dot(s) after the measurement will be different than before the tunnel couplings were turned on. This probability is suppressed by the charging energy of the MZM island, but it is not zero. If the charge of the quantum dot(s) is different after the measurement than it was before the measurement, then QPP has occurred (the MZM island was poisoned by the dot(s)). To correct this error, one could repeat the measurement until the final dot occupations are as desired. The chance of such a QPP event can be reduced by tuning the quantum dot(s) far away from resonance before disconnecting the couplings.

\subsection{Projective measurement of two-MZM parity}
\label{sec:2MZMmodel}

We first discuss the case of two MZMs $\gamma_1$ and $\gamma_2$ coupled to a single quantum dot as shown in the left panel of Fig.~\ref{fig:mst}.  While coupling to a single MZM does not provide any information on fermion parity, non-local coupling to two or more MZMs may contain this information~\cite{Fu10}. The coupling of quantum dots
to MZMs was first discussed in Ref.~\onlinecite{Flensberg11}, which considered the case of a grounded superconductor (i.e., $E_C=0$).

When the tunneling amplitudes are zero, the MZM island and the quantum dot are decoupled.
In that case, the Hamiltonian for the MZM island is
\begin{equation}
\label{eq:H0}
H_0 = H_{\text{BCS}}+H_C,
\end{equation}
where $H_{\text{BCS}}$ is the BCS Hamiltonian for an s-wave superconductor coupled to multiple semiconductor nanowires, and $H_{C}$ is the charging energy Hamiltonian for the MZM island. In the low-energy approximation when energies are much smaller than the superconducting gap $\Delta$, the low-energy subspace contains only MZMs. (See Appendix~\ref{app:MZMmodels} for a more detailed discussion.) We neglect the length-dependent energy splitting of MZMs, unless otherwise stated. We also assume charging energies of tetrons and hexons are large compared to temperature.
The corresponding charging energy Hamiltonian is
\begin{equation}
\label{eq:HC}
H_C = E_C\left( \hat{N}_{S}-N_{g}\right)^2.
\end{equation}
The operator $\hat{N}_{S}$ counts the combined charge of the nanowire-superconductor island in units of the electron charge $e$ { and has integer eigenvalues $N_S$}.  The induced (dimensionless) charge on the island $N_g$ is controlled by the gate voltage. Henceforth, we consider the limit when the charging energy on the island is large compared to temperature ($E_C \gg T$), so that its charge does not change during the course of a measurement. For simplicity, we will assume that $|N_g|\ll 1$, so that the ground-state configuration has an average charge $\langle \hat{N}_{S}\rangle=0$ and energy $E_0=E_C  N_g^2$. For $\Delta\gg E_C$ the two lowest excited states $\ket{N_{S}=\pm 1}$ have energies $E_{1}=E_C \left( 1 - N_g\right)^2$ and $E_{2}=E_C \left( 1 + N_g\right)^2$. Thus, for $E_C \gg T$ the corresponding excitation energies are much larger than the temperature.

We assume that the semiconductor quantum dot is in a few-electron occupancy regime. The corresponding Hamiltonian is given by
\begin{align}
H_{\text{QD}}=\sum_{\alpha}h_{\alpha} f^\dag_{\alpha}f_{\alpha}+\varepsilon_C \left(\hat{n}-n_g \right)^2,
\end{align}
where $\alpha$ indexes the electron orbitals of the quantum dot, $f_{\alpha}$ and $f_{\alpha}^\dag$ are the corresponding fermionic annihilation and creation operators, respectively, and $\hat{n}=\sum\limits_{\alpha} f^\dag_{\alpha}f_{\alpha}$ is the total occupation operator. The $h_{\alpha}$ are the corresponding orbital energies and $\varepsilon_C$ is the charging energy.
Here, we assume that quantum dot is in the spinless regime due to the large magnetic field necessary to drive the semiconductor nanowires into the topological phase~\cite{Lutchyn10,Oreg10}. We consider the low temperature limit such that the charging energy $\varepsilon_C$ and the level spacing in the dot are much larger than the temperature. The regime of interest is when the quantum dot is tuned to be near the charge-degeneracy point for $n$ and $n+1$ electrons. In this case, one approximates the above Hamiltonian by an effective one corresponding to a single spinless fermion level
\begin{equation}
\label{eq:HQD}
H^{\rm eff}_{\text{QD}}=h \hat{n}_f +\varepsilon_C \left(\hat{n}_f-n_g \right)^2,
\end{equation}
{ where the operator $\hat{n}_f = f^\dagger f$ has integer eigenvalues $n_f$}. The two relevant low-energy states of the dot are defined by ${\ket{n_f=0,1}}$.  This approximation is justified as long as the dot charging energy is the largest relevant energy scale in the system, i.e., much larger than the charging energy of the superconducting island, $\varepsilon_C \gg E_C$. The charge-degeneracy point $n_g^*$ is defined by the condition  $\epsilon_1(n_g^*) = \epsilon_0(n_g^*)$, where
\begin{align}
\epsilon_1 (n_g) &=\varepsilon_C\left(1-n_g\right)^2+h,\\
\epsilon_0 (n_g) &= \varepsilon_C n_g^2.
\label{eq:energies}
 \end{align}

When $T \ll \Delta, E_C$, we can consider the low energy approximation where one writes $H_0$ in terms of the MZMs on the island. Provided that the tunneling matrix elements between the quantum dot and the MZM island are smaller than
the induced superconducting gap in the nanowires and the charging energy, $t_j \ll \Delta, E_C$, one can write the effective tunneling Hamiltonian~\cite{Fu10, Flensberg11} as
\begin{equation}
\begin{split}
\label{eq:Htunn}
H_{\text{tunn}}&= -i\frac{e^{-i\phi/2}}{2} \left( t_1f^\dagger \gamma_1 +t_2 f^\dagger \gamma_2\right) +\text{h.c.},
\end{split}
\end{equation}
where $t_1$ and $t_2$ correspond to tunneling between the quantum dot and $\gamma_1$ and $\gamma_2$, respectively, and  $e^{i\phi/2}$ is the shift operator which adds an electron to the island $e^{i\phi/2}\ket{N_S}=\ket{N_S+1}$.  Finally, the total Hamiltonian for the coupled system is given by
\begin{equation}
\label{eq:Htot}
H_{\text{tot}}= H_0+H^{\rm eff}_{\text{QD}}+H_{\text{tunn}}.
\end{equation}

The effect of $H_{\text{tunn}}$ is to allow fermions to tunnel between the quantum dot and MZM island.  We assume that the charging energy on the island is large at all times. Therefore, all electron charging processes are virtual, i.e.,  any fermion that hops onto the MZM island must hop back to the dot and vice versa. As shown below, such virtual transitions perturb the ground state energies in a parity-dependent manner.  Consider first the case where $n_f=1$ when the tunneling amplitudes are turned off.  Turning on the $t_j$ allows a fermion to tunnel from the dot into the MZM and then tunnel back onto the dot through a possibly different MZM.  This process mixes the ground state ${\ket{N_{S}=0}\otimes \ket{n_f=1}}$ with the excited state ${\ket{N_{S}=1}\otimes \ket{n_f=0}}$, resulting in a shifted ground state energy (to the lowest order in $|t_j|/E_C$)
\begin{equation}
\label{eq:eps1}
\varepsilon^{\rm tot}_{1} = E_C N_g^2 +\epsilon_1 -\frac{ |t_1|^2+|t_2|^2 +i p_{12} \left( t_1^* t_2 -t_1 t_2^*\right)}{4\left(E_C(1-2N_g)+\epsilon_0 -\epsilon_1\right)}.
\end{equation}
{ Here, $p_{jk}$ is the eigenvalue of $i\gamma_j \gamma_k$, the fermion parity of the two MZMs coupled to the quantum dot.  In other words, this calculation applies to both $p_{12}=\pm 1$ initial ground states.} This parity dependence originates from elastic co-tunneling through the corresponding pair of MZMs.

Alternatively, if the quantum dot is unoccupied when ${t_j=0}$, then when the $t_j$ are turned on, an electron can tunnel from a MZM onto the dot, then tunnel into a (possibly different) MZM, mixing the ground state ${\ket{N_{S}=0}\otimes \ket{n_f=0}}$ with the excited state ${\ket{N_{S}=-1}\otimes \ket{n_f=1}}$.  The corresponding shifted ground state energy is (to lowest order in $|t_j|/E_C$)
\begin{equation}
\label{eq:eps2}
\varepsilon^{\rm tot}_{0} = E_C N_g^2+\epsilon_0 -\frac{|t_1|^2+|t_2|^2-i p_{12} \left( t_1^* t_2 -t_1 t_2^*\right)}{4\left(E_C(1+2N_g)+\epsilon_1-\epsilon_0\right)}.
\end{equation}
In both Eqs.~(\ref{eq:eps1}) and (\ref{eq:eps2}), the parity dependence arises from the coupling between the quantum dot and MZMs. Indeed, by setting either $t_1$ or $t_2$ to zero one finds a correction to the quantum dot ground-state energy that is independent of $p_{12}$.  { At the charge degeneracy point $n_g^*$ of the quantum dot, the parity dependence of $\varepsilon_1^{\text{tot}}-\varepsilon_0^{\text{tot}}$ scales as $\text{Im}[t_1^*t_2/E_C]$.}

It is also important to observe that the parity dependence disappears if both $t_1$ and $t_2$ are real, even if both quantities are finite. Since time-reversal symmetry is broken, this is not generic.  However, for spinless fermions one may introduce an artificial anti-unitary symmetry $\mathcal{T}$ that squares to $+1$~\cite{Fidkowski11a}. Since a bilinear coupling between $\gamma_1$ and $\gamma_2$ is precluded, $t_{1}$ and $t_{2}$ are necessarily real.  Fortunately, $\mathcal{T}$ is most certainly \emph{not} a microscopic symmetry of our setup. However, the parity dependence of the shifted ground state energies may be ``accidentally'' weak for non-generic tunneling amplitudes. We comment further on this issue in Section~\ref{sec:Conclusions}.

\subsection{Projective measurement of four-MZM parity}
\label{sec:4MZMmodel}

In order to describe the device configuration shown in the right panel of Fig.~\ref{fig:mst}, the Hamiltonian of Eq.~(\ref{eq:Htot}) is modified to include two superconducting islands (four MZMs) and two quantum dots.  The decoupled MZM island Hamiltonian $H_0$ becomes a sum of Hamiltonians for the left and right MZM islands (labeled $a=1$ and $2$, respectively). The two islands may have different charging energies and induced charges, so the total (decoupled) charging energy Hamiltonian is the sum of those of the two islands:
\begin{eqnarray}
\label{eq:HCj}
H_{C}&=& \sum_{a=1,2} H_{C,a} , \\
H_{C,a} &=& E_{C,a} \left( \hat{N}_{S,a}-N_{g,a}\right)^2.
\end{eqnarray}
For simplicity, we again assume that $|N_{g,a}|\ll 1$ for both islands, so the ground state of the decoupled MZM islands has energy ${E_0=E_{C,1}N_{g,1}^2+E_{C,2}N_{g,2}^2}$.  In general, the charging energies and induced charges of the two quantum dots can also be different. For simplicity, we consider the case in which they are the same.
The effective Hamiltonian for the two semiconductor QDs may be written as
\begin{equation}
\label{eq:HQD2}
\begin{split}
H^{\rm eff}_{\text{QD}}&= \sum_{a=1,2} h_a \hat{n}_{f,a} +\varepsilon_{C,a} \left(\hat{n}_{f,a}-n_{g,a} \right)^2
\\ &\quad +\varepsilon_M \left( \hat{n}_{f,1}-n_{g,1}\right)\left( \hat{n}_{f,2}-n_{g,2}\right).
\end{split}
\end{equation}
The first term in Eq.~(\ref{eq:HQD2}) is simply the sum of the effective Hamiltonians of the two decoupled QDs, while the last term describes a mutual charging energy between the two quantum dots. We consider the case when $\varepsilon_M \ll \varepsilon_{C,a}$. The mutual charging energy may be appreciable for the geometry shown in the right panel of Fig.~\ref{fig:mst}, but can be neglected in other measurements of the joint parity of four MZMs (e.g., measurements involving MZMs on opposite sides of the two-sided hexon shown in Fig.~\ref{fig:insects}).  For simplicity, we will henceforth set ${h_1=h_2\equiv h}$ and ${\varepsilon_{C,1}=\varepsilon_{C,2}\equiv \varepsilon_{C}}$. We assume that there is no direct tunneling from one dot to the other; the only way for
an electron to tunnel between quantum dots is through a superconducting island.

The tunneling Hamiltonian now involves four MZMs, taking the form
\begin{equation}
\label{eq:Htun4}
\begin{split}
H_{\text{tunn}} &= -\frac{ie^{-i\phi_1/2}}{2} \left( t_1 f_1^\dagger \gamma_1 +t_2 f_2^\dagger \gamma_2\right)
\\ &\quad -\frac{e^{-i\phi_2/2}}{2}\left( t_3 f_1^\dagger \gamma_3 +t_4 f_2^\dagger \gamma_4 \right) +\text{h.c.},
\end{split}
\end{equation}
where the upper and lower quantum dots are labeled 1 and 2, respectively, so that $f_1$, $f_1^\dagger$, $f_2$, and $f_2^\dagger$ are their corresponding annihilation and creation operators. $e^{-i\frac{\phi_{1}}{2}}$ and $e^{-i\frac{\phi_{2}}{2}}$ are the electron shift operators for left and right islands, respectively.

As we saw for two MZMs, nonzero tunneling amplitudes mediate virtual transfer of fermions between the MZM islands and the quantum dot, thereby shifting the spectrum from that of the decoupled system.  Crucially, the perturbed energies depend on the joint parity of the two MZM islands $p=p_{12}p_{34}$ and does not depend on $p_{12}$ or $p_{34}$ individually.  This dependence can be intuitively understood by considering the tunneling paths a fermion can take: it either travels partway around the loop and  then backtracks (thereby only picking factors of $p_{jk}^0$ or $p_{jk}^2$, both of which equal one), or it makes a full loop (picking up a factor of $p_{12}p_{34}$). These arguments can be generalized to higher orders in perturbation theory where multiple loops are allowed. The resulting energy shifts only depend on the joint parity in any order of perturbation theory.

More quantitatively, the total Hamiltonian
\begin{equation}\label{eq:Htot4}
H^{\rm eff}_{\text{tot}}= H_{C}+H_{\text{BCS}}+H^{\rm eff}_{\text{QD}}+H_{\text{tunn}}
 \end{equation}
has four low-energy states for given values of $p_{12}$ and $p_{34}$, which we label ${\beta=0,1,2,3}$, with corresponding energies $\varepsilon^{\text{tot}}_{\beta}$. When $t_j=0$ and $N_{g,a}=0$, these four states are those
in which the occupancies $(n_{f,1},n_{f,2})$ of the two dots are $(0,0)$, $(1,0)$, $(0,1)$, and $(1,1)$, and which have the respective energies $\epsilon_0$, $\epsilon_1$, $\epsilon_2$, and $\epsilon_3$, which are defined in Eqs.~(\ref{eq:epsilon_0})-(\ref{eq:epsilon_3}).

Consider the case where both islands have equal charging energy, $E_{C,a}=E_C$.  When $t_j \neq 0$, the states $\beta=0$ and $3$, corresponding to quantum dot occupancies $(0,0)$ and $(1,1)$, do not hybridize.  The tunneling Hamiltonian allows fermions to tunnel into and out of the same MZM, resulting in the perturbed energies given by
\begin{eqnarray}
\label{eq:veps0}
\varepsilon^{\text{tot}}_{0} &=& \epsilon_{0} - \frac{1}{4}\left( \frac{|t_1|^2+|t_3|^2}{E_C+\epsilon_1 -\epsilon_{0}} +\frac{|t_2|^2+|t_4|^2}{E_C +\epsilon_2 -\epsilon_{0}}\right), \\
\varepsilon^{\text{tot}}_{3} &=& \epsilon_{3} - \frac{1}{4}\left( \frac{|t_1|^2+|t_3|^2}{E_C+\epsilon_1 -\epsilon_{3}} +\frac{|t_2|^2+|t_4|^2}{E_C +\epsilon_2 -\epsilon_{3}}\right),
\label{eq:veps3}
\end{eqnarray}
to leading order in $t_j/E_C$. These energies are clearly independent of the MZM parities.

In contrast, nonzero $t_j$ hybridizes the $\beta=1$ and $2$ states, corresponding to quantum dot occupancies $(1,0)$ and $(0,1)$. The second order perturbation theory Hamiltonian for these two states can be written as
\begin{equation}
\label{eq:HPTPauli}
H^{(0)}+H^{(2)}= B_0 \openone +B_x \sigma_x +B_y \sigma_y +B_z \sigma_z,
\end{equation}
where the Pauli matrices $\sigma_\mu$ act in the basis of the quantum dot states $(1,0)$ and $(0,1)$. We find diagonal elements
\begin{align}
B_0 &= \frac{\epsilon_1+\epsilon_2}{2} \nonumber
\\ &  -\frac{1}{8} \Big( \left( |t_1|^2+|t_3|^2 \right) \left( \frac{2E_C +\epsilon_0 +\epsilon_3 -\epsilon_1 -\epsilon_2}{\left( E_C +\epsilon_0 -\epsilon_1 \right) \left( E_C +\epsilon_3 -\epsilon_2 \right)} \right) \nonumber
\\ & +\left( |t_2|^2+|t_4|^2 \right) \left( \frac{ 2 E_C +\epsilon_0 +\epsilon_3 -\epsilon_1 -\epsilon_2}{\left( E_C +\epsilon_3 -\epsilon_1\right) \left( E_C +\epsilon_0 -\epsilon_2 \right)}\right) \Big)
\\ B_z &= \frac{\epsilon_1-\epsilon_2}{2} \nonumber
\\ &-\frac{1}{8} \Big( \left( |t_1|^2+|t_3|^2 \right) \left( \frac{\epsilon_3-\epsilon_2-\epsilon_0+\epsilon_1}{\left( E_C+\epsilon_0-\epsilon_1\right)\left(E_C+\epsilon_3-\epsilon_2\right)}\right) \nonumber
\\ &+\left( |t_2|^2+|t_4|^2 \right) \left( \frac{\epsilon_0-\epsilon_2-\epsilon_3+\epsilon_1}{\left( E_C+\epsilon_3-\epsilon_1\right)\left( E_C+\epsilon_0-\epsilon_2\right)}\right)\Big)
\end{align}
and off-diagonal matrix elements
\begin{align}
B_x &= \text{Re}\Big[p_{12} t_1 t_2^* +p_{34} t_3 t_4^* \Big]\times \nonumber
\\ & \quad \frac{1}{8}\Big( \frac{2 E_C +\epsilon_0+\epsilon_3-2\epsilon_2}{\left(E_C +\epsilon_0 -\epsilon_2 \right)\left( E_C +\epsilon_3 -\epsilon_2 \right)} \nonumber
\\ &\quad + \frac{2E_C +\epsilon_0+\epsilon_3 -2\epsilon_1}{ \left( E_C +\epsilon_0-\epsilon_1 \right) \left(  E_C +\epsilon_3 -\epsilon_1\right)} \Big)
\\ B_y &=\text{Im}\Big[p_{12} t_1 t_2^* +p_{34} t_3 t_4^* \Big]  \times \nonumber
\\ & \quad \frac{1}{8}\Big( \frac{2 E_C +\epsilon_0+\epsilon_3-2\epsilon_2}{\left(E_C +\epsilon_0 -\epsilon_2 \right)\left( E_C +\epsilon_3 -\epsilon_2 \right)} \nonumber
\\&\quad + \frac{2E_C +\epsilon_0+\epsilon_3 -2\epsilon_1}{ \left( E_C +\epsilon_0-\epsilon_1 \right) \left(  E_C +\epsilon_3 -\epsilon_1\right)} \Big).
\end{align}
The latter correspond to elastic co-tunneling processes mediated by different pairs of MZMs. The energy eigenvalues of Eq.~\eqref{eq:HPTPauli} are given by
\begin{eqnarray}
\label{eq:veps1}
\varepsilon^{\text{tot}}_{1} &=& B_0 - \sqrt{B_x^2+B_y^2+B_z^2},\\
\varepsilon^{\text{tot}}_{2} &=& B_0 + \sqrt{B_x^2+B_y^2+B_z^2}.
\label{eq:veps2}
\end{eqnarray}
Clearly, the parity dependence in these energies comes from $B_x^2+B_y^2$ and results in a term under the square root in Eqs.~(\ref{eq:veps1}) and (\ref{eq:veps2}) that is proportional to
\begin{equation}
|p_{12} t_1 t_2^* + p_{34} t_3 t_4^*|^2=|t_1|^2|t_2|^2+|t_3|^2|t_4|^2+2 p\, \text{Re}\left( t_1 t_2^* t_3^* t_4\right).
\end{equation}
Thus, the only MZM parity dependence of the energies is on the total parity $p=p_{12}p_{34}$ of the four MZMs, arising from fermions tunneling around the entire loop. In Appendix~\ref{app:MZMmodels}, we discuss the more general dependence of eigenvalues on parameters and compare results from exact diagonalization to the perturbative approximation of Eqs.~\eqref{eq:veps1} and \eqref{eq:veps2}.

In the lower panel of Fig.~\ref{fig:Energy}, we plot the eigenvalues of Eq.~(\ref{eq:Htot4}) as a function of the induced charge $n_{g,1}$ on the top quantum dot.  Notice that the parity dependence of $\varepsilon^{\text{tot}}_{1}$ and $\varepsilon^{\text{tot}}_{2}$ is strongest for $n_{g,1}=n_{g,2}$, where charge fluctuations are strongest. Experimentally it would therefore be best to tune to a regime where the $(1,0)$ and $(0,1)$ states are resonant (and lower in energy than the $(0,0)$ and $(1,1)$ states). The corresponding stability diagram for the ground state of the decoupled double dot system is shown in the upper panel of Fig.~\ref{fig:Energy}.

The energy dependence on the four-MZM joint parity $p$ could also be achieved with a single quantum dot.  The right panel of Fig.~\ref{fig:mst} can be modified by removing the lower dot and directly coupling MZMs $\gamma_2$ and $\gamma_4$. Such a system sacrifices some of the tunability of the double quantum dot system and could introduce complications from low-lying excited states in the semiconductor wire segment connecting $\gamma_2$ and $\gamma_4$. Nonetheless, if a single-dot system were substantially easier to realize, it could prove to be more advantageous to achieve the same projective measurement of four-MZM parity in this way. Similarly, the two-MZM parity measurements of Section~\ref{sec:2MZMmodel} could also be performed using two quantum dots instead of one.

\begin{figure}
	\includegraphics[width=0.95\columnwidth]{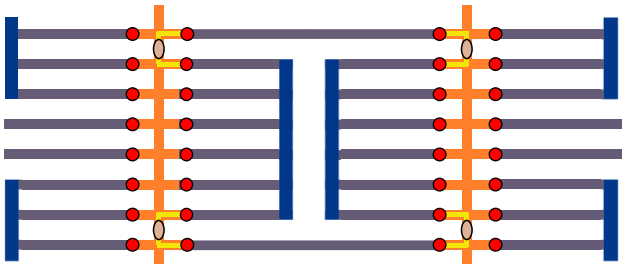}
	\caption{An example configuration for a joint parity measurement of 8 MZMs involving two one-sided hexons and two coherent links (using the same legend as Fig.~\ref{fig:mst}). Four of the MZMs involved in the measurement are associated with coherent links and are used to facilitate the measurement of the other four MZMs, which are associated to the hexons. The resulting measurement can provide a two-qubit entangling operation on the two hexons.}
	\label{fig:nMZMmst}
\end{figure}

\begin{figure}[t!]
	\includegraphics[width=0.8\columnwidth]{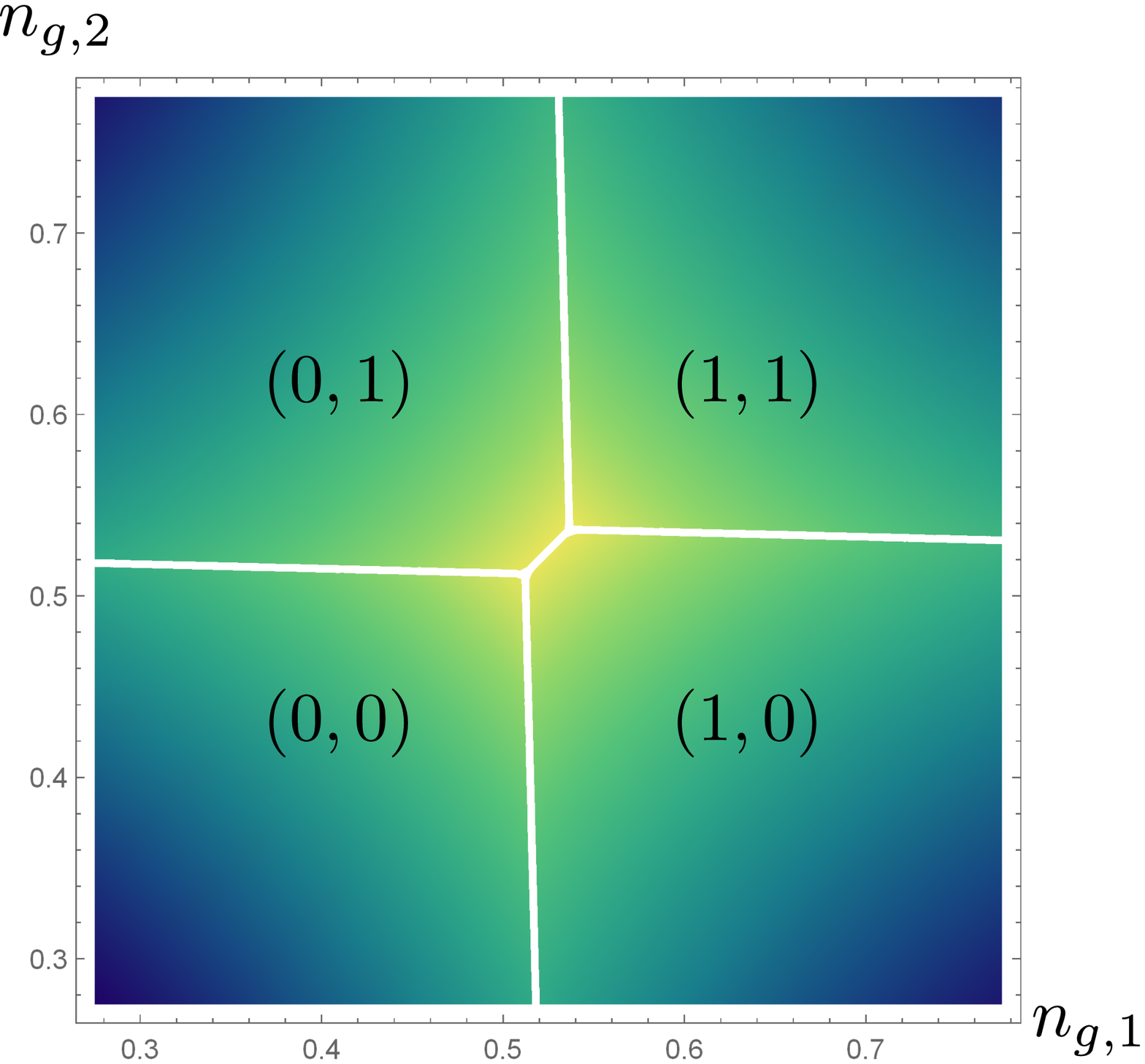}

\vspace{3mm}
	\includegraphics[width=0.95\columnwidth]{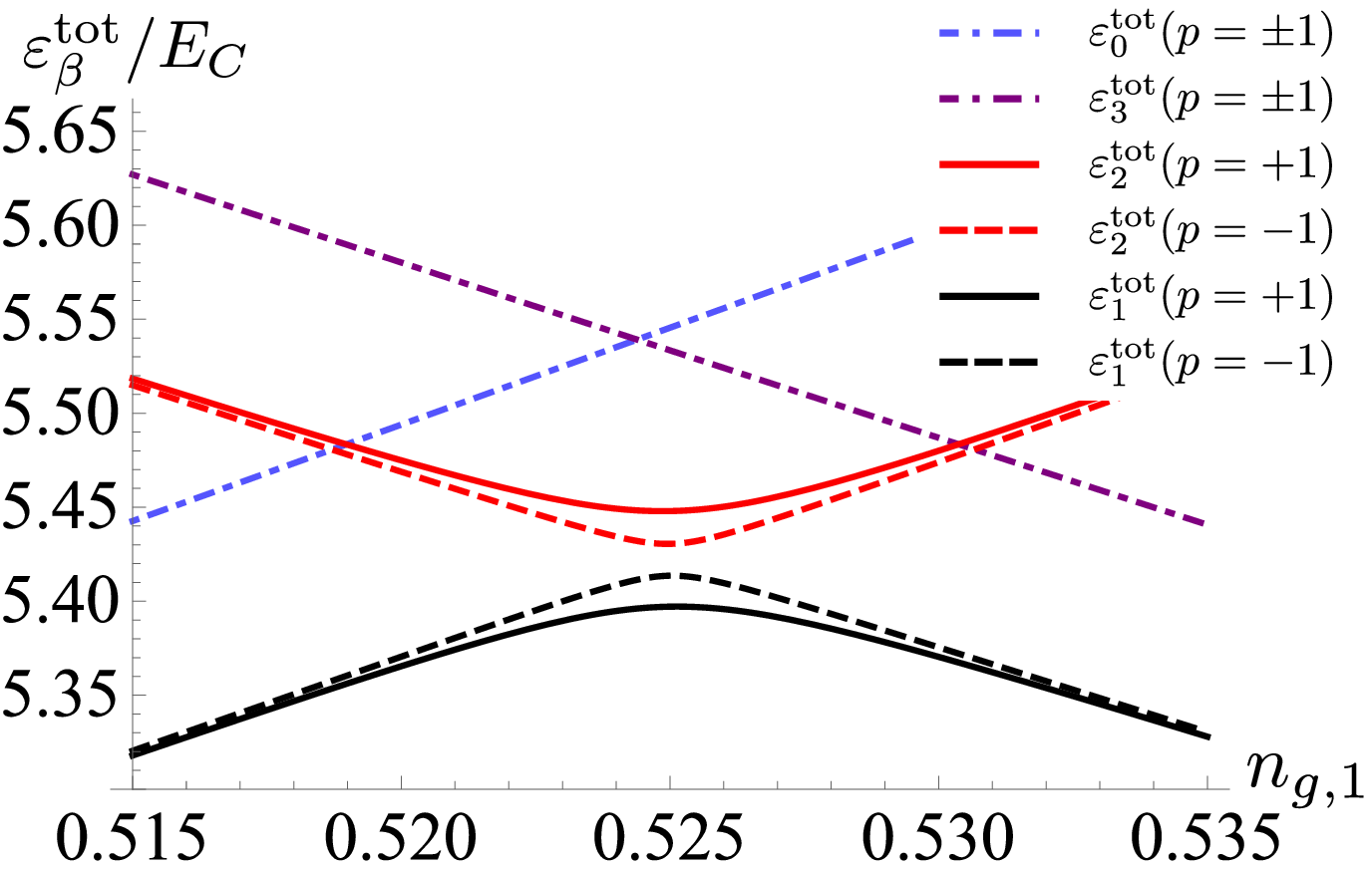}
	\caption{Energy as a function of dimensionless induced charges on the quantum dots for the system shown in the right panel of Fig.~\ref{fig:mst}.  {\it Top panel}: Stability diagram for the decoupled system ($t_j=0$) as a function of the occupation numbers $(n_{f,1},n_{f,2})$ of the double quantum dot system in the ground state. The color scale refers to the ground state energy, whose precise values away from zero (indicated by white) are unimportant for the current discussion. {\it Bottom panel}: The four lowest energies $\varepsilon_{\beta}^{\text{tot}}/E_C$ as a function of $n_{g,1}$ for $n_{g,2}=(1+h/\varepsilon_C)/2$ with  tunneling amplitudes $t_1=0.1 E_C$ and $t_{j\neq 1}=0.2 E_C$. We use the parameter values $N_{g,a}=0$, $\varepsilon_C=10 E_C$, $h=  E_C/2$, and $\varepsilon_M=E_C/2$. For non-vanishing tunneling amplitudes, the quantum dot states $(1,0)$ and $(0,1)$ hybridize.  The symmetric combination of the $(1,0)$ and $(0,1)$ states has energy $\varepsilon^{\text{tot}}_2$ (shown in red) and the antisymmetric combination has energy $\varepsilon^{\text{tot}}_1$ (shown in black). These energies $\varepsilon^{\text{tot}}_1$ and $\varepsilon^{\text{tot}}_2$ depend on the joint parity $p$ of the four MZMs; the solid curves correspond to even parity $p=1$ and dashed curves to odd parity $p=-1$.  As our model only considers two quantum dot levels, the states $(0,0)$ and $(1,1)$ do not hybridize.  These states have corresponding  parity-independent energies $\varepsilon^{\text{tot}}_0$ (shown as the blue dot-dashed curve) and $\varepsilon_3^{\text{tot}}$ (shown as the purple dot-dashed curve), respectively.  From the stability diagram (top panel), we see that the mutual charging energy $\varepsilon_M$ increases the range of $n_{g,1}$ and $n_{g,2}$ for which the parity-dependent energy $\varepsilon^{\text{tot}}_{1}$ is the ground state.}
	\label{fig:Energy}
\end{figure}

Finally, the above analysis is easily generalized to measure the joint parity of any even number of MZMs.  { Whenever gate voltages are tuned such that the tunneling connections create a single closed loop path for electrons that traverses $2n$ MZMs, the energy of the system will depend on the $2n$-MZM parity.  An example configuration for a multiple-MZM measurement using an array of one-sided hexons is shown in Fig.~\ref{fig:nMZMmst}.  In practice, the measurement visibility will decrease with each additional MZM pair, so it is important to utilize measurements involving the smallest number of MZMs possible.}

\subsection{Experimental proposals for MZM parity measurements}
\label{sec:measurement}

The parity-dependent energy shift, discussed in the previous subsections, can be observed using energy level spectroscopy, quantum dot charge, or differential capacitance measurements. We now briefly discuss these different measurements and consider specific proposals which differ in their speed and sensitivity to noise. Each such measurement is designed to project the system to a definite parity state (of
two MZMs in Section~\ref{sec:2MZMmodel} and four MZMs in Section~\ref{sec:4MZMmodel}).
For concreteness, we focus henceforth on the four-MZM case (right panel of Fig.~\ref{fig:mst}); the discussion generalizes straightforwardly to the two-MZM case (left panel of Fig.~\ref{fig:mst}).

We assume the double quantum dot system is properly tuned such that the relevant states are those sensitive to the parity of the MZMs, that is, the relevant states have one electron shared between the two quantum dots.  Moreover, we focus on the regime in which the system has only weakly occupied excited states, so that the system can be described by the ground state with corresponding energy $\varepsilon^{\rm tot}_1$. When the double dot system is tuned close to resonance, the gap to the lowest excited state is of the order of $|t|^2/E_C$ when $|t_j|\sim |t|$. In order to have an appreciable difference between the occupation of the ground and excited states, we require that $T \ll |t|^2/E_C$~\footnote{For the parameters of Fig.~\ref{fig:Energy}, with $E_C=1~$K, the gap is of the order of $50~$mK, which would suffice for typical temperatures $T\sim 20~$mK.}. Away from resonance, the condition on temperature can be relaxed at the cost of reducing the visibility (see Fig.~\ref{fig:Energy}). Similarly, finite temperature effects are negligible for single quantum dot measurements in which the first excited state is separated from the ground state by an energy on the order of $E_C$.

Let us first consider energy level spectroscopy. The dependence of the ground-state energy on parity is shown in Fig.~\ref{fig:Energy}.  One possible spectroscopic measurement is done by coupling the system (MZM island and quantum dots) to a superconducting transmission line resonator.  The resonator frequency will have a parity-dependent frequency shift $\Delta\omega$ which can be detected using the reflectometry technique~\cite{Majer07}. We find that for the four-MZM device discussed in Section~\ref{sec:4MZMmodel}, the frequency shift is given by
\begin{equation}
\label{eq:transmonfreq}
 \Delta\omega \sim \frac{g^2}{4\delta\omega^2}\frac{t^2}{E_C},
\end{equation}
where $g$ is the coupling between the resonator and the quantum dot and $\delta\omega$ is the detuning, i.e., the frequency difference between $\varepsilon^{{\rm tot}}_{2}-\varepsilon^{{\rm tot}}_{1}$ at the degeneracy point $n_g^*$ and the resonator frequency. Here, we have assumed that all of the tunneling matrix elements
are comparable to $t$ (see Appendix~\ref{app:transmon} for details). Using realistic parameters defined in Fig.~\ref{fig:Energy}, frequency estimates given in Ref.~\onlinecite{Knapp16}, and ${E_C=160}$~$\mu$eV (see Ref.~\onlinecite{Albrecht16}), we estimate $\Delta\omega \sim 100$~MHz. This frequency shift falls well within the range of transmon sensitivity. Spectroscopy with a transmission line resonator benefits from a fairly short measurement time on the order of $1 \mu$s. However, the resonator will have to operate in large magnetic fields, so one would need to adapt this technology to such conditions.

The main drawback of this proposal is that while this measurement technique is suitable for a small number of qubits, it may become problematic when scaling to a two-dimensional array of qubits. This is because the resonators need to be taken off the plane containing the topological qubits, since there is no room
for them in the planar layout shown in Fig.~\ref{fig:1}. Coupling out-of-plane resonators to qubits is an open experimental problem.

\begin{figure}
	\includegraphics[width=0.9\columnwidth]{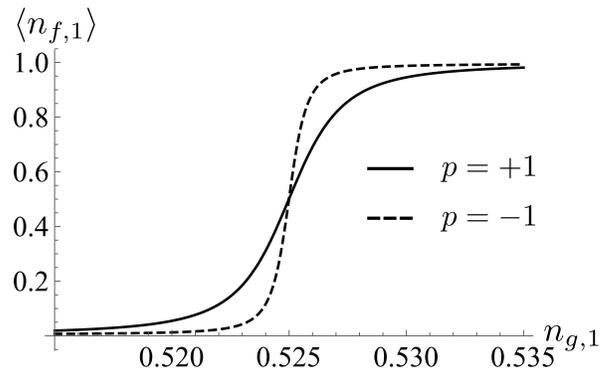}
	\caption{Average charge (in units of electron charge) on the upper quantum dot as a function of the dimensionless induced charge $n_{g,1}$ for the system shown in the right panel of Fig.~\ref{fig:mst}.  We assume the system is in the ground state, and plot the average charge for both even parity (solid curve) and odd parity (dashed curve). We use the parameter values $N_{g,a}=0$, $\varepsilon_C=10 E_C$, $h=  E_C/2$, and $\varepsilon_M=E_C/2$.}
	\label{fig:Charge}
\end{figure}

Another way of performing a joint parity measurement is to detect the average charge on a quantum dot. Indeed, the charge $n_{f,1}$ on the upper dot is related to the energy by
\begin{equation}
\langle n_{f,1} \rangle \approx n_{g,1} - \frac{1}{2 \varepsilon_C} \left( \frac{\partial E_{\text{GS}}}{\partial n_{g,1}}- \frac{\varepsilon_M}{2 \varepsilon_C } \frac{\partial E_{\text{GS}}}{\partial n_{g,2}}\right),
\end{equation}
where $E_{\text{GS}}$ denotes the ground state energy of the system. In this expression, we have neglected $O(\varepsilon_M^2/\varepsilon_C^2)$ terms. The dependence of the average charge on the joint parity of MZMs is shown in Fig.~\ref{fig:Charge}. Quantum charge fluctuations broaden the step function in a manner that depends on the joint fermion parity of MZMs. Hence, measurement of the charge on the dot allows one to distinguish different parity states.  Given that the average charge on the dot can be measured very accurately at low temperatures, i.e., up to roughly $10^{-3} \, e /\sqrt{\tau_{\text{int}}}$ where $\tau_{\text{int}}$ is the integration time~\cite{Lehnert03, Reilly07, Barthel10} , we believe that our predictions are within experimental reach. Charge measurements are very fast and accurate. This technique is well understood in the semiconductor community and is compatible with large magnetic fields.  While the inclusion of SETs in the qubit plane makes the design somewhat more complicated, it does not preclude scaling the system up to a two-dimensional array of qubits.

Finally, we discuss the third proposal -- a differential capacitance (also referred to as the quantum capacitance) measurement~\cite{Duty05, Ota10, Petersson10,Colless13}. The differential capacitance of the upper quantum dot is given by
\begin{equation}
\label{eq:CQ}
\frac{C_{\text{diff}}}{C_{\Sigma,D}} =-\left(\frac{C_g}{C_{\Sigma,D}}\right)^2 \frac{\partial(\langle n_{f,1}\rangle-n_{g,1})}{\partial n_{g,1}},
\end{equation}
where $C_g$ is the capacitance between the gate and the upper quantum dot, and ${C_{\Sigma,D}\equiv e^2/2\varepsilon_C}$ is the total capacitance of the dot. When the system is tuned close to resonance of the two quantum dots, the energy is sensitive to changes in $n_{g,1}$, making the differential capacitance become appreciable. We can use rf-reflectometry to measure the differential capacitance of the upper
quantum dot by coupling the gate voltage $V_{g,1}= e n_{g,1}/C_{g}$ directly to an LC circuit.  The circuit's resonant frequency will depend on the differential capacitance, which, in turn, depends on the joint parity of the four MZMs.  Thus, the reflection of an rf-signal sent through the circuit can be analyzed to infer the parity state of the system. The frequency of the rf-signal will have to be properly engineered. If the frequency of the rf-signal is lower than the excitation gap near the resonance (i.e.,  near ${n_{g,1}=n_{g}^*}$, which is the location of the anti-crossing in Fig.~\ref{fig:Energy}), the system will remain in the ground state, and the differential capacitance will contain information about the ground state curvature at this point. However, if the frequency is too large, the system will undergo a Landau-Zener transition at the resonance (transitioning from one of the lower curves in Fig.~\ref{fig:Energy} to one of the upper ones), and the reflected signal will not contain information about the ground state curvature, resulting in a vanishingly small differential capacitance. Since differential capacitance is peaked at the degeneracy point, thermal fluctuations or gate-voltage fluctuations will broaden the signal. In order to suppress the effect of thermal fluctuations, we require that $|t|^2/E_C \gg T$. Provided this broadening is smaller than the parity-dependent differential capacitance difference, the projective measurement can be efficiently performed.

\begin{figure}
	\includegraphics[width=0.9\columnwidth]{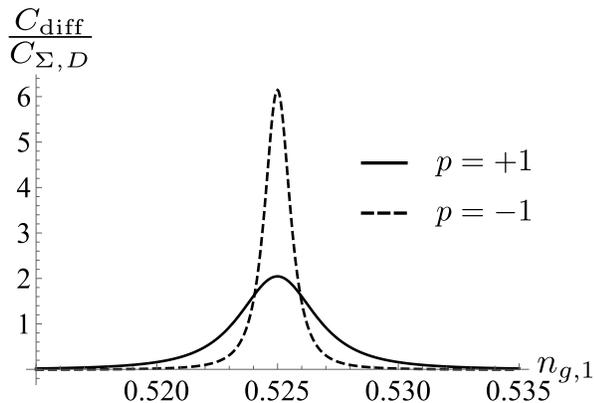}
	\caption{Differential capacitance of the ground state $C_{\text{diff}}$ (in units of $C_{\Sigma,D}$) as a function of the dimensionless induced charge $n_{g,1}$ for the system shown in the right panel of Fig.~\ref{fig:mst}.  Both even parity (solid curve) and odd parity (dashed curve) are shown. We use the parameter values $N_{g,a}=0$, $\varepsilon_C=10 E_C$, $h=  E_C/2$, $\varepsilon_M=E_C/2$, and $C_g/C_{\Sigma,D}=0.1$.}
	\label{fig:CQ}
\end{figure}

Assuming that the quantum dot charging energy $\varepsilon_C\sim 1-10$~K, which corresponds to the total capacitance  ${C_{\Sigma,D}\sim 10^2-10^3}$~aF, the change of the differential capacitance for different parity states should be ${\delta C_{\rm diff}\sim 10^2-10^3}$~aF (see Fig.~\ref{fig:CQ}).
Note that this curve is the derivative of the charge as a function of the $n_{g,1}$ curve shown in Fig.~\ref{fig:Charge}. That is, it involves the second derivatives (rather than the first derivatives) of the energy with respect to $n_{g,1}$ and $n_{g,2}$. The curves in Fig.~\ref{fig:CQ} are peaked
where the curves in Fig.~\ref{fig:Charge} are steepest.
Reflectometry experiments in quantum dot systems have measured differential capacitances of the order of 10~aF in 40~$\mu$s~\cite{Colless13}. Therefore, we believe that the joint parity state should be measurable through the differential capacitance even when the tunnel couplings are not optimized. The gates needed for the reflectometry measurement are already necessary in the system in order to define the quantum dots (see Fig.~\ref{fig:1}), and the LC circuits are can be moved off the plane of the MZM islands.

{ Both charge-sensing and reflectometry detection of differential capacitance have the attractive feature of being measurements of ground state properties.  Up to exponentially small thermal corrections, there is no decoherence in the ground state; as such, the visibility of these measurements will not decrease significantly over time.  We elaborate on this statement in Appendix~\ref{app:dephasing} for the charge measurement. }

\section{Clifford-complete Majorana architectures}
\label{sec:architectures}

We now show how the projective measurements of the previous section may be used in combination with MZM-based qubits to implement the complete set of multi-qubit Clifford gates in a topologically protected manner.

\subsection{Hexon architectures}

In this section, we describe the three different hexon architectures, an example of which is shown in Fig.~\ref{fig:1}.  Six is the smallest number of MZMs that supports the combination of one computational qubit (encoded in four of the MZMs) and one ancillary pair of MZMs.  This combination is particularly useful because the presence of the ancillary pair makes it possible to generate the braiding transformations of the topological qubit without physically transporting the MZMs.  That is, sequences of topological charge measurements can generate the braiding transformations on the qubit states encoded in the MZMs~\cite{Bonderson08b, Bonderson08c}.
The topological charge of an even number of MZMs is their joint electron number parity.
In this paper, we focus on measurement-based protocols.  However, the braiding transformations can equivalently be performed using similar methods that instead utilize adiabatic tuning of couplings between MZMs~\cite{Bonderson13a} or hybrid protocols that use both nearly-adiabatic tuning and measurement~\cite{Knapp16}. Furthermore, an entangling gate can be implemented with the addition of a joint parity measurement of four MZMs from neighboring hexons, two MZMs from each hexon.  Thus, by using the hexon together with the ability to perform joint parity measurements, one generates all multi-qubit Clifford gates with topological protection, while simultaneously protecting the qubit from QPP errors.

\subsubsection{Quantum information basics}
\label{sec:hexon_qi_basics}

\begin{figure}
\includegraphics[width=0.55\columnwidth]{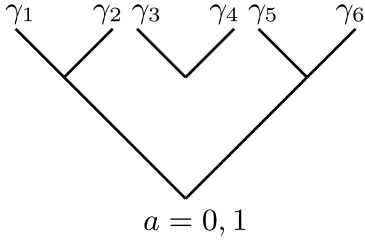}
              \caption{Diagrammatic representation of the topological states (degenerate ground states) of a hexon. The center two MZMs $\gamma_3$ and $\gamma_4$ fuse to even fermion parity, forming the ancillary pair of MZMs.  The left and right pairs of MZMs both fuse to $a=0$ or $1$, which correspond to even or odd fermion parity, respectively.  These outer pairs of MZMs form the computational qubit.  The fusion channel $a$ labels the qubit basis state.}
              \label{fig:qubit}
\end{figure}

The full set of single-qubit Clifford gates can be generated on the computational qubit encoded in a single hexon given an appropriate minimal set of joint parity measurements of pairs of MZMs. We can diagrammatically represent the topological state of a hexon as shown in Fig.~\ref{fig:qubit}. We label the MZMs $\gamma_j$ with $j=1,\ldots,6$ from left to right. The diagram may be interpreted as follows: The center two MZMs $\gamma_3$ and $\gamma_4$, forming the ancillary pair, fuse to even fermion parity ($p_{34} =-1$).  The left-most and the right-most pairs of MZMs, $\gamma_1$ and $\gamma_2$, and $\gamma_5$ and $\gamma_6$, respectively, forming the computational qubit, have the same fusion channel $a=0$ (even fermion parity) or $1$ (odd fermion parity).  That is, the fusion channel $a$ labels the qubit basis states
{ \begin{eqnarray}
|0\rangle &=& |p_{12} =p_{56} =-1\rangle  \\
|1\rangle &=& |p_{12} = p_{56} =+1\rangle
.
\end{eqnarray}}
The total fusion channel of the four MZMs forming the computational qubit is even fermion parity ($p_{12}p_{56} =1$).

In Section~\ref{sec:measurement}, we explained how joint fermion parity measurements may be implemented using dispersive transmon measurements, charge-sensing, or reflectometry.  While the outcomes of quantum measurements are inherently probabilistic, for our purposes, we can use a ``forced-measurement'' protocol~\cite{Bonderson08b} to obtain the desired measurement outcome of a particular step of the measurement-only protocol. This is a repeat-until-success protocol involving alternating measurements between the pair of MZMs that is to become ancillary and the pair that was ancillary, until the desired outcome is achieved. As such, the encoded computational state information is preserved and this allows us to think in terms of projectors, rather than projective measurements.

\begin{figure}
              \includegraphics[width=0.55\columnwidth]{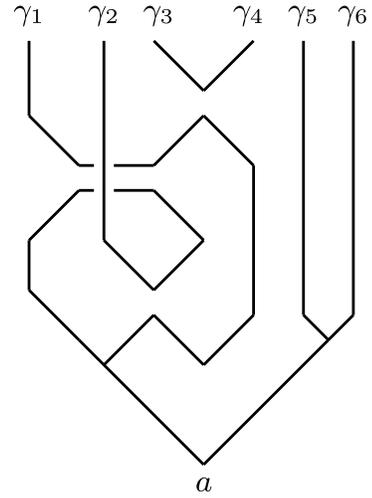}
              \caption{Diagrammatic representation of $\Pi_0^{(34)}\Pi_0^{(13)}\Pi_0^{(23)}\Pi_0^{(34)}$ applied to the topological state of a hexon qubit.  Pairs of MZMs are projected to the vacuum (even fermion parity) fusion channel to perform anyonic teleportations on the topological state space of the MZMs.  The series of projections has the same effect as exchanging the positions of MZMs 1 and 2, i.e., it generates the braiding operator $R^{(12)}$. This provides a diagrammatic proof of Eq.~(\ref{eq:R12}), as originally given in Ref.~\onlinecite{Bonderson08b}.}
              \label{fig:R12}
\end{figure}

Let $\Pi_0^{(jk)}= \frac{1 - i \gamma_j \gamma_k}{2}$ project MZMs $j$ and $k$ to the vacuum (even fermion parity) channel.  Braiding operations can be implemented through the application of a series of such projectors.  For instance, the following sequence of projections generates the braiding transformation corresponding to exchanging the first and second MZMs
\begin{equation}
\label{eq:R12}
\Pi_0^{(34)}\Pi_0^{(13)}\Pi_0^{(23)}\Pi_0^{(34)} \propto R^{(12)} \otimes \Pi_0^{(34)},
\end{equation}
where ${R^{(12)}\equiv (1 + \gamma_1\gamma_2)/\sqrt{2}}$ is the braiding transformation for exchanging MZMs 1 and 2. Whether the operator $R^{(12)}$ describes a clockwise or counterclockwise exchange of the MZMs is a matter of convention since the $\gamma_i$ operators can be changed by a sign via a gauge transformation. Here, we define it as a counterclockwise exchange as diagrammatically represented in Fig.~\ref{fig:R12}. This choice determines whether the projector $\Pi_0^{(13)}$ is interpreted in the diagrammatic representation as an over-crossing or under-crossing with respect to the $\gamma_2$ charge line.

We note that the above convention should be fixed with respect to a particular measurement setup (defined by the complex couplings $t_j$ of MZMs to quantum dots). The effect of a change of the measurement setup during the calculation (e.g., by deciding to measure a certain pair of MZMs differently than in the initial definition) can be tracked by a bookkeeping of phase changes~\cite{Alicea11,Freedman11,Halperin12}.

While Eq.~\eqref{eq:R12} has the elegant diagrammatic representation shown in Fig.~\ref{fig:R12}, which makes the relation to braiding apparent, it can also be derived explicitly in terms of Majorana operators
\begin{eqnarray}
&& \!\!\!\!\!\!\!\! \Pi_0^{(34)}\Pi_0^{(13)}\Pi_0^{(23)}\Pi_0^{(34)} \notag \\
&=& \frac{1 - i \gamma_3 \gamma_4}{2} \frac{1 - i \gamma_1 \gamma_3}{2} \frac{1 - i \gamma_2 \gamma_3}{2} \frac{1 - i \gamma_3 \gamma_4}{2} \notag \\
&=&2^{-4} (1 - i \gamma_3 \gamma_4) (1 - i \gamma_1 \gamma_3 - i \gamma_2 \gamma_3 + \gamma_1 \gamma_2) (1 - i \gamma_3 \gamma_4) \notag \\
&=& 2^{-4} (1 + \gamma_1 \gamma_2) (1 - i \gamma_3 \gamma_4)^2\notag \\
&=& 2^{-3} (1 + \gamma_1 \gamma_2)  (1 - i \gamma_3 \gamma_4) \notag \\
&=& 2^{-3/2} R^{(12)} \otimes \Pi_0^{(34)},
\end{eqnarray}
where we used the fact that $(1 - i \gamma_3 \gamma_4) i \gamma_a \gamma_3 (1 - i \gamma_3 \gamma_4) = i \gamma_a \gamma_3(1 + i \gamma_3 \gamma_4) (1 - i \gamma_3 \gamma_4)  = 0$ for $a\neq 3,4$. The way the projectors are written in terms of MZMs is again a choice of convention.

A sufficient gate set for generating all single-qubit Clifford gates is given by the two (intra-hexon) braiding transformations, $R^{(12)}$ and $R^{(25)}$, which, up to an overall phase, respectively correspond to the computational gates
\begin{eqnarray}
R^{(12)} &=& \left(
\begin{array}{cc}
1 & 0 \\
0 & -i
\end{array}
\right) , \\
R^{(25)} &=& \frac{1}{\sqrt{2}}\left(
\begin{array}{cc}
1 & i \\
i & 1
\end{array}
\right)
,
\end{eqnarray}
using the qubit basis. Note that the Hadamard gate is given by ${H=R^{(12)} R^{(25)} R^{(12)}}$.

The braiding transformation $R^{(25)}$ may be implemented using the following sequence of projections
\begin{equation}\label{eq:R25}
\Pi_0^{(34)} \Pi_0^{(35)} \Pi_0^{(23)} \Pi_0^{(34)} \propto R^{(25)}\otimes \Pi_0^{(34)}.
\end{equation}

In order to have a multi-qubit Clifford-complete gate set, we only need to add the ability to perform an entangling two-qubit Clifford gate between neighboring computational qubits. Similarly labeling the MZMs of a second hexon by $j=7,\ldots,12$, we find that the following sequence of projective parity measurements on two and four MZMs in two hexons
\begin{equation}
\label{eq:Xop}
\Pi_0^{(34)} \Pi_0^{(35)} \Pi_0^{(5678)} \Pi_0^{(45)}\Pi_0^{(34)} \propto W^{(5678)} \otimes \Pi_0^{(34)},
\end{equation}
generates $W^{(5678)}\equiv \left(1+i\gamma_5\gamma_6\gamma_7\gamma_8\right)/\sqrt{2}$.  In terms of the topological qubit basis states of the two hexons, this yields the two-qubit entangling Clifford gate
\begin{equation}
\label{eq:Xdef}
W =\left(
\begin{array}{cccc}
1 & 0 & 0 & 0 \\
0 & i & 0 & 0 \\
0 & 0 & i & 0 \\
0 & 0 & 0 & 1 \\
\end{array}
\right)
,
\end{equation}
up to an overall phase. Proofs of Eq.~(\ref{eq:Xop}) in terms of both the diagrammatic calculus and the MZM operators are given in Appendix~\ref{app:Hexdetails}. Note that the controlled-$Z$ gate is given by ${C(Z) = R^{(12)} R^{(78)} W^{(5678)}} $.

We emphasize that this two-hexon operation respects the fermion parity of each hexon separately, so it is compatible with the protection from QPP afforded by the Coulomb charging energy on each superconducting island. We also note that, as long as one is able to perform the appropriate measurements of MZMs, one ancillary pair of MZMs on an island is sufficient for implementing entangling gates between two computational qubits on separate islands (i.e.,  between two hexon qubits), without the need of extra ancillary MZMs. In the above example, the only ancillary pair needed was MZMs 3 and 4.

Taken together, Eqs.~(\ref{eq:R12}), (\ref{eq:R25}), and (\ref{eq:Xop}) reveal a sufficient set of measurements that allow us to generate all multi-qubit Clifford gates.
While the operational efficiency may be improved if we are able to perform measurements on additional groups of MZMs, practical constraints may limit which groups of MZMs we can jointly measure, as we will discuss below. In our proposed hexon architectures, we find that we are always able to perform measurements that are Clifford complete.

The three different hexon architectures that are described in the following subsections
support several additional operations that make computations more efficient.  For instance, it is convenient to be able to shuttle the computational MZMs, so that they are adjacent to each other.  This shuttling can be achieved with a series of projective fermion parity measurements of pairs of MZMs, as shown in Appendix~\ref{app:Hexdetails}.  From the Gottesman-Knill theorem, we know that Clifford operations can be efficiently modeled on a classical computer.  This can be used to transfer some of the computation from quantum operations, such as those described above, to classical simulation, by appropriately keeping track of which gates have been performed.  These ``Pauli frame changes'' are discussed in further detail for the tetron architectures in Section~\ref{sec:Tetrons}.

\subsubsection{One-sided Hexon}
\label{sec:combs}

\begin{figure}
	\includegraphics[width=0.85\columnwidth]{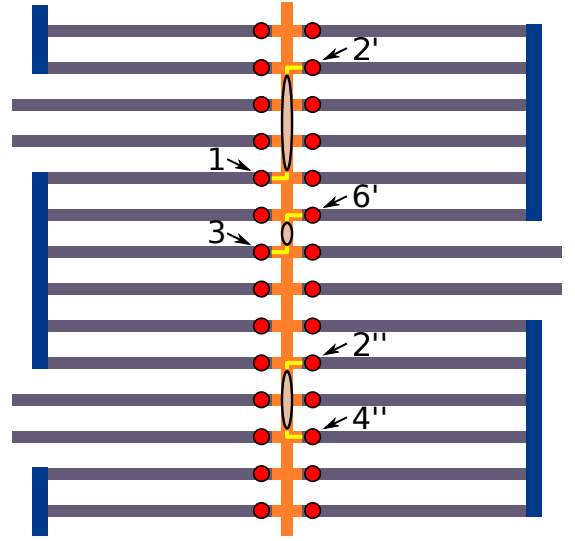}
	\caption{An example generalizing the measurements of MZMs from Fig.~\ref{fig:mst} (using the same legend). The upper region shows a four-MZM measurement of the joint parity operator $-\gamma_1\gamma_3\gamma_{2'}\gamma_{6'}$. The lower region shows a two-MZM measurement of the parity operator $i\gamma_{2''}\gamma_{4''}$. The quantum dots (gray ellipses) and their couplings (yellow lines) to MZMs are defined by appropriately tuning a set of underlying gates (see Fig.~\ref{fig:1}).  { Note: the illustration is not drawn to scale.  In practice, the length (horizontal direction on the figure) is much larger than the width $L \gg w$, so as to simultaneously optimize topological protection due to the length of the 1DTSs and suppression of QPP error rates by large charging energies.  As a practical constraint, in order for the quantum dots connecting MZMs to remain coherent, the vertical separation of the MZMs connected to the same quantum dot must be shorter than the effective coherence length of that quantum dot.  The same principles apply to subsequent figures. }
}
	\label{fig:operations}
\end{figure}

The main operational principles of one-sided hexon architectures are discussed in Section~\ref{sec:overview}. Here, we provide further details. Figure~\ref{fig:operations} gives examples of defining connections and quantum dots in the semiconducting structure that is coupled to MZMs for possible two-MZM measurements and four-MZM measurements. With obvious generalizations of the depicted two-MZM measurement, it is possible to measure the parity of an arbitrary pair of MZMs inside a hexon. Together with a set of four-MZM measurements between neighboring hexons, this design allows for more than enough measurements to achieve Clifford completeness (see Section~\ref{sec:hexon_qi_basics}).

{ Measurement of the joint parity of vertically-separated MZMs places a practical constraint on the width $w$ of the one-sided hexon (i.e., the length of the the backbone).  A quantum dot coupled to the top and bottom MZMs of a given hexon must remain coherent for the measurement to be successful.  Thus, the width of the one-sided hexon must be smaller than the effective coherence length of the quantum dots.  Furthermore, as discussed in Section~\ref{sec:overview}, simultaneously optimizing charging energy and suppressing hybridization of the computational MZMs (i.e., $L\gg \xi$) implies that it is beneficial to design the one-sided hexon so that it is much longer than it is wide ($L\gg w$).  The same two principles apply to all qubit designs presented in this paper.  For ease of illustrating the important features of the qubit designs, the corresponding figures are not drawn to scale.

The one-sided hexon has an additional constraint on the width compared to the alternative hexon designs presented in the following sections: the 1DTSs should have vertical separation less than $\xi$ in order to strongly hybridize the MZMs at the backbone side of the device.  When this condition is satisfied, the one-sided hexon provides topological protection corresponding to MZM separation distances of $2L$ for 1DTSs of length $L$.  This property should also enable the one-sided hexon design to realize a better optimal combination of topological protection and protection from QPP granted by the charging energy.  As discussed in Section~\ref{sec:overview}, we roughly expect the charging energy of a hexon to have $1/L$ dependence for $L \gg w$. Garnering topological protection for MZM separations of $2L$ makes it more endurable to decrease $L$ for the trade-off of increasing the charging energy and its corresponding QPP protection.  Another potential trade-off involved in decreasing $L$ is a reduced visibility for MZM parity measurements. This is because the MZM parity dependent terms in the shifted ground state energies of the hexon coupled to quantum dots depend inversely on excited state energies that increase as $L$ decreases, { see e.g., Eqs.~(\ref{eq:eps1}) and (\ref{eq:eps2}).}
}

A possible challenge for one-sided hexons could arise if the energy splitting due to hybridization of the MZMs at the backbone side of the wires is small for some reason. When the device has the
$\mathcal{T}^2 = +1$ symmetry mentioned at the end of Section~\ref{sec:2MZMmodel}, these energy splittings will vanish.
Generically, this symmetry is not present, but it can occur when the cubic Rashba couplings vanish
and the Zeeman field is perfectly aligned with the wires. If the symmetry is only weakly broken, then
some of these energy splittings will be small. When the energy splittings are smaller than the temperature, there will be fluctuating low energy degrees of freedom in the superconducting backbone. For the purpose of protecting the information stored in MZMs at the non-backbone side of the hexon, the relevant length scale for topological protection is then reduced from $2L$ to $L$, the distance separating the MZMs from the backbone. Similar arguments would apply for low energy states induced by disorder at the backbone-1DTS interface. We briefly return to this issue in
Section~\ref{sec:Conclusions}.

\subsubsection{Two-sided hexon}

\begin{figure}
	\includegraphics[width=0.95\columnwidth]{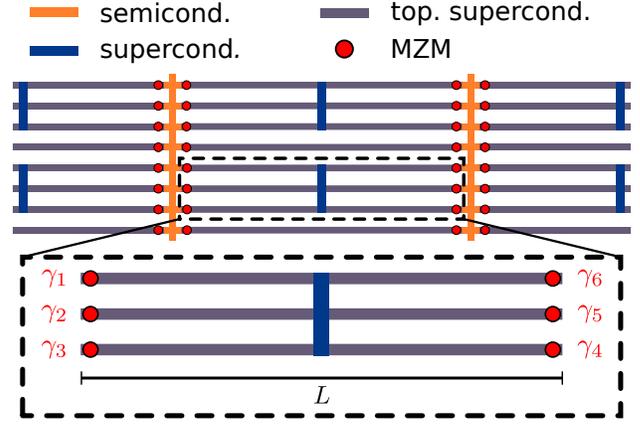}
	\caption{A two-sided hexon architecture. { Note: the illustration is not drawn to scale for the same reason as Fig.~\ref{fig:operations}.} The magnification shows a single two-sided hexon. Additional topological superconducting links and semiconducting structures allow appropriate measurements to manipulate and entangle two-sided hexons.}
	\label{fig:insects}
\end{figure}

A two-sided hexon consists of three 1DTSs joined by a superconducting backbone, as depicted schematically in the magnification of Fig.~\ref{fig:insects}. In contrast to the one-sided hexon, the backbone is located far away from the MZMs at the ends of the 1DTS. Similar to the one-sided hexon, it is straightforward to measure the parity of any pair of the three MZMs at a given side (left or right) of the two-sided hexon. However, achieving single qubit Clifford completeness requires the ability to measure the parity for at least two distinct pairings of MZMs involving one MZM from each side of the hexon. For example, enumerating the MZMs 1-6 as shown in Fig.~\ref{fig:insects}, we see that Eqs.~\eqref{eq:R12} and \eqref{eq:R25} utilize the measurements $\Pi_0^{(34)}$ and  $\Pi_0^{(35)}$. Due to the large distance $L$ between the MZMs on the left and right sides of the hexon, such measurements require long coherent links between both sides. These can be provided by floating topological superconductors, as in the case of the inter-hexon links in the one-sided design. Due to the connectivity of all the MZMs to the semiconducting structure at each corresponding side of the two-sided hexon, adding a single link of length $L$ to each hexon is sufficient to perform arbitrary two-MZM measurements within the hexon.

Entangling four-MZM measurements between horizontally adjacent two-sided hexons can be implemented in a manner similar to those in the one-sided hexon case (cf. Fig.~\ref{fig:operations}). For vertically adjacent two-sided hexons, entangling operations could be performed by defining a quantum dot in each of the semiconducting structures to the left and to the right of the hexons and connecting each dot to each of the two hexons. To avoid unwanted two-MZM measurements, each dot at the left side should have exactly one connection to the left side of each of the hexons involved, and similarly each dot at the right side should have exactly one connection to the right side of each hexon.

The main differences from the one-sided hexon designs are that the connectivity graph of the hexon qubits (linking pairs that can be directly acted on by entangling operators) is now rectangular (4-valent), rather than hexagonal, and that the relevant distance for MZM hybridization is $L$, rather than $2L$. In order to attain the same level of topological protection, two-sided hexons will generally be more elongated than their one-sided counterparts. We therefore expect the two-sided hexons to have a smaller charging energy (roughly half as large) than the one-sided hexons, for the same level of topological protection. Note that the presence of accidental low energy states, e.g., due to disorder, at the backbone-1DTS interfaces might further reduce the length scale of the topological protection to $L/2$.

\subsubsection{Linear hexon}
\label{sec:longsnakes}

\begin{figure}
	\includegraphics[width=0.95\columnwidth]{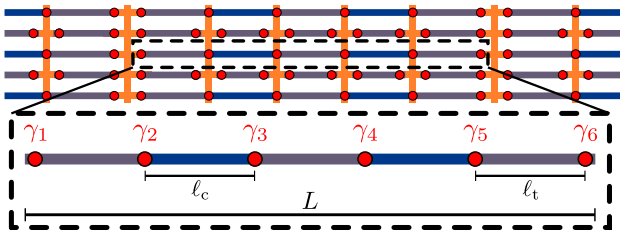}
	\caption{A linear hexon architecture.  { Note: the figure is not drawn to scale for the same reason as in Fig.~\ref{fig:operations}.}  The length $\ell_\text{c}$ of the non-topological segments is much larger than the corresponding coherence length $\xi_\text{c}$ of the non-topological regions and the length $\ell_\text{t}$ of the topological segments is much larger than the coherence length $\xi$ of the topological regions. The legend used is the same as in Fig.~\ref{fig:insects}. The magnification shows a single linear hexon. Additional topological superconducting links and semiconducting structures allow appropriate measurements to manipulate and entangle linear hexons.}
	\label{fig:longsnakes}
\end{figure}

A linear hexon consists of a single 1DTS wire of length $L$, where two segments of length $\ell_\text{c}$ are tuned to be in a normal superconducting state (for example, by gating), leaving three topological segments of length $\ell_\text{t}$. This is depicted schematically in the magnification in Fig.~\ref{fig:longsnakes}. Since topological regions are joined by the same superconducting shell, this construction does not require additional superconducting backbones to define an island hosting six MZMs. This simplifies the fabrication of linear hexons. { On the other hand, no pairs of MZMs within a single linear hexon can be simultaneously connected to a single quantum dot. As such, this design requires a more elaborate measurement apparatus to enable measurements within a hexon. As in the other hexon designs, we envision floating topological superconductors as coherent links that can bridge longer distances. Each MZM measurement in a linear hexon array involves a combination of such links and quantum dots.}

{ We arrange the hexons in a rectangular array.  Between each vertical row of hexons, we arrange a vertical row of coherent links, where five links are used to span the length of one hexon.  Quantum dots exist in the orange regions of Fig.~\ref{fig:longsnakes} and can be controlled by gates.  The use of quantum dots is completely analogous to those depicted in Fig.~\ref{fig:operations}.  The dots can be tunably coupled to any adjacent MZM independently.  In this way, any pair of MZMs connected by an orange region can be simultaneously coupled to a quantum dot.  Two-MZM measurements within a given hexon are performed using the coherent links spanning that hexon.  It is possible to perform all two-MZM measurements within a hexon, which grants single-qubit Clifford completeness.

Entangling operations on vertically-adjacent linear hexons also works similar to the examples depicted in Fig.~\ref{fig:operations}.  As discussed in Section~\ref{sec:combs}, there will be a maximum vertical distance between MZMs that can be simultaneously coupled to a given quantum dot.  We assume this distance allows the dots to vertically reach at least two rows apart (i.e., at least between neighboring rows of hexons) in either direction.  A greater reach can reduce the need for some operations, such as Swap gates, but is not necessary to achieve Clifford completeness.

Entangling operations on horizontally-adjacent linear hexons require the use of  links to facilitate coherent transport between distant MZMs.  In principle, the linear hexon design allows the joint measurement of any four MZMs within a given horizontal row of hexons, where two of the measured MZMs belong to one hexon and two belong to another hexon, by using multiple coherent links to couple distant pairs of MZMs. However, practical constraints of the measurement visibility will limit the number of links that can be used in a given measurement.  Fortunately, Clifford completeness can be achieved with measurements that require at most two links per measurement.}

In order to attain good topological protection, both $\ell_\text{c}$ and $\ell_\text{t}$ should be much longer than the corresponding coherence lengths ($\xi_\text{c}$ and $\xi$) in the conventional and topological superconducting regions. Assuming similar length scales for the latter, the relevant scale of the topological protection is given by $L/5$ in terms of the length of the parent 1DTS. Linear hexon designs therefore require much larger $L$ as compared to the other hexon designs. We expect this also leads to the smallest hexon charging energy (and, hence, the worst QPP protection) of the three designs.

\subsection{Tetron architectures}
\label{sec:Tetrons}

In this section, we describe the architectures of tetrons, which are topological qubits composed of four MZMs, examples of which are shown in Figs.~\ref{fig:shortsnakes} and \ref{fig:mammals}.
Four is the smallest number of MZMs for which a sector of fixed total fermion parity supports a qubit, i.e., a two-dimensional Hilbert space.
The absence of the extra ancillary pair of MZMs that were present in the hexon designs { results in two important differences. The first is that we have only two main tetron designs; the tetron analog of the one-sided hexon design cannot be scaled into a two-dimensional array, as each qubit can only connect to its vertically adjacent neighbors.  The second difference is that} we are not able to generate topologically protected single
qubit Clifford gates via operations acting on only one tetron.
Instead, the Clifford gates are generated either by joint parity measurements on a pair of tetron qubits or by
``Pauli frame changes.'' In the following section, we show how to perform the desired gates using a limited set of measurements; in subsequent sections, we detail various designs, some of which will allow more variety in the possible measurement operations. The more limited set of operations will require a more complicated construction of the Clifford gates requiring additional resources and operations, so there is a trade-off to minimizing the number of different measurements implemented.

\subsubsection{Quantum information basics}
\label{sec:tetron-qi-basics}

Consider a system of qubits arranged in a plane in a square lattice{, with each qubit labeled by its integer horizontal and vertical coordinates in the lattice.
Assume that one has the ability to make a limited set of measurements that we call the elementary operations: between any pair of qubits $j$ and $k$ separated by a displacement $(0,\pm 1)$, one can make measurements of the operators $Y_j Y_k$, $Z_j Z_k$, $X_j X_k$, $X_j Z_k$, and $Z_j X_k$; between any pair of qubits $j$ and $k$ separated by a displacement $(\pm 1,0)$, one can make measurements of either the operator $Z_j Z_k$ or $Y_j Y_k$ (either one suffices).  Assume also that we can make single qubit measurements of the operators $X_j$, $Y_k$, and $Z_l$.}

We first explain how this set of measurements provides a Clifford-complete set of operations for this system, up to Pauli frame changes.
Subsequently, we explain how a smaller set of measurements may generate Cilfford completeness by creating ``standards,'' which removes the need for the single qubit measurements and the vertical $Y_jY_k$ measurements.
In a final reduction, we show that Clifford completeness may be attained
even if the only available operations are vertical
$X_j X_k$ measurements and horizontal $Z_j Z_k$ measurements.
{ In the basic architectures utilizing these methods, we break the qubits into a ``checkerboard" arrangement, using one color as data qubits and one as ancillary qubits.  That is, we designate a qubit as a data qubit when the sum of its coordinates is even, and as an ancillary qubit when the sum of its coordinates is odd. }

The available measurements described in this section differ by a notational choice from those described in later sections by a permutation of the $X$, $Y$, and $Z$ operators, which amounts to a ``Pauli frame change.''  The reason is that, here, $X$ and $Z$ are a natural pair of measurements to use to build CNOT gates.

{\it Pauli Frame Changes---}
Pauli frame changes~\cite{Knill05} refers to the idea of not performing certain single qubit Clifford gates such as $X$, $Z$, and $H$ in a quantum circuit, but instead modifying subsequent measurements accordingly. The idea is that, given a sequence composed of single qubit operators $X$, $Z$, and $H$, and single-qubit Pauli measurements, we classically track the total single qubit operation and perform the appropriately conjugated measurements.  Thus, for a sequence such as:
measure $Z_j Z_k$, apply $H_k$, measure $Z_k Z_l$, apply $X_k$, and measure $Z_j Z_k$,
we commute the operators $X$ and $H$ through the measurements by appropriately changing what measurements we perform.
Using $Z_k Z_l H_k=H_k X_k Z_l$ and $Z_j Z_k X_k H_k=-X_k H_k Z_j X_k$, we find that this sequence of operations is equivalent
to the sequence: measure $Z_j Z_k$, measure $X_k Z_l$, measure $-Z_j X_k$, and apply $X_k H_k$. The final single qubit Clifford gates do not need to be performed if they follow all measurements in the circuit.

One can also commute the single qubit Clifford gates through CNOT gates. Since magic state injection is performed using CNOT gates, single qubit Clifford gates are not necessary even when the circuit includes $T$ gates,
where
\begin{equation}
\label{eq:Tdef}
T \equiv \left( \begin{array}{cc} 1 & 0 \\ 0 & e^{i\pi/4} \\ \end{array}\right).
\end{equation} 

The effect of Pauli frame changes is to permute the set of two-qubit measurements. This may change the set of available measurements if the set of elementary operations does not include all two-qubit measurements.
For this reason, we will avoid Pauli frame changes corresponding to commuting the Clifford phase gate
\begin{equation}
S \equiv \left( \begin{array}{cc} 1 & 0 \\ 0 & i \\ \end{array}\right)
\end{equation}
through other operations. This allows the set of available vertical measurements needed to remain fixed throughout the computation.

As we describe specific operations that we build out of the elementary measurements, we will sometimes say that we can perform an operation ``up to $\{ X,Z,H \}$" or ``up to $\{X,Z \}$,'' describing the possible frame change on the qubits. The particular frame change that is implemented is determined by the measurement outcomes.  An operation up to $\{ X,Z \}$ may map $Z \rightarrow \pm Z$ and $X \rightarrow \pm X$ (with the mapping on $Y$ determined by the mapping of $X$ and $Z$). An operation up to $\{ X,Z,H \}$ may additionally map $Z \rightarrow \pm X$ and $X \rightarrow \pm Z$.

{\it  Vertical Teleportation---} Using measurements $X_k X_l$ and $Z_k Z_l$ { between a pair of qubits with displacement ${(0,\pm1)}$,} one can create an EPR pair of the qubits $k$ and $l$.  A further pair of measurements $X_j X_k$ and $Z_j Z_k$ will teleport the state of qubit $j$ to qubit $l$.  This teleportation is up to $\{ X,Z \}$ on qubit $l$.

\begin{figure}
\includegraphics[width=.99\columnwidth]{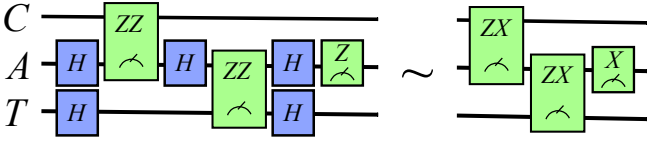}
\caption{{ Two equivalent circuits implementing the CNOT gate. The control, ancillary, and target qubits are labeled $C$, $A$, and $T$, respectively. Gates labeled $H$ are Hadamard gates, the other boxes correspond to one and two qubit measurements as indicated by the corresponding Pauli operators.  The left circuit implements a CNOT up to $\{ X,Z\}$ on qubits $C$ and $T$.  As explained in text, the Hadamard operators can be commuted through to yield the simplified circuit shown on the right, up to \{$X,Z,H$\}. }  }
\label{CNOT}
\end{figure}

{\it Vertical CNOT and Swap---} We can also apply a CNOT gate, up to $\{ X,Z \}$, { on two qubits separated by $(0,\pm2)$,} e.g., two data qubits separated vertically by one ancillary qubit in between them.
For this, we can use the { left} circuit of Fig.~\ref{CNOT} (this circuit is the same as in Fig.~2 of
Ref.~\onlinecite{Xue13}), where the control, ancillary, and target qubits are labeled $C$, $A$, and $T$, respectively. Qubit $A$ is initialized in an eigenstate of $Z$. We take qubits $C$ and $T$ to be on the even sublattice, separated in the vertical direction with $A$ the ancillary qubit in between them. This circuit gives a CNOT up to $\{ X,Z \}$ on qubits $C$ and $T$.

The Hadamard operators in this circuit can be commuted through the measurements to the end of the circuit, resulting in the CNOT gate up to $\{ X,Z,H \}$. The resulting sequence of operations in the simplified circuit is: measure $Z_C X_A$, measure $Z_A X_T$, and measure $X_A$. 

The ability to perform CNOT gates in both directions on a pair of qubits allows one to Swap the pair of qubits (through the application of three alternating CNOT gates). This allows arbitrary motion of the data qubits in the vertical direction.

{\it Hadamard Gate Without Pauli Frame Change and Single Qubit $X$ Measurement---}
The method of Pauli frame changes above is the most efficient method to implement a single qubit Clifford gate, as no actual operations need to be performed on the qubits.
However, switching between different Pauli frames may change the set of available operations.  The Hadamard gate does not affect the set of available operations in the vertical direction. Thus, if we only consider vertical measurements, we can perform Hadamard gates by frame changes.  However, suppose that we wish to perform a Hadamard gate followed by a measurement of $Z_j Z_k$ of qubits separated { by $(\pm 1,0)$}.  In this case, the new frame requires a measurement of $Z_j X_k$, which is not an elementary operation for those qubits.

In order to overcome this, we show how to perform a Hadamard gate while only utilizing Pauli frame changes that are up to $\{ X,Z \}$, as such Pauli frame changes will leave the set of available horizontal operations unchanged.
Let $\text{Swap}_{jk}$ swap qubits $j$ and $k$.
Consider the operation
\begin{equation}
U=\text{Swap}_{jl} H_l \text{Swap}_{jl} H_l,
\end{equation}
for a pair of data qubits $j$ and $l$ that are { separated by $(0,\pm 2)$ (i.e., vertically nearest-neighbor data qubits with one ancillary qubit in between them)}. Our implementation of $\text{Swap}_{jl}$ is up to $\{ X,Z \}$. Since $H_l$ appears twice in this operation, the net frame change in performing $U$ is still up to $\{ X,Z\}$; that is, it will not interchange $X_k \leftrightarrow Z_k$. As an operator, $U=H_j H_l$ applies the Hadamard gate to each of the two qubits.

An alternative way to implement a Hadamard gate is to use the following variant of the vertical teleportation protocol.
Use measurements $X_k Z_l$ and $Z_k X_l$ between a pair of { qubits separated by $(0,\pm 1)$}, e.g., one data qubit and one ancillary qubit, to create an EPR pair up to the Hadamard gate on $l$. Then measure $X_j X_k$ and $Z_j Z_k$ to teleport the state of qubit $j$ to qubit $l$ while performing a Hadamard on the encoded state. Since teleportation may be used to route qubits, this allows the Hadamard gate to be performed ``for free" at the same time as a teleportation.

{\it Horizontal CNOT and Swap---}
Using the method described above to generate a Hadamard gate without frame change,
the horizontal measurements of $Z_j Z_k$ can be conjugated to become measurements of $Z_j X_k$ or $X_j X_k$.
We thereby obtain a CNOT gate acting on a { pair of qubits separated by $(\pm 2,0)$} by using only $Z_j Z_k$ measurements horizontally. If instead we have only $Y_j Y_k$ measurements horizontally, we can use an $S$ gate (which we explain how to implement below) to conjugate them to become measurements of $X_j X_k$.
Since a Swap is generated from three alternating CNOT gates, we now have the ability to perform horizontal Swaps of second nearest neighbor pairs of qubits, using the intermediate qubit as an ancillary qubit to facilitate the operation.

{\it $S$ gate---} An $S$ gate can be implemented without frame change by utilizing state injection of a $+1$ eigenstate of $Y$.  Such a state can be produced by measuring a single qubit $Y$ operator.

{
Note that instead of implementing a standard state injection using unitary gates (e.g., a CNOT gate), a measurement-based injection is more tailored for our architectures.  In particular, a shorter circuit for implementing an $S$ gate (up to $Z$ gates on the source) is given by the sequence of operations: prepare an ancillary qubit in the $+1$ eigenstate of $Y$, measure the operator $ZX$, where $Z$ is on the data qubit and $X$ is on the ancillary qubit, and then measure $Z$ on the ancillary qubit.}

{\it $Y_j Y_k$ Measurement---}
We can measure $Y_j Y_k$ between any pair of qubits $j$ and $k$ that are { separated by $(0,\pm 2)$} using only the other elementary operations, through the following sequence: apply a CNOT gate from $j$ to $k$, apply a Hadamard gate on qubit $j$,
apply a CNOT gate from $j$ to $k$, measure $Z_k$, apply a CNOT gate from $j$ to $k$, apply a Hadamard gate on qubit $j$,
and apply a CNOT gate from $j$ to $k$.
One may verify that the result of this sequence of operations is equal to a measurement of $-Y_j Y_k$.

{\it Living without Single-Qubit Measurements: Using ``Standards"---}
If it is not possible to perform single-qubit measurements, but only two-qubit measurements, it is still possible to generate a Clifford complete set of operations.  By measuring $X_j X_k$, $Y_j Y_k$, or $Z_j Z_k$, a qubit state that is an eigenvector of $X$, $Y$, or $Z$ can be copied indefinitely.
We call such a qubit a ``standard." 

{ To achieve Clifford completeness without single-qubit measurements, we store standards in every data qubit with odd horizontal coordinate.  The data qubits now have coordinates $(2n,2m)$ in the lattice, for $n,m\in\mathbb{Z}$ (i.e., there are now three ancillary qubits per data qubit).  With this arrangement, one can perform single qubit measurements on qubits with even horizontal coordinate.  } 
In fact, which eigenstate of $X$, $Y$, or $Z$ we choose for the standard is arbitrary, as the choice has no effect on measurements, when restricting to Clifford operations. If magic
state injection is performed, the choice of eigenstate used for the $Y$ standard becomes important. In this case, magic state injection
can be used to identify the choice of $Y$ standard (see the discussion on page 38 of Ref.~\onlinecite{Preskill04}).

{\it Restricted Two-Qubit Operations---}
Now suppose that we can measure $X_j$ or $Z_j$ on any single qubit, but we can only perform the limited set of
two-qubit measurements: $Z_j Z_k$ between a pair of vertically-separated
qubits and $X_j X_k$ between a pair of horizontally-separated qubits.
This is still sufficient to build a universal quantum computer if we can produce an approximate magic state.
While this is not likely to be a practical architecture and all architectures we describe have more than this set of measurements,
it is interesting that this restricted set of operations remains universal.
The following discussion of operations will be up to $\{ X,Z \}$.


Using the same circuit shown in Fig.~\ref{CNOT}, we can perform a CNOT between two qubits separated by a displacement $(\pm 1,\pm 1)$.  For example, to generate a CNOT gate with the $(0,0)$ qubit as the control and the $(1,1)$ qubit as the target, we use the following sequence: prepare the ancillary qubit $A$ in an $X$ eigenstate, measure $Z_C Z_A$, and  measure $X_A X_T$. Given the ability to perform CNOT gates, we can perform Swap. In the above example, the $(1,1)$ qubit is a standard. Applying multiple Swap operations allows the data qubits to move arbitrarily within the data qubit sublattice while leaving the standards intact.

However, we do not yet have the ability to perform the full Pauli group with this restricted set, since we do not have the ability to perform the Hadamard gate. These gates cannot be implemented through Pauli frame changes as we have a smaller set of elementary operations.
Suppose, however, that we could produce many $Y$ standards, either $Y=+1$ or $Y=-1$ eigenstates.  Using this $Y$ standard and state injection,
we can perform an $S$ gate.  Once we have an $S$ gate, we can also measure $Y_j X_k$, $X_j Y_k$, and $Y_j Y_k$ between any two horizontally separated qubits.  Thus, we have the ability to perform all the elementary operations described at the beginning of this section, but with $Y$ and $Z$ operators interchanged and horizontal and vertical directions interchanged. We therefore have operations that are Clifford complete up to $\{ X,Z \}$.

If we can produce approximate $Y$ standards, we can distill them using methods similar to, but simpler than the methods of Ref.~\onlinecite{Bravyi12}. For this, we can use any CSS code that allows transversal $S$ gates, such as the $7$-qubit code~\cite{Gottesman98}. Using a CSS code allows us to check the stabilizers of the code using only CNOT gates and measurement and preparation of qubits in $Z$ and $X$ eigenstates. We note that we can generate $Y$ standards, for example, if we can generate an approximate $S$ gate. Of course, if we can generate approximate $T$ gates, then we can produce approximate $S=T^2$ gates. Similarly, if we can produce approximate magic states, we can use them to produce approximate $Y$ standards.

\subsubsection{Linear tetron}

\begin{figure}
	\includegraphics[width=0.95\columnwidth]{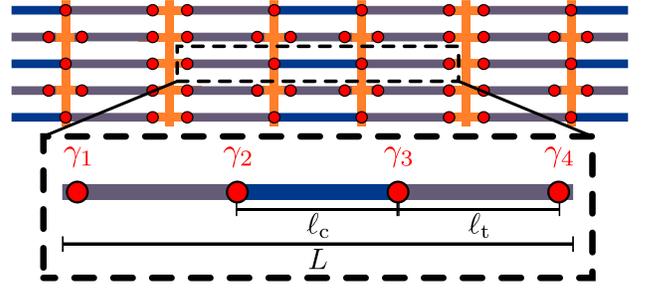}
	\caption{A linear tetron architecture. { Note: the illustration is not drawn to scale for the same reason as in Fig.~\ref{fig:operations}.} The length $\ell_\text{c}$ of the non-topological segments is much larger than the corresponding coherence length $\xi_\text{c}$ of the non-topological regions and the length $\ell_\text{t}$ of the topological segments is much larger than the coherence length $\xi$ of the topological regions. The legend used is the same as in Fig.~\ref{fig:insects}. The magnification shows a single linear tetron. Additional topological superconducting links (gray) and semiconducting structures (orange) allow appropriate measurements to manipulate and entangle linear tetrons.}
	\label{fig:shortsnakes}
\end{figure}

A linear tetron consists of a single 1DTS wire in which a middle segment of length $\ell_\text{c}$ has been tuned to be in a normal superconducting state (for example, by gating), leaving two topological segments of length $\ell_\text{t}$. This is depicted schematically in the magnification in Fig.~\ref{fig:shortsnakes}. As a result, there are four MZMs, one at each end of the wire and one at each of the two boundaries between topological
and normal superconducting regions. The linear tetron is, in some sense, the simplest of our qubit designs. However, this simplicity of the single qubit is somewhat offset by the complexity of the associated measurement apparatus, i.e.,  the array of quantum dots and floating topological superconductor links that are needed for measurements, which we now describe (see Fig.~\ref{fig:shortsnakes}).

{ We arrange the tetrons in a rectangular array. Between each vertical row of tetrons, we arrange a vertical row of coherent links, where three links are used to span the length of one tetron. These links can be provided by floating topological superconductors, as in the case of the linear hexon.  Measurements of linear tetrons are done in a similar manner to measurements of linear hexons.  Any pair of MZMs connected by an orange region of Fig.~\ref{fig:shortsnakes} can be simultaneously coupled to a quantum dot.  As discussed in Section~\ref{sec:combs}, there will be a maximum vertical distance between MZMs that can be simultaneously coupled to a given quantum dot.  We assume this distance allows quantum dots to span the separation between neighboring rows of hexons.  This is sufficient to perform the measurements used in the protocols of Section~\ref{sec:tetron-qi-basics}.  Just as for linear hexons, a greater reach can reduce the need for some operations.  
}

Let us label the MZMs on a given tetron as $\gamma_1$, $\gamma_2$, $\gamma_3$, and $\gamma_4$, from left to right.
We required the total fermion parity of a tetron to be even (e.g., by using charging energy),
{ ${p_{12}p_{34} = 1}$. The qubit basis states are then defined to be
\begin{eqnarray}
|0\rangle &=& | p_{12} = p_{34} =-1\rangle  \\
|1\rangle &=& |p_{12} =p_{34}=+1\rangle
.
\end{eqnarray}}
The Pauli operators on the qubit are represented in terms of MZM operators as
\begin{eqnarray}
X &=& i\gamma_2 \gamma_3 = i\gamma_1 \gamma_4 ,\\
Y &=& i\gamma_1 \gamma_3 = -i\gamma_2 \gamma_4 , \\
Z &=& i\gamma_1 \gamma_2 = i \gamma_3 \gamma_4,
\end{eqnarray}
up to an overall phase.

In order to distinguish different tetrons, we label each tetron and its operators by its (integer-valued) coordinate $(j,k)$ in the two-dimensional array.

Measurements of $Z^{(j,k)} Z^{(j,k+1)}$, $X^{(j,k)} X^{(j,k+1)}$, and $Y^{(j,k)} Y^{(j,k+1)}$ between vertically-neighboring tetrons
can be performed by turning on the couplings of the corresponding MZMs to the adjacent quantum dots located between the two tetrons, and then probing these quantum dots by measuring the shift of the capacitance or charge, as discussed in Section~\ref{sec:measurement}. More specifically, quantum dots connecting MZMs of vertically-neighboring tetrons can be directly coupled to the pairs $ \gamma_i^{(j,k)}$ and $\gamma_i^{(j,k+1)}$ for $i=1,2,3,4$.
By turning on two such pairs of couplings, we can measure the claimed two-qubit operators.

Measurements of $Z^{(j,k)} Z^{(j+1,k)}$ between horizontally-neighboring tetrons further require the use of links to facilitate coherent transport between distant MZMs.
In some cases, we need to use multiple links in order to couple more distant MZMs, as discussed for the linear hexon design in Section~\ref{sec:longsnakes}. 
To be more specific, a quantum dot that sits between two horizontally-neighboring tetrons at $(j,k)$
and $(j+1,k)$ can be directly coupled to both $\gamma_4^{(j,k)}$ and $\gamma_1^{(j+1,k)}$.
Using the combination of two coherent links and three quantum dots, we can couple this unit to both $\gamma_3^{(j,k)}$ and $\gamma_2^{(j+1,k)}$.
We can think of this combination of coherent links and dots as an effective quantum dot, to relate to the measurement discussion of Section~\ref{sec:4MZMmodel}.
The ``quantum dot'' energy levels now depend on the joint parity $p = p_{(j,k)}p_{(j+1,k)}$, which is the eigenvalue of the operator
$Z^{(j,k)} Z^{(j+1,k)} = - \gamma_3^{(j,k)} \gamma_4^{(j,k)} \gamma_1^{(j+1,k)} \gamma_2^{(j+1,k)}$.

In this way, we can perform all the two-qubit measurements assumed in the previous section.
These two-qubit measurements are sufficient for Clifford-complete operations, as described in
Section~\ref{sec:tetron-qi-basics}. { However, similar to the case of linear hexons, we can also perform single qubit measurements and other two-qubit (entangling) measurements by using
the links to facilitate coherent transport across longer distances. For example, an effective quantum dot (composed of links and quantum dots) can be coupled to any two different MZMs
from the same tetron, so that the dot's energy levels depend on the parity of these two
MZM operators, in other words, on the eigenvalue of the corresponding Pauli operator.}

\subsubsection{Two-Sided tetron}

\begin{figure}
	\includegraphics[width=0.95\columnwidth]{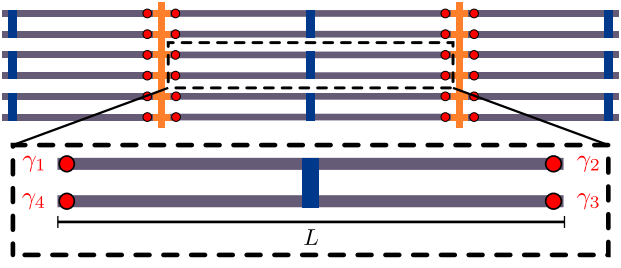}
	\caption{A two-sided tetron architecture. { Note: the illustration is not drawn to scale for the same reason as in Fig.~\ref{fig:operations}.} The legend used is the same as in Fig.~\ref{fig:insects}. The magnification shows a single two-sided tetron. Additional topological superconducting links and semiconducting structures allow appropriate measurements to manipulate and entangle two-sided tetrons.}
	\label{fig:mammals}
\end{figure}

A two-sided tetron consists of two 1DTSs joined by a superconducting backbone, as depicted schematically in the magnification in Fig.~\ref{fig:mammals}. The backbone is located in the middle of the wires, far away from any of the four MZMs at the ends of the 1DTSs. This design facilitates joint measurements between horizontally-neighboring
tetrons and somewhat complicated measurements between vertically-neighboring tetrons, which are
analogous to those depicted in Fig.~\ref{fig:operations} for hexons.  { One could include links to increase the variety of operations, but it is instructive to consider the architecture design with no links.}

Let us label the MZMs of a two-sided tetron as $\gamma_1$, $\gamma_2$, $\gamma_3$, and $\gamma_4$, in clockwise order starting from the upper left.

A joint Pauli operator on horizontally-neighboring tetrons can be measured if one quantum dot is coupled to
$\gamma_2^{(j,k)}$ and  $\gamma_1^{(j+1,k)}$ and a second quantum dot is coupled to
$\gamma_3^{(j,k)}$ and  $\gamma_4^{(j+1,k)}$. This coupling configuration allows for a measurement of $X^{(j,k)} X^{(j+1,k)} = - \gamma_2^{(j,k)} \gamma_3^{(j,k)} \gamma_1^{(j+1,k)} \gamma_4^{(j+1,k)}$.

The simplest measurement of vertically-neighboring tetrons is given by coupling one quantum dot
to $\gamma_1^{(j,k)}$ and $\gamma_4^{(j,k+1)}$ and a second quantum dot to
$\gamma_2^{(j,k)}$ and $\gamma_3^{(j,k+1)}$. This coupling configuration allows for a measurement of $Z^{(j,k)} Z^{(j+1,k)} = - \gamma_1^{(j,k)} \gamma_2^{(j,k)} \gamma_3^{(j,k+1)} \gamma_4^{(j,k+1)}$.

Using the semiconductor wires (orange in Fig.~\ref{fig:mammals}), we can perform additional measurements that require coupling MZMs over a slightly more extended range.  { The length scale of the quantum dots constrains the distance over which a measurement may be performed, see the discussion in Section~\ref{sec:combs}.}  For example, we can couple one quantum dot
to $\gamma_1^{(j,k)}$ and $\gamma_4^{(j,k+1)}$ and instead couple the second quantum dot to
$\gamma_3^{(j,k)}$ and $\gamma_3^{(j,k+1)}$. This coupling configuration allows for a measurement of $Y^{(j,k)} Z^{(j+1,k)} = - \gamma_1^{(j,k)} \gamma_3^{(j,k)} \gamma_3^{(j,k+1)} \gamma_4^{(j,k+1)}$.
The other two-qubit joint Pauli measurements of vertically-neighboring tetrons can be similarly implemented.

The only single-qubit measurement that is possible in this architecture (without introducing links) is a measurement of $X$, which can be implemented by
coupling a quantum dot to $\gamma_1^{(j,k)}$ and $\gamma_4^{(j,k)}$ or to $\gamma_2^{(j,k)}$ and $\gamma_3^{(j,k)}$.
As described in Section~\ref{sec:tetron-qi-basics}, if we only have this restricted set of measurements, we can still achieve Clifford complete operations through the use of ``standards.''

As in the other architectures, we can increase the set of possible operations by introducing horizontal links provided by floating topological superconductors, that facilitate coherent transport across the length of the hexons. In particular, this enables measurement of all single-qubit Pauli operators measurements. Such structures were previously considered in Ref.~\onlinecite{Plugge16b}.

{
\subsection{Design summary}
\label{sec:design-summary}

The different hexon and tetron architectures presented in this paper have different advantages and challenges.  A priori, it is difficult to make quantitative performance estimates and rankings between the designs.  In this section, we summarize the common principles that apply to all the presented designs, as well as their differences.

\subsubsection{Common design principles}

{ The common design principles we used to protect the encoded quantum information are \footnote{Note that we also avoided using tunable Josephson junctions in our designs.   Tunable Josephson junctions provide the ability to cut and reconnect the superconducting backbone of the designs, which allows for additional flexibility in coupling distant MZMs. In order to reach both regimes of coupled and fully decoupled MZMs, the Josephson junction would need to be tunable over a wide range $E_J\sim E_C \dots E_J\gg E_C$, which might be difficult in practice.}:
(1)  magnetic field alignment in the direction of the 1DTSs to maximize the gap; 
(2)  avoiding the use of topological T-junctions to avoid low energy modes close to the MZMs; 
(3) finite charging energy of the individual qubit units (hexons or tetrons) to suppress QPP at low temperature; 
(4)  long 1DTSs to suppress hybridization of the computational MZMs;
and
(5) the ability to perform a sufficient set of measurements to achieve a topologically protected Clifford complete gate set.

Design principles (1)-(5) lead to exponential suppression of errors in the qubit architectures.  More explicitly, error rates from QPP and thermally excited quasiparticles are exponentially suppressed by the ratios $E_C/T$ and $\Delta/T$, respectively.  Errors due to MZM hybridization are exponentially small in $L/\xi$.   Furthermore, the fidelity of manipulating quantum information in the presented measurement-based scheme scales exponentially with the integration time of the measurement~\footnote{The fidelity of manipulating quantum information in a measurement-only protocol is set by the experimental certainty for distinguishing between two measurement outcomes.}. }

\subsubsection{Design Differences}
\label{sec:design-diff}

We qualitatively compare the different designs using the four axes mentioned in the introduction:
(1) QPP time, (2) signal visibility, (3) fabrication simplicity, and (4) computational efficiency.

(1) The larger the charging energy $E_C$ of each individual qubit unit (hexon or tetron), the stronger the suppression of QPP. As discussed in Section~\ref{sec:overview}, when the length $L$ of a qubit unit (along the direction of the 1DTS wires) is much larger than its width $w$, we expect the geometric capacitance of a qubit unit to depend roughly linearly on $L$ and sub-linearly on $w$. Thus, there is a trade-off between shorter $L$, which provides better protection against QPP, and longer $L$, which provides better protection against hybridization of the MZMs. The maximum combined protection is achieved at some optimal value of $L$, where the corresponding error rates are equal. This is roughly when $E_C / T \approx L_{\text{h}} / \xi$, where $L_{\text{h}}$ is the effective distance between MZMs. For the one-sided hexons, two-sided hexons, linear hexons, linear tetrons, and two-sided tetrons, we roughly have $L_{\text{h}} \approx 2L$, $L$, $L/5$, $L/3$, and $L$. Assuming that the coherence length $\xi$ and the $w$ dependence of $E_C$ is approximately the same for all the qubit designs, we rank their relative error protection (combined protection from QPP and MZM hybridization) from largest to smallest as: one-sided hexons, two-sided tetrons, two-sided hexons, linear tetrons, and linear hexons. Note that this ordering assumes that there are no low energy states from weakly hybridized MZMs at the superconducting backbone of the one-sided hexons. When this assumption is not valid, $L_{\text{h}} \approx L$ for the one-sided hexons, and so its ranking will drop to below the two-sided hexons.

(2) Since the MZM measurements rely on fermion parity dependent energies of the system when MZMs are coupled to quantum dots, the visibility of such measurements will be lower when the charging energy $E_C$ is larger [see (1)]. This is because the parity dependent terms in these energies depend inversely (to lowest order in perturbation theory) on excited state energies that are of the order of the charging energy. This effect can be compensated to some degree by increasing the tunneling amplitudes $t_j$. Another aspect that influences the visibility is the separation distance between the MZMs being measured. Longer distance measurements require more coherent links, which will decrease visibility. In this regard, the linear hexons and tetrons require more coherent links than the other designs. It is difficult to precisely estimate the effects of all these factors on the visibility in order to produce a meaningful ranking of designs.

(3) The simplicity of fabricating different designs will ultimately be decided experimentally. Here, we mention qualitative differences in the fabrication of the designs. While it is clear that fabricating a tetron is slightly easier than fabricating a corresponding hexon, we do not expect qualitative differences in the fabrication difficulty and focus on hexons in the following discussion. One important challenge for the one-sided and two-sided hexons will be the deposition of the superconducting backbone, as this must be done without disturbing the underlying 1TDSs too much. Attaining the larger charging energy of the one-sided hexon additionally requires  sufficient hybridization of the MZMs in the superconducting backbone. The linear hexons have the advantage of not requiring any such superconducting backbones. The drawback is the presence of more coherent links and the requirement of tuning larger regions of the 1DTS out of the topological regime.

(4) A full analysis of the complexity of all designs will be published elsewhere~\cite{Hastings16}. Here, we provide some general remarks about the computational efficiency of the designs.
Hexons are computationally more efficient than tetrons. With six MZMs for each qubit, it is possible to do all single qubit Clifford operations within a hexon.  Furthermore, adjacent qubits may be entangled without any additional ancillary hexons. In contrast, even in the most efficient tetron design, roughly half of the tetrons are required to be ancillary in order to provide the full set of Clifford gates, thus requiring a total of eight MZMs for each qubit. The computational efficiency is further reduced when limiting the number of allowed single qubit Pauli measurements, as in the case of the two-sided tetron without coherent links. Clever algorithms using standards (described in Section~\ref{sec:tetron-qi-basics}) still allow realization of all Clifford operations. However, the computational efficiency is reduced, since this scheme requires 3/4 of the tetrons to be ancillary (as one fourth of the tetrons are used to encode standards), leaving only one fourth as computational data qubits. On top of the increased hardware requirements, the preparation and distribution of appropriate standards requires additional applications of gate operations.
}

\section{Universal quantum computing}
\label{sec:universal}

\subsection{$T$ gate}

Nearly all of our discussion so far has focused on achieving topologically protected Clifford complete operations via an adaptive sequence of measurements. However, the Gottesman-Knill theorem demonstrates that Clifford operations can be efficiently modeled on a classical computer by updating the list of stabilizers that define the ground state vector at each computational step~\cite{Gottesman98a}. Thus, a computing device that is only Clifford complete can be classically simulated. Nevertheless, Clifford operations are valuable because they can be augmented by a single additional (non-Clifford) gate to become a universal model for quantum computation (BQP complete)~\cite{Boykin99}, { and they play a significant role in prominent error-correction protocols}~\cite{Gottesman97}. In the designs presented in this paper, this augmentation process can be achieved with the \emph{identical} hardware described for Clifford complete computation by using a more elaborate classical control protocol than is required to implement the Clifford complete operations.

While any additional (non-Clifford) gate in principle suffices~\cite{Nebe00, Nebe06,Campbell12},
an attractive choice is the $T$ gate (also known as the $\pi/8$ phase gate) of Eq.~(\ref{eq:Tdef}).
Given the ability to perform Clifford gates and perform measurements, the ability to apply a $T$ gate is equivalent to the ability to generate ancillary qubits in a ``magic state,'' such as
\begin{equation}
\frac{1}{\sqrt{2}} \left( \ket{0}+e^{i\pi/4}\ket{1}\right).
\end{equation}
The fundamental virtue of magic states is that they may be distilled using only Clifford operations~\cite{Bravyi05}. In its original formulation, magic state distillation is a process that consumes 15 low fidelity approximations to the magic state and produces one copy of the magic state with improved fidelity, using only Clifford operations. This { procedure} requires only very modest fidelity of $1-\epsilon$, where $\epsilon \lesssim 0.14$, for the 15 input magic states to commence and asymptotically yields a magic state with fidelity of $1-\text{const}\times \epsilon^3$. Much work has since been done on optimizing distillation protocols and related strategies for crossing the divide between Clifford completeness to universal quantum computation, see Refs.~\onlinecite{Duclos12}, \onlinecite{Bravyi12}, \onlinecite{Campbell16}. Note that the distillation can either be performed on the physical topological qubits for low-depth circuits, or at the level of logical qubits in error-correcting codes. In the following we will focus on magic state preparation and distillation for the physical topological qubits. Once an approximate magic state can be prepared on the level of physical qubits, the high fidelity Clifford gates also allow preparation of an approximate logical magic state.

Magic state distillation will constitute the bulk of the work of any quantum computer of a few hundred qubits with topologically protected Clifford gates. For larger quantum computers, the cost of communication, i.e., use of Swap gates, could rival distillation as an expense until long-range communication is properly addressed.  For this reason, the circuit depth for magic state distillation is a good surrogate for the overall efficiency of the layout of a quantum computer.   { A detailed study examining the efficiency of distillation in the five design layouts detailed in this paper will be published elsewhere~\cite{Hastings16}. Such an analysis will quantify the computational efficiency axis (4) discussed in Section~\ref{sec:design-diff}.}

{ Magic state distillation is often presented without a concrete qubit architecture in mind, where all unitary Clifford gates and measurements are possible on all qubits and all pairs of qubits (a complete graph model).  
In the five planar design layouts that we have presented, magic state distillation may be efficiently synthesized using the combinations of measurements that each layout permits (see Section~\ref{sec:architectures}).   This is a concrete implementation of measurement-only quantum computation as described in Refs.~\onlinecite{Bonderson08b}, \onlinecite{Bonderson13a}. In the case of the tetron designs, Clifford completeness requires at least half of the tetrons to be ancillary (more if links are not used). In all our designs, an additional portion of the hexon or tetron qubits will need to be dedicated to the preparation of approximate magic states.
}

{ We now sketch} how the same classical control electronics used to produce Clifford operations can instead be used to produce approximate magic states. Details will be explained in Ref.~\onlinecite{Karzig_in_prep}, which describes an extension and combination of two antecedents already in print: an adiabatic protocol which produces high fidelity magic states via a dynamic decoupling that exploits topologically protected regions of the single qubit Bloch sphere~\cite{Karzig15a}, and a hybrid adiabatic-measurement protocol that utilizes measurement to suppress diabatic errors~\cite{Knapp16}.

As described in Ref.~\onlinecite{Karzig15a}, a MZM based qubit state adiabatically evolved around a closed loop in the Bloch sphere picks up a relative phase of $\alpha$ between the even and odd fermion parity sectors, where $\alpha$ is the solid angle enclosed by the loop, i.e.,  the geometric phase. This evolution is performed by changing couplings to ancillary MZMs. Because MZM couplings drop off exponentially in system parameters, combinations in which one (or two) couplings are zero constitute topologically protected great circle paths on the Bloch sphere (formed by the boundaries of octants of the sphere). It is the protected nature of these paths, specifically the closed path enclosing one octant of the Bloch sphere, that leads to the topologically protected phase gate $S$, for which $\alpha = \pi/2$. Producing the $T$ gate with $\alpha = \pi/4$ is not protected, but one may cancel low-frequency errors by defining a particular loop contour $c$ on the Bloch sphere. This contour is ``snake-like,'' consisting of different vertical sweeps from the north pole to the equator, some partial evolution along the equator, and sweeps back to the north pole (see Fig.~4 in Ref.~\onlinecite{Karzig15a}). Optimal cancellation is achieved by selecting Chebyshev roots as turning points along the equator.

One may adapt this idea to a measurement-based scheme in the same spirit as measurement-only topological quantum computation~\cite{Bonderson08b, Bonderson08c}, in which the unitary gate implemented by an elementary braid exchange is instead implemented by a composition of measurements, each incorporating one fixed anyon of an ancillary pair. In the present context, a similar sequence of measurements that now project on the turning points of $c$ can produce the same relative phase between the different qubit states as the earlier adiabatic protocol~\cite{Karzig15a}. This yields a measurement-based implementation of the $T$ gate. As in earlier measurement-only schemes, recovery from unwanted measurement outcomes must be addressed.  In the case of hexons, one may use the incorporated ancillary MZMs { for this recovery}. In the case of tetrons, one must instead utilize one of the nearby ancillary tetrons.

Whereas all measurements used in our constructions of Clifford gates involved creating a single closed loop through MZM islands and quantum dot tunnel junctions, in order to create the projections required to simulate adiabatic evolution along the Bloch sphere contour $c$, this is not sufficient. Just as $c$ explores along the $X-Y$ equator, we will need to simultaneously turn on tunneling at the junctions used to implement an $X$  measurement and the junctions used to implement a $Y$ measurement. Moving along this equator corresponds to tuning the relative tunneling amplitudes between these two sets of junctions. Tuning these ratios will require uniformity of manufacture and careful calibration of each junction, but given the mathematical ability of dynamic decoupling to remove low frequency errors, we have some confidence in this procedure, at least to generate magic states accurate enough for distillation.

\subsection{Quantum Error Correction}

{ Combining the topologically protected implementation of the Clifford gates with the ability to produce and distill magic states, we expect the architectures proposed in Section~\ref{sec:architectures} can lead to a high fidelity quantum computer that allows for many gate operations before decohering. For low-depth quantum computing (with magic state distillation), this approach might even be sufficient without quantum error correction~\cite{Moore98}. Large-scale quantum computing, however, will still require embedding the presented architectures into an error correcting superstructure. A detailed discussion of error correction is beyond the scope of this paper and will instead be addressed in a future work~\cite{Hastings16}. We note that, since our designs allow for a complete set of high fidelity Clifford operations, any error correcting stabilizer code can be implemented on a software basis without changing the presented designs. Finding an optimized hardware for error correction is an interesting subject for future research.

}

\section{Conclusions and Near-term Directions}
\label{sec:Conclusions}

Experimental explorations of MZMs in nanowire devices have evolved to an impressive degree over the past few years, with synergistic breakthroughs on the fabrication and characterization fronts~\cite{Mourik12,Rokhinson12,Deng12,Churchill13,Das12,Finck12,Chang14,Higginbotham15,Krogstrup15,Albrecht16,Zhang16}.  In parallel, new theory insights have emerged that appear auspicious for eventual quantum computing applications.  Notable examples include anticipation of the virtues of charging energy, both for protecting quantum information and facilitating Majorana measurements~\cite{Fu10,Landau16,Plugge16a,Starkpatent16,Vijay16b,Plugge16b}; measurement-only topological quantum computation~\cite{Bonderson08b,Bonderson08c}; improved modeling of microscopic details of Majorana systems~\cite{Nijholt16,Stanescu11,Prada12,Rainis13,Stanescu13a,Soluyanov16,Vuik16,Winkler16,Cole16}; a quantitative understanding of braiding ``speed limits''~\cite{Karzig13,Scheurer13,Karzig14,Karzig15} and the role of measurement in saturating them~\cite{Knapp16}; and improved methods of producing phase gates which, together with topologically protected operations, enable computational universality~\cite{Karzig15a}.

In this paper, we have considered these new theoretical developments in the
context of realistic experimental implementations in order to design scalable MZM-based quantum computing architectures with a variety of possible modules. All of our designs feature $(i)$ parallel topological wires connected into units with appreciable charging energy and $(ii)$ common measurement-based approaches that use proximate quantum dots and/or interferometry to enact all operations necessary for achieving fault-tolerant universal quantum computation.  These architectures display some similarities to the surface-code setups introduced recently in Refs.~\onlinecite{Terhal12,Landau16,Plugge16a}, but seek to leverage the topological quantum information processing afforded by MZMs, rather than pursuing active error correction.  Our study is instead closer in spirit to the parallel works of Refs.~\onlinecite{Vijay16b,Plugge16b}, but goes beyond these works in designing two-dimensional arrays, rather than few-qubit arrangements.

While we primarily focused on issues pertinent for long-term circuit designs, there are many interesting shorter-term goals for investigating the basic operating principles in relatively simple setups.  Demonstrating the ability to perform fermion parity measurements of MZMs presents one notable target given the prominence of measurements in our proposed schemes.  In this regard, it is worth commenting that relying only on parallel topological wires entails a potential challenge: the measurement visibility for certain MZM pairs within a given unit can be ``accidentally'' low, as discussed in Section~\ref{sec:2MZMmodel}.  We stress, however, that many factors (e.g., higher-order band structure corrections, orbital magnetic field effects, or additional spin-orbit couplings) are expected to alleviate this issue.  Optimizing the visibility for such cases poses a worthwhile problem both for theory and experiment.  Other measurement issues also warrant further attention.  Contrary to the topological qubits themselves, interferometric measurements are not immune to dephasing, which can hamper visibility.  {  It is important to establish the length scales over which 1DTSs can serve as effective coherent links, the time scales required to perform a measurement, and the length and time scales over which we can resolve the state of MZMs coupled through intervening quantum dots.}

\begin{figure}
	\includegraphics[width=0.9\columnwidth]{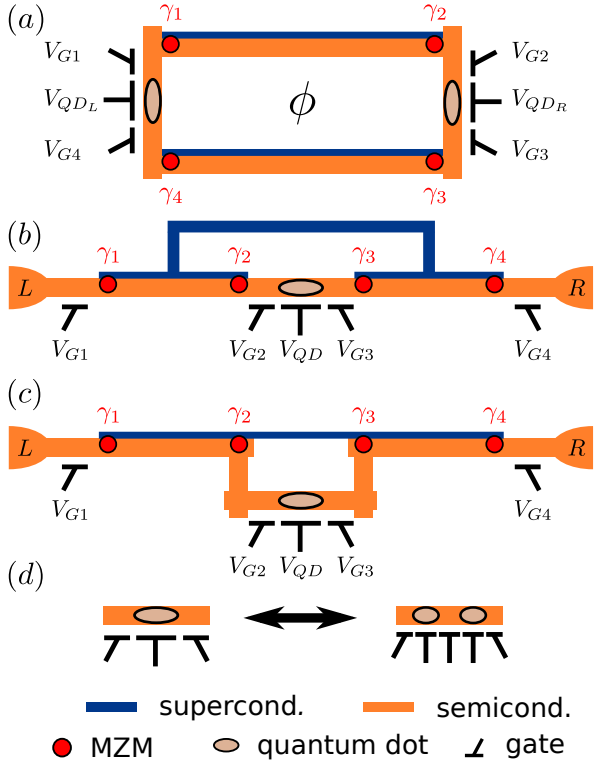}
	\caption{Examples of designs for experiments that demonstrate some of the basic operating principles in our scalable quantum computing architectures. (a) Experimental test of long distance coherent transport through floating 1DTSs. { Two long wires are coated with a superconductor and tuned into the topological regime.} If single electron transport is coherent, the hybridization of left and right quantum dots should show Aharonov-Bohm oscillations when changing the enclosed flux $\phi$. (b) and (c) { Experimental test of QPP rate, MZM hybridization, and measurement functionality of a topological qubit.} These { single tetron} configurations contain the minimal structure for a topological qubit. The left and right leads can be used to first tune the system into the topological regime by checking for zero bias peaks associated with $\gamma_1$ and $\gamma_4$. The central dot allows one to measure $i\gamma_2\gamma_3$, e.g., using charge sensing. The apparatuses shown in (b) and (c) differ in fabrication details. In (b) the quantum dot is defined on the same nanowire as the MZMs in a region where the superconducting shell was etched away. A superconducting bridge joins the two superconducting shells. In (c) no superconducting bridge is needed. A non-topological region between $\gamma_2$ and $\gamma_3$ is created by etching away the semiconducting part of the nanowire, or possibly by gating. The quantum dot is defined in a nearby nanowire connected to the 1DTS composed of four joined wires~\cite{BakkersLeo}. Note that the distance between $\gamma_2$ and $\gamma_3$ has to be much larger than the superconducting coherence length. (d) For enhanced measurement flexibility, additional gates allow one to replace single dot configurations by double dot configurations.}
	\label{fig:experiments}
\end{figure}

These measurement-centric issues can be addressed in the framework of experiments relevant for quantum information.  We partition these experiments into groups involving progressively more complex device geometries:

\emph{Two separate floating 1DTSs.} In order to test the length scales over which floating 1DTSs can be used as coherent links, a device with two floating nanowires, as shown in Fig.~\ref{fig:experiments}(a), could be used to realize an interference experiment that is conceptually similar to the proposal in Ref.~\onlinecite{Fu10}. Two 1DTSs act as the arms of an Aharonov-Bohm interferometer. If coherence is maintained, the tunneling between a quantum dot on the left and one on the right is modulated by the enclosed flux. Although this device cannot access properties of topological qubits, since the MZM parities are fixed by the individual 1DTSs' charging energies, such an experiment is a crucial test of the concept of using floating 1DTSs as coherent links.  { Moreover, such a system can provide strong evidence for topological superconductivity by observing $\pi$ shifts in the interference pattern when the dimensionless gate voltage applied to one 1DTS changes by $1$ (indicating the parity of the 1DTS has flipped).  }

\emph{Single tetron.} A device with four MZMs on a single superconducting island with charging energy, i.e., a tetron, features the minimal number of MZMs that yields a ground-state degeneracy and constitutes a single topological qubit. Figs.~\ref{fig:experiments}(b) and (c) show devices designed for first-generation experiments.  A wealth of information can already be gleaned from such  systems.  For example, one can partly characterize the qubit's stability by continuously measuring the parity of a given MZM pair. In the devices shown in Figs.~\ref{fig:experiments}(b) and (c), the parity $i\gamma_2\gamma_3$ can be measured via the central quantum dot. QPP events would manifest as telegraph noise in the signal of such a continuous measurement, provided subsequent instances of QPP events are separated by sufficiently long times. The extracted parity lifetimes would help quantify the suppression of QPP events by charging energy, as well as limitations from thermally-excited or non-equilibrium excited quasiparticles within a tetron. In the regime of very small QPP rates, the central quantum dot allows us to quantify the hybridization between MZM pairs.

A measurement of $i\gamma_2\gamma_3$ initializes the qubit in an $X$ eigenstate. After performing such a measurement and turning off the couplings to the quantum dot, hybridization due to tunneling between $\gamma_1$ and $\gamma_2$ or between $\gamma_3$ and $\gamma_4$ will split the ground state degeneracy, acting as a perturbation to the Hamiltonian proportional to $Z$. Thus, the probability distribution of outcomes for subsequent measurements of $i\gamma_2\gamma_3$ will reveal this energy splitting by varying the intermittent time intervals between measurements.
{ In Appendix~\ref{app:dephasing}, we elaborate on the details of how the coherence times may be extracted from such an experiment, as well as the effect of noise on the measurements.} If the quantum dot in the middle is replaced by two quantum dots, as depicted in Fig.~\ref{fig:experiments}(d),
we will have more flexibility in performing the measurement. For instance, it will be possible to turn on the couplings of the MZMs to the quantum dots
for a very short time interval and then subsequently measure their effect on the double dot system over a longer time interval.

{ By adapting the protocols from Ref.~\onlinecite{Aasen16}, tetrons can be used to detect the nontrivial fusion rules of MZMs. The implementation is especially straightforward if the setup allows the individual measurements of both $i\gamma_1 \gamma_2$ and $i \gamma_2 \gamma_3$ (this would be the case for either of the tetron designs of Sec.~\ref{sec:Tetrons}).  A measurement-only approach to fusion-rule detection can then be viewed as follows.  Measurement $i\gamma_{1} \gamma_2$ projects onto a particular fusion channel, initializing the topological qubit in a fixed direction of the Bloch sphere.  Subsequently measuring $i\gamma_{2}\gamma_3$ then projects the topological qubit onto an axis of the Bloch sphere rotated from the previous state's direction by $\pi/2$. This should yield equal probability of the two possible measurement outcomes, reflecting the two accessible fusion channels.

In principle, it is also possible to detect fusion rules in the devices show in Figs.~\ref{fig:experiments}(b) and (c), provided one can control the coupling between MZMs $\gamma_1$ and $\gamma_2$ or between $\gamma_3$ and $\gamma_4$, for example by tuning the topological gap via the external magnetic field. Once the coupling becomes appreciable, the environment { will relax the system to the ground state, effectively performing the projection into a fixed fusion channel.}}

Finally, { these devices allow one to implement an approximate $T$ gate. With a  well-timed pulse changing $V_{G,2}, V_{Q,D},$ and $V_{G,3}$, $\gamma_2$ and $\gamma_3$ can be coupled temporarily such that the dynamical phase accrued by the state of the system is $\frac{\pi}{8}p_{23}$.  }

\emph{Single hexon.}  Adding two more MZMs to the above module, i.e., building a hexon, provides the minimal architecture to test the measurement-only implementation of braiding transformations, as explained in Section~\ref{sec:hexon_qi_basics}. Moreover, this setup allows for more advanced approaches to implementing $T$ gates~\cite{Karzig15a}.

\emph{Two tetrons.}  The final basic operation needed for quantum computing, namely four-MZM measurement, can be demonstrated with two tetrons each realizing a single topological qubit.  Experimental validation requires some care to ensure that the implementation of the measurement does not unintentionally probe the state of any MZM pairs within the quartet of MZMs whose joint parity is being measured.  For example, suppose that we wish to measure $-\gamma_1^{(1)} \gamma_2^{(1)} \gamma_1^{(2)}\gamma_2^{(2)} = Z^{(1)} Z^{(2)}$, but inadvertently project onto an eigenstate of $i\gamma_1^{(1)}\gamma_2^{(1)}=Z^{(1)}$ in the process. This error can be detected by initializing the system in the state $\frac{1}{2}\left( \left| 0 \right\rangle + \left| 1 \right\rangle \right)\left( \left| 0 \right\rangle + \left| 1 \right\rangle \right)$, performing the measurement (intended to be) of $Z^{(1)} Z^{(2)}$, and then performing a measurement of $Z^{(1)}$. If the final measurement does not yield both possible outcomes with equal probability, then it indicates that the measurements are not performing as intended. A battery of similar tests may be used to more precisely characterize errors in the measurements.

Together, these experiments test much of the physics underlying our scalable designs.  Outcomes of even the simplest tests should discriminate among the various possible { qubit designs} that we proposed and inform inevitable refinements.  Yet another issue that should factor into eventual designs is circuit calibration, in the sense of ensuring that each individual 1DTS wire resides in its topological phase for systems supporting a large number of qubits. One could, of course, view successful implementation of the preceding experiments as calibration, though coarser methods that merely indicate the presence of MZMs, rather than information about their quantum states, are clearly desirable.  We expect that the interferometric measurements involving quantum dots, which we invoked for computation, also suffice for this purpose, though detailed studies would be certainly be useful.

A further research topic is to understand how much advantage can be gained by using architectures that allow more general measurements. Even using only a fairly limited set of two-qubit measurements, we were able to generate universal Clifford operations.  However, more general measurements simplify the implementation of certain computational operations and, thus, might allow quantum algorithms to be implemented using fewer measurements in total.  This leads to a trade-off worth investigating further.

{ Finally, we note that tailored algorithms for our measurement-based architectures can significantly increase the efficiency of our designs. Efficient algorithms will, in general, differ from the standard literature, which usually relies on a set of unitary Clifford gates and single-qubit measurements. For example, instead of using several CNOT gates for state injection or swap operations, where each CNOT gate requires a set of single-qubit and multi-qubit measurements, it will be more efficient to directly perform the desired operation without using CNOT gates (see the example in Sec.~\ref{sec:tetron-qi-basics} -- \emph{S gate}). The search for such tailored algorithms will be an important subject of future research.}

\section*{Acknowledgments}

It is a pleasure to acknowledge inspiring conversations with Anton Akhmerov, Alexander Altland, David Clarke, Mingtang Deng, Reinhold Egger, Joshua Folk, Leo Kouwenhoven, Deividas Sabonis, Eran Sela, Saulius Vaitiekenas, and David Wecker. We acknowledge the Aspen Center for Physics, where parts of this work where performed and which is supported by National Science Foundation grant PHY-1066293.  J.A.~gratefully acknowledges support from the National Science Foundation through grant DMR-1341822; the Caltech Institute for Quantum Information and Matter, an NSF Physics Frontiers Center with support of the Gordon
and Betty Moore Foundation through Grant GBMF1250; and the Walter Burke Institute for Theoretical Physics at Caltech.  C.K.~acknowledges support by the National Science Foundation
Graduate Research Fellowship Program under Grant No.
DGE 1144085. Y.O.~acknowledges support by the Israel Science Foundation (ISF), Deutsche Forschungsgemeinschaft (Bonn) within the network CRC TR
183, and the European Research Council under the European Community’s Seventh Framework Program (FP7/2007-2013)/ERC Grant agreement No. 340210. K.F. and S.P. acknowledge funding by the Danish National Research Foundation and from the Deutsche Forschungsgemeinschaft (Bonn) within the network CRC TR 183.
C.M.M.~thanks the Danish National Research Foundation and Villum Foundation for support.

\appendix

\section{Parity-dependence of MZM island-quantum dot energies}
\label{app:MZMmodels}

In this appendix, we discuss the model considered in Section~\ref{sec:4MZMmodel}. Let us consider the following Hamiltonian for the two MZM islands
\begin{equation}
H_0 = \sum_{a=1,2} H_{\text{BCS},a}+H_{C,a}
,
\end{equation}
where $a=1$ and $2$ label the two islands. The first term describes quasiparticle excitations in the island and the second term corresponds to the charging energy. In the low-energy approximation (i.e.,  energies much smaller than the superconducting gap), one can write the effective Hamiltonian $H_0$ in terms of the MZMs. In order to demonstrate this fact, we consider a toy model for a MZM island in which it is written as a collection of discretized Majorana wires with the same superconducting phase $\phi$.  In the dimerized limit, the Hamiltonian for each wire is given by~\cite{Kitaev01}
\begin{equation}
H_{\text{wire}} = -\Delta_P \sum_{k=1}^{M-1} \left( c_k^\dagger -e^{i\phi}c_k\right)\left( c_{k+1} +e^{-i\phi}c_{k+1}^\dagger\right)
\end{equation}
where $c$ and $c^\dagger$ respectively correspond to the fermion annihilation and creation operators in the wire and $\Delta_P$ is the induced p-wave superconductor gap.  The operator $e^{i\phi}$, where $\phi$ is the phase of the superconductor, adds a Cooper pair to the MZM island.

In order to carefully track phase factors, it is helpful to employ the number-conserving formalism from Ref.~\onlinecite{Chew16}.  { (For other number-conserving descriptions of MZMs see, e.g., Refs.~\onlinecite{Fu10,Flensberg11,Zazunov11,Fidkowski11b,Hutzen12,Sau11b,Beri12,Beri13,Altland13,Lutchyn16}.)} In particular, two manifestly physical operators that commute with $H_{\text{wire}}$ are
\begin{align}
\Gamma_{c,1}^\dagger &= c_1^\dagger+e^{i\phi}c_1
\\ \Gamma_{c,M}^\dagger &= i\left( c_M^\dagger -e^{i\phi}c_M\right).
\end{align}
The operators $\Gamma_{c,1}^\dagger$ and $\Gamma_{c,M}^\dagger$ add a charge to the MZM island and, therefore, do not commute with the number-conserving Hamiltonian $H_{C,a}$ given in Eq.~(\ref{eq:HCj}).  However, the charge-neutral combination
\begin{equation}
\Gamma_{c,1}^\dagger \Gamma_{c,M} = e^{-i\phi}\Gamma_{c,1}^\dagger \Gamma_{c,M}^\dagger
\end{equation}
does commute with $\hat{N}_{S,a}$.  Furthermore, as $i\Gamma_{c,1}^\dagger \Gamma_{c,M}$ squares to identity and anticommutes with $\Gamma_{c,1}$ and $\Gamma_{c,M}$, it counts the fermion parity of the MZM island.  

For the system depicted in the right panel of Fig.~\ref{fig:mst}, we label the four 1DTSs $1,2,3,4$, matching the corresponding MZM labels in the figure. We write the corresponding fermion operators of the 1DTSs as $c_{k}^{(j)}$ and $c_{k}^{(j)\dagger}$.  Then we can define
\begin{align}
\Gamma_1 \equiv \Gamma_{c^{(1)},M}, && \Gamma_2 \equiv \Gamma_{c^{(2)},M}, && \Gamma_3 \equiv \Gamma_{c^{(3)},1}, && \Gamma_4 \equiv \Gamma_{c^{(4)},1}.
\end{align}
{ We define fermion parities on the left and right MZM islands as eigenvalues of the operators $i\Gamma_1^\dagger\Gamma_2$ and $i\Gamma_3^\dagger\Gamma_4$.  That is, 
\begin{align}
i\Gamma_1^\dagger \Gamma_2\ket{p_{12}}=p_{12}\ket{p_{12}}, &&  i\Gamma_3^\dagger \Gamma_4\ket{p_{34}}=p_{34}\ket{p_{34}}.
\end{align}}

Let us now derive the tunneling Hamiltonians given in Secs.~\ref{sec:2MZMmodel} and \ref{sec:4MZMmodel} for tunneling between the quantum dots and MZMs.  A natural model of the coupling between the MZM islands and the quantum dots is that fermions from the dot can hop into the fermionic mode at the end of the 1DTSs.  This corresponds to a Hamiltonian of the form
\begin{equation}
\label{eq:Htunncd}
\begin{split}
H_{\text{tunn}} &= -\left( t_1 f_1^\dagger c_{M}^{(1)} +t_2 f_2^\dagger c_{M}^{(2)} +t_3 f_1^\dagger c_{1}^{(3)} +t_4 f_2^\dagger c_{1}^{(4)} \right)
\\ &\quad +\text{h.c.}
\end{split}
\end{equation}
Recall from Eq.~(\ref{eq:HQD2}) that the quantum dot fermionic operators are $f_a$ and $f_a^\dagger$.  Eq.~(\ref{eq:Htunncd}) reduces to Eq.~(\ref{eq:Htun4}) when we project the Hamiltonian to the low-energy subspace described by the operators $\Gamma_j$
\begin{align}
c_{M}^{(1)}& \to \frac{i}{2} \Gamma_1 , & c_{M}^{(2)} &\to \frac{i}{2}\Gamma_2 , & c_{1}^{(3)} &\to \frac{1}{2}\Gamma_3 , & c_{1}^{(4)} &\to \frac{1}{2}\Gamma_4 ,
\end{align}
and then rewrite the $\Gamma_j$ operators in terms of the conventional MZM operators~\cite{Read00}
\begin{equation}
\gamma_j = e^{-i\phi_j/2}\Gamma_{j}^\dagger.
\end{equation}

The Hamiltonian of the total system is
\begin{equation}
H_{\text{tot}}=H_0+H_{\text{QD}}+H_{\text{tunn}}.
\end{equation}
{ It is convenient to use the basis of lowest-energy eigenstates of the decoupled MZM islands.  We write this basis as $\ket{N_{S,1},N_{S,2};p_{12};p_{34}}$.  For a given ground state $\ket{0,0;p_{12},p_{34}}$, the five lowest-energy states related through $H_{\text{tot}}$ are
\begin{align}
\ket{0} &= \ket{0, 0;p_{12},p_{34}}
\\ \ket{1} &= \ket{1, 0; -p_{12}, p_{23}} 
= \Gamma_1^\dagger \ket{0}= i p_{12} \Gamma_2^\dagger\ket{0}
\\ \ket{2} &= \ket{-1, 0; -p_{12}, p_{34}} 
= \Gamma_1 \ket{0}=ip_{12} \Gamma_2 \ket{0}
\\ \ket{3} &= \ket{0,1; p_{12}, -p_{34}} 
= \Gamma_3^\dagger \ket{0} = ip_{34} \Gamma_4^\dagger\ket{0}
\\ \ket{4} &= \ket{0, -1; p_{12}, -p_{34}} 
= \Gamma_3 \ket{0} = ip_{34} \Gamma_4 \ket{0}.
\end{align}
We write the lowest-energy quantum dot states in the basis $\ket{n_{f,1},n_{f_2}}$.  The four lowest-energy quantum dot states are
\begin{align}
\ket{\tilde{0}} &=\ket{0, 0}
\\ \ket{\tilde{1}}&= \ket{1, 0}
\\ \ket{\tilde{2}}&= \ket{0, 1}
\\ \ket{\tilde{3}}&= \ket{1, 1}.
\end{align}
}

In this basis, $H_0$ and $H_{\text{QD}}$ take the form
\begin{align}
H_0 &= \sum_\mu E_\mu \ket{\mu}\bra{\mu} \otimes \tilde{\openone}
\\ H_{\text{QD}}& = \openone \otimes \sum_\beta \epsilon_\beta \ket{\tilde{\beta}}\bra{\tilde{\beta}}
\end{align}
where energies $E_\mu$ of the decoupled MZM islands are
\begin{align}
E_0 &= E_{C,1} N_{g,1}^2+\!E_{C,2} N_{g,2}^2 ,  \\
E_{1} &= E_{C,1} \left(1-N_{g,1}\right)^2 +E_{C,2}   N_{g,2}^2 , \\
E_{2} &= E_{C,1} \left( 1+N_{g,1}\right)^2 +E_{C,2}   N_{g,2}^2 , \\
E_{3} &= E_{C,1} N_{g,1}^2 +E_{C,2} \left( 1 -N_{g,2}\right)^2 , \\
E_{4} &= E_{C,1} N_{g,1}^2 +E_{C,2} \left( 1 +N_{g,2}\right)^2
\end{align}
and energies $\epsilon_\beta$ of the decoupled double-dot system are
\begin{align}
\label{eq:epsilon_0}
\epsilon_0 &= \varepsilon_{C,1} n_{g,1}^2 +\varepsilon_{C,2} n_{g,2}^2+\varepsilon_M n_{g,1} n_{g,2}
\\ \epsilon_1 &= \varepsilon_{C,1} \left(1- n_{g,1}\right)^2+h_1 +\varepsilon_{C,2} n_{g,2}^2 -\varepsilon_M(1-n_{g,1})n_{g,2}
\\ \epsilon_2 &= \varepsilon_{C,1} n_{g,1}^2 +\varepsilon_{C,2}\left( 1- n_{g,2}\right)^2+h_2-\varepsilon_Mn_{g,1}(1-n_{g,2})
\\ \epsilon_3&= \varepsilon_{C,1} \left(1-n_{g,1}\right)^2 +h_1+\varepsilon_{C,2} \left(1- n_{g,2}\right)^2+h_2 \nonumber
\\ &\quad  +\varepsilon_M (1-n_{g,1})(1-n_{g,2}).
\label{eq:epsilon_3}
\end{align}
For brevity, we now define the shorthand notation ${E_{\mu,\tilde{\beta}}= E_{\mu}+\epsilon_\beta}$.

The tunneling Hamiltonian in Eq.~(\ref{eq:Htun4}) becomes
\begin{align}
H_{\text{tunn}} &= - \frac{it_1}{2} \left( \ket{2}\bra{0}+\ket{0}\bra{1} \right) \otimes \left(\ket{\tilde{1}}\bra{\tilde{0}}+\ket{\tilde{3}}\bra{\tilde{2}} \right) \nonumber
\\ &\quad - p_{12}\frac{t_2}{2} \left( \ket{2}\bra{0} -\ket{0}\bra{1} \right) \otimes \left(\ket{\tilde{2}}\bra{\tilde{0}}-\ket{\tilde{3}}\bra{\tilde{1}} \right)
\nonumber\\ &\quad - \frac{t_3}{2} \left(\ket{4}\bra{0}+\ket{0}\bra{3}  \right) \otimes \left( \ket{\tilde{1}}\bra{\tilde{0}}+\ket{\tilde{3}}\bra{\tilde{2}}\right)
\nonumber\\ &\quad + p_{34}\frac{it_4}{2} \left( \ket{4}\bra{0}-\ket{0}\bra{3}\right) \otimes \left( \ket{\tilde{2}}\bra{\tilde{0}}-\ket{\tilde{3}}\bra{\tilde{1}} \right) +\text{h.c.}
\end{align}

\begin{figure}
	\includegraphics[width=.9\columnwidth]{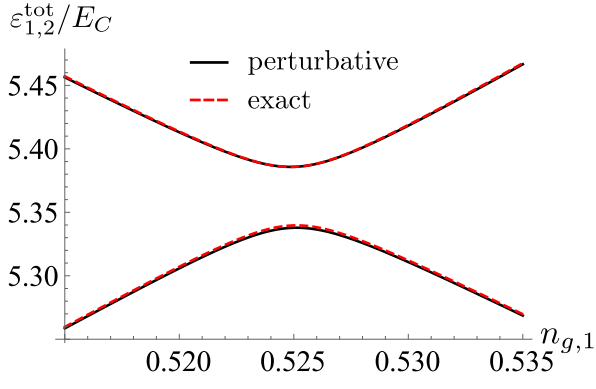}
	\caption{Agreement between perturbative (solid black) and exact (dashed red) parity-dependent energies $\varepsilon^{\text{tot}}_{1}$ and $\varepsilon^{\text{tot}}_{2}$ in units of $E_C$.  We use the parameter values $N_{g,a}=0$, $n_{g,2}=n_{g}^*$,  $p=p_{12}p_{34}=1$,  $\varepsilon_{C,a}=10 E_C$, $h=E_C/2$, $\varepsilon_M=E_C/2$, $t_1=0.1 E_C$, and $t_{j \neq 1}=0.2 E_C$.
	}
	\label{fig:PT}
\end{figure}

{ As discussed in the main text, for given values of $p_{12}$ and $p_{34}$ there are four states in the low-energy subspace when the quantum dots are tuned near their degenerate point.  Two of the states are superpositions of $\ket{0}\otimes\ket{\tilde{0}}$ and $\ket{0}\otimes\ket{\tilde{3}}$.  The energies of these two states are independent of the MZM parity to second order $t/E_C$, see Eqs.~(\ref{eq:veps0}) and (\ref{eq:veps3}).  

The other two states are superpositions of $\ket{0}\otimes\ket{\tilde{1}}$ and $\ket{0}\otimes\ket{\tilde{2}}$.  The corresponding energies are parity dependent, see Eqs.~(\ref{eq:veps1}) and (\ref{eq:veps2}).  We can write down an effective Hamiltonian for this parity-dependent low-energy state subspace to second order in $t/E_C$ as
\begin{equation}\label{eq:Hpert}
H_{\text{eff}} = H^{(0)}+H^{(2)}
\end{equation}
where $H^{(0)}$ is the zeroth-order Hamiltonian
\begin{equation}
H^{(0)}= \left( \begin{array}{cc}  E_{0,\tilde{1}} & 0 \\ 0 & E_{0,\tilde{2}} \\ \end{array} \right)
\end{equation}
and $H^{(2)}$ is the second-order Hamiltonian
\begin{widetext}
\begin{equation}\label{eq:HPertGen}
H^{(2)} = \frac{1}{4} \left( \begin{array}{cc}   \frac{|t_1|^2}{E_{0,\tilde{1}}-E_{1,\tilde{0}}} +\frac{|t_2|^2}{E_{0,\tilde{1}}-E_{2,\tilde{3}}}+\frac{|t_3|^2}{E_{0,\tilde{1}}-E_{3,\tilde{0}}} +\frac{|t_4|^2}{E_{0,\tilde{1}}-E_{4,\tilde{3}}} &
a \\
 a^* &
  \frac{|t_1|^2}{E_{0,\tilde{2}}-E_{2,\tilde{3}}} +\frac{|t_2|^2}{E_{0,\tilde{2}}-E_{1,\tilde{0}}}+\frac{|t_3|^2}{E_{0,\tilde{2}}-E_{4,\tilde{3}}} +\frac{|t_4|^2}{E_{0,\tilde{2}}-E_{3,\tilde{0}}} \\
\end{array}\right)
\end{equation}
where the off-diagonal elements are
\begin{equation}
\begin{split}
& a = -  \frac{p_{12} t_1 t_2^*}{2}\left( \frac{ 2E_{0,\tilde{1}}-E_{1,\tilde{0}}-E_{2,\tilde{3}} }{\left(E_{0,\tilde{1}}- E_{1,\tilde{0}}\right)\left(E_{0,\tilde{1}}-E_{2,\tilde{3}}\right)} + \frac{ 2E_{0,\tilde{2}}-E_{2,\tilde{3}}-E_{1,\tilde{0}}}{\left(E_{0,\tilde{2}}-E_{2,\tilde{3}}\right) \left(E_{0,\tilde{2}}-E_{1,\tilde{0}} \right)}\right)
\\ &\quad \quad
- \frac{p_{34} t_3 t_4^*}{2} \left( \frac{2E_{0,\tilde{1}}-E_{3,\tilde{0}}- E_{4,\tilde{3}}  }{\left(E_{0,\tilde{1}}-E_{3,\tilde{0}}\right) \left(E_{0,\tilde{1}}-E_{4,\tilde{3}} \right)}+ \frac{2E_{0,\tilde{2}}-E_{4,\tilde{3}}-E_{3,\tilde{0}} }{\left( E_{0,\tilde{2}}-E_{4,\tilde{3}}\right)\left( E_{0,\tilde{2}}-E_{3,\tilde{0}}\right)}  \right) .
\end{split}
\end{equation}
\end{widetext}
}
The perturbatively computed energies of Eq.~(\ref{eq:Hpert}) are in good agreement with the energy values obtained numerically by exact diagonalization of Eq.~(\ref{eq:Htot}), as shown in Fig.~\ref{fig:PT}.

\begin{figure}
	\begin{centering}
		\includegraphics[width=.9\columnwidth]{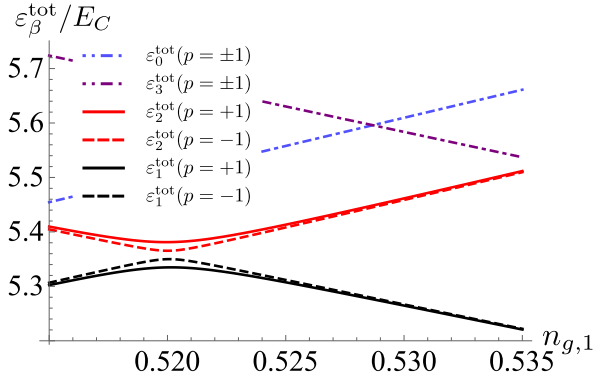}
		\par\end{centering}
	\caption{Energy $\varepsilon^{\text{tot}}_\beta$ as a function of $n_{g,1}$ for $n_{g,2}=n_g^*-0.01$. We use the parameter values $N_{g,a}=0$, $\varepsilon_C=10 E_C$, $h=E_C/2$, $\varepsilon_M=E_C/2$, $t_1=0.1 E_C$, and $t_2=0.2 E_C$. The parity-independent energies $\varepsilon^{\text{tot}}_0$ (shown in blue dot-dashed) and $\varepsilon^{\text{tot}}_3$ (shown in purple dot-dashed).  Parity-dependent energies $\varepsilon^{\text{tot}}_1$ (black) and $\varepsilon^{\text{tot}}_2$ (red) are shown with solid curves for even parity and dashed curves for odd parity.}
	\label{fig:dot2}
\end{figure}

We now discuss dependence of the energy spectrum on different physical parameters.  Let $n_g^*=\left( 1+h/\varepsilon_C\right)/2$, the charge-degenerate point when $\varepsilon_M=0$.  When $N_{g,1} \neq 0$ and $n_{g,2}=n_g^*$, the parity-dependent energies are no longer symmetric about the point $n_{g,1}=n_g^*$. Since $E_C$ is chosen to be much smaller than $\varepsilon_C$, this asymmetry is small.  When $N_{g,1}=0$ and $n_{g,2}\neq n_g^*$, as shown in Fig.~\ref{fig:dot2}, the crossing of the parity-independent energies $\varepsilon^{\text{tot}}_{0}$ and $\varepsilon^{\text{tot}}_{3}$ shifts horizontally with respect to the avoided level crossing of the parity-dependent energies $\varepsilon^{\text{tot}}_{1}$ and $\varepsilon^{\text{tot}}_{2}$.

When $\varepsilon_M=0$, as is appropriate for four-MZM parity measurements involving MZMs from two different hexons, the level crossing of $\varepsilon^{\text{tot}}_{0}$ and $\varepsilon^{\text{tot}}_{3}$ shifts down and the avoided crossing of $\varepsilon^{\text{tot}}_{1}$ and $\varepsilon^{\text{tot}}_{2}$ shifts up, so that both crossings are at the same energies.  In this situation, it is important to ensure that either the ground state corresponds to the parity-dependent energy $\varepsilon^{\text{tot}}_1$, or that the states with parity-independent energies $\varepsilon^{\text{tot}}_{0}$ and $\varepsilon^{\text{tot}}_{3}$ are inaccessible.

\section{Transmon measurement}
\label{app:transmon}

Following the discussion in Ref.~\onlinecite{Knapp16}, a projective parity measurement using a transmon-type dispersive readout would require resolving a frequency of
\begin{equation}\label{eq:deltaomega}
\begin{split}
\Delta\omega &= \frac{g^2}{2}\Big( \frac{1}{\delta \omega +\varepsilon^{\text{tot}}_1(p=+1) -\varepsilon^{\text{tot}}_1(p=-1)}
\\ &\quad- \frac{1}{\delta \omega +\varepsilon^{\text{tot}}_2(p=+1)-\varepsilon^{\text{tot}}_2 (p=-1)}\Big).
\end{split}
\end{equation}
Using the perturbative expressions for $\varepsilon^{\text{tot}}_{1}$ and $\varepsilon^{\text{tot}}_{2}$, given in Eqs.~(\ref{eq:veps1}) and (\ref{eq:veps2}), with tunneling amplitudes ${|t_2|=|t_3|=|t_4|=t>0}$ and ${|t_1|=0.5t}$, we find the maximal dispersive shift (when $t_1^*t_2t_3t_4^*$ is real) to be
\begin{equation}
\Delta\omega=\frac{g^2}{4\delta\omega^2} \frac{t^2}{E_C}.
\end{equation}
For the frequency estimates given in Ref.~\onlinecite{Knapp16}, ${g/2\pi \approx 40}$~MHz, ${\delta\omega/2\pi \approx 200}$~MHz, and estimating ${E_C \approx 160~\mu}$eV and $t=0.2 E_C$, Eq.~(\ref{eq:deltaomega}) gives  $\Delta\omega \approx 100$~MHz, which falls well within the range of transmon sensitivity.

\section{ Measurement procedure and dephasing in a single module with four MZMs}
\label{app:dephasing}

{ In this appendix, we elaborate on how to measure the MZM hybridization in the devices shown in Fig.~\ref{fig:experiments}(b) and (c), and we discuss the general effect of noise on the charge sensing measurements.  We first consider the case without noise.

\subsection{Measurement of the MZM hybridization}

We assume that one of the devices shown in Fig.~\ref{fig:experiments}(b) and (c) has been tuned into the topological phase, such that there are two MZMs on either leg of the qubit.  The gates $V_{G1}$ and $V_{G4}$ are tuned such that the qubit is decoupled from the left and right leads.  The gates $V_{G2}$ and $V_{G3}$ respectively control the tunneling amplitude $t_2$  and $t_3$ for an electron to tunnel between the quantum dot and the corresponding MZM $\gamma_2$ or $\gamma_3$.

The system is described by the Hamiltonian 
\begin{equation}
\label{eq:Hloop}
H=H_C+H_{\text{tunn}}+H_{\text{hyb}}.
\end{equation}
For simplicity, we consider the system at $N_g=0$ and at the charge degenerate point $n_g=n_g^*$.  We add a constant to the charging energy Hamiltonian to ignore the (constant) energy contribution from the quantum dot:
\begin{equation}
\begin{split}
H_C&=E_C\hat{N}_S^2+\varepsilon_C\left(\hat{n}_f-n_g^*\right)^2+h \hat{n}_f
-\varepsilon_C {n_g^*}^2
\\ &=E_C \hat{N}_S^2.
\end{split}
\end{equation}
 The tunneling Hamiltonian is given by
\begin{equation}
H_{\text{tunn}}= -\frac{i}{2} t_2 f^\dagger e^{-i\phi/2}\gamma_2 -\frac{1}{2}t_3 f^\dagger e^{-i\phi/2}\gamma_3 +\text{h.c.},
\end{equation}
where, as in Appendix~\ref{app:MZMmodels}, the operator $e^{i\phi/2}$ adds an electron to the MZM island and $f$ is the annihilation operator for the quantum dot.  

The last term in $H$ describes the hybridization between the MZMs. For simplicity, we assume that the direct overlap between $\gamma_2$ and $\gamma_3$ can be neglected, so the hybridization Hamiltonian takes the form 
\begin{equation}
H_{\text{hyb}}\approx \delta E_{12} i\gamma_1\gamma_2+\delta E_{34} i\gamma_3\gamma_4.
\end{equation}
The states of the system are spanned by the basis $\ket{N_s;p_{23},p_{14}}\otimes\ket{n_f}$.

The hybridization energies $\delta E_{12}$ and $\delta E_{34}$ are exponentially suppressed in the separation between MZMs $\gamma_1$ and $\gamma_2$ and MZMs $\gamma_3$ and $\gamma_4$, respectively. The combined MZM hybridization $B_x=\delta E_{12}+\delta E_{34}$ will lead to oscillations between the two low-energy states of the qubit (which are degenerate ground states for $B_x=0$).

The goal of the experiment is to quantify $B_x$ via the following protocol:
(1) initialize the system with a measurement of $i\gamma_2 \gamma_3$, (2) turn off the tunneling amplitudes $t_2$ and $t_3$ for a time $\tau_0$, (3) turn on the tunneling amplitudes $t_2$ and $t_3$ for a time $\tau_1$ and remeasure the system, e.g., with charge sensing.  We now demonstrate that this sequence will return a charge expectation value for the quantum dot $\langle \hat{n}_f\rangle$ that depends on the quantity $B_x$.  We further show that charge noise acting on the quantum dot does not cause the $B_x$ dependence of $\langle \hat{n}_f \rangle$ to decay in $\tau_1$. { In this analysis, we neglect the effect of exponentially suppressed thermal corrections.}

Assume that the system is initialized in the state ${\ket{0;-1,-1}\otimes\ket{1}}$ at time $\tau =0$.  If we assume ideal step functions for turning off and on the tunneling amplitudes, then the charge expectation value will be
\begin{equation}
\bra{\psi(\tau_0+\tau_1)} \hat{n}_f\ket{\psi(\tau_0+\tau_1)}
\end{equation}
where the state at $\tau_0+\tau_1$ is given by 
\begin{equation}
\ket{\psi(\tau_0+\tau_1)}=e^{-i H \tau_1}e^{-i \left( H_{\text{hyb}}+H_C\right)\tau_0} \ket{0;-1,-1}\otimes\ket{1}.
\end{equation}
Note that $H_C\ket{0;-1,-1}\otimes\ket{1}=H_C\ket{0;1,1}\otimes\ket{1}=0$.  We can write
\begin{equation}
\label{eq:slow_rabi}
\begin{split}
&e^{-i\left(H_{\text{hyb}}+H_C\right)\tau_0}\ket{0;-1,-1}\otimes \ket{1}
\\ &= e^{-iH_{\text{hyb}}\tau_0}\ket{0;-1,-1}\otimes\ket{1}
\\ &= \cos\left(B_x\tau_0 \right)\ket{0;-1,-1}\otimes\ket{1}-i\sin\left(B_x\tau_0\right)\ket{0;1,1}\otimes\ket{1}.
\end{split}
\end{equation}

 We choose $\tau_1$ such that $\tau_1\ll B_x^{-1}$.  This is easy to achieve as $B_x$ is an exponentially suppressed parameter.    Given this choice, we may safely neglect $H_{\text{hyb}}$ for the duration of time $\tau_1$ when the tunneling amplitudes are on.  For simplicity, until otherwise stated, we take the regime where the parity dependence of $H_{\text{tunn}}$ is strongest, corresponding to ${t_2=t_3=t}$, for which $\ket{0;-1,-1}\otimes\ket{1}$ is maximally coupled to the excited state $\ket{1;1,-1}\otimes\ket{0}$ and $\ket{0;1,1}\otimes \ket{1}$ is decoupled from $\ket{1;-1,1}\otimes\ket{0}$.  At this special point,
\begin{equation}
\label{eq:Ht-special}
\begin{split}
H_{\text{tunn}}&=- \frac{it}{2} f^\dagger e^{-i\phi/2} \left( \gamma_2-i\gamma_3\right)+\text{h.c.}
\\ &= -it f^\dagger e^{-i\phi/2} c_{23}+it^* e^{i\phi/2} c_{23}^\dagger f.
\end{split}
\end{equation}
From the second equality, we can see that ${H_{\text{tunn}}\ket{0;1,1}\otimes\ket{1}=0},$ as ${f^\dagger\ket{0;1,1}\otimes\ket{1}}={c_{23}^\dagger\ket{0;1,1}\otimes\ket{1}}{=0}$.

Diagonalizing Eq.~(\ref{eq:Ht-special}), we have
\begin{equation}\label{eq:HCHt}
\begin{split}
H_C+H_{\text{tunn}} =& \varepsilon_-\ket{g}\bra{g}
+\varepsilon_+\ket{e}\bra{e}
\\ & +\varepsilon_-'\ket{g'}\bra{g'}+\varepsilon_+'\ket{e'}\bra{e'},
\end{split}
\end{equation}
where $\ket{g}$ and $\ket{e}$ are the ground and excited states in the sector spanned by $\{ \ket{0;-1,-1}\otimes\ket{1}, \, \ket{1;1,-1}\otimes\ket{0} \}$ and $\ket{g'}$ and $\ket{e'}$ are the ground and excited states in the sector spanned by $\{ \ket{0;1,1}\otimes\ket{1}, \, \ket{1;-1,1}\otimes\ket{0} \}$. 

At the fine-tuned point $t_2=t_3 = t$, the primed sector is unaffected by the tunneling Hamiltonian, thus giving $\ket{g'}=\ket{0;1,1}\otimes\ket{1}$ and $\ket{e'}=\ket{1;-1,1}\otimes\ket{0}$, with corresponding energies $\varepsilon_-'=0$ and $\varepsilon_+'=E_C$.  In the unprimed sector, the ground and excited states will be
\begin{align}
\ket{g}=\frac{1}{\sqrt{\varepsilon_+^2+|t|^2}}\left( \varepsilon_+ \ket{0;-1,-1}\otimes\ket{1}-it^*\ket{1;1,-1}\otimes\ket{0} \right), \\
\ket{e}=\frac{1}{\sqrt{\varepsilon_+^2+|t|^2}}\left( -it \ket{0;-1,-1}\otimes\ket{1}+\varepsilon_+\ket{1;1,-1}\otimes\ket{0} \right).
\end{align}
with the corresponding energies
\begin{equation}
\varepsilon_{\mp}=\frac{1}{2}\left(E_C\mp \sqrt{E_C^2+4|t|^2}\right)
.
\end{equation}

It follows that the state at time $\tau_0+\tau_1$ may be written (neglecting the very small corrections depending on $B_x \tau_1$) as
\begin{equation}
\begin{split}
\ket{\psi(\tau_0+\tau_1)}&=\cos\left(B_x\tau_0\right) a_g(\tau_1)\ket{g}
\\ &\quad +\cos\left(B_x\tau_0\right)a_e(\tau_1)\ket{e}
\\ &\quad -i\sin\left(B_x\tau_0\right)\ket{g'}
\end{split}
\end{equation}
where the coefficients are given to leading order in $t/E_C$ by
\begin{align}
a_g(\tau_1)&=\frac{\varepsilon_+}{\sqrt{\varepsilon_+^2+|t|^2}}e^{-i\varepsilon_-\tau_1} \simeq \left( 1-\frac{1}{2} \frac{|t|^2}{E_C^2}\right)e^{-i\varepsilon_-\tau_1},
\\ a_e(\tau_1) &= \frac{it^*}{\sqrt{\varepsilon_+^2+|t|^2}}e^{-i\varepsilon_+\tau_1}\simeq \frac{it^*}{E_C}e^{-i\varepsilon_+\tau_1}
.
\label{eq:coefficients}
\end{align}

Thus, the quantum dot charge expectation value for the measured state is
\begin{equation}\label{eq:Qexp}
\begin{split}
&\bra{\psi(\tau_0+\tau_1)}\hat{n}_f \ket{\psi(\tau_0+\tau_1)}
\\ & =\cos^2\left(B_x\tau_0\right)\Big( |a_g(\tau_1)|^2 \bra{g}\hat{n}_f\ket{g} +|a_e(\tau_1)|^2 \bra{e}\hat{n}_f\ket{e}
\\ &\quad+ a_g(\tau_1)a_e^*(\tau_1) \bra{e}\hat{n}_f\ket{g} +a_g^*(\tau_1)a_e(\tau_1)\bra{g}\hat{n}_f\ket{e} \Big)
\\ &\quad+\sin^2\left(B_x\tau_0\right)
,
\end{split}
\end{equation}
where the charge matrix elements are given to leading order in $t/E_C$ by
\begin{align}
\bra{g}\hat{n}_f\ket{g}&= \frac{\varepsilon_+^2}{\varepsilon_+^2+|t|^2}\simeq \left( 1- \frac{|t|^2}{E_C^2}\right),
\\ \bra{e}\hat{n}_f\ket{e} &=\frac{|t|^2}{\varepsilon_+^2+|t|^2}\simeq \frac{|t|^2}{E_C^2},
\\ \bra{g}\hat{n}_f\ket{e} &= \bra{e}\hat{n}_f\ket{g}^*=\frac{-it\varepsilon_+}{\varepsilon_+^2+|t|^2} \simeq \frac{-it}{E_C}.
\end{align}

Therefore, the expectation value is
\begin{equation}
\begin{split}
&\bra{\psi(\tau_0+\tau_1)}\hat{n}_f \ket{\psi(\tau_0+\tau_1)} \\ 
&=1- \frac{2|t|^2}{E_C^2+4|t|^2}\cos^2\left(B_x\tau_0\right)
\\ &\quad +\frac{2|t|^2}{E_C^2+4|t|^2}\cos^2\left(B_x\tau_0\right)\cos\left(\sqrt{E_C^2+4|t|^2}\tau_1\right).
\end{split}
\end{equation}

\subsection{The effect of noise}

Turning on a finite tunneling amplitude essentially couples the MZM island to a charge qubit given by the occupation of the quantum dot.   The dominant source of dephasing and relaxation in charge qubits originates from the electrostatic coupling of the  charge to fluctuations in the background charges of the substrate~\cite{Paladino13}. 
We expect this to be the dominant source of noise for the experiment considered in the systems of Fig.~\ref{fig:experiments}(b) and (c) to come from the charge noise coupled to the dipole moment of the quantum dot-MZM island system.  The noise has both diagonal and off-diagonal components in the energy basis.  The corresponding Bloch equations, ignoring the exponentially small corrections from temperature, tell us that Eq.~(\ref{eq:coefficients})  should be modified in the presence of charge noise as
\begin{align}
|a_e(\tau_1)|^2&\to |a_e(\tau_1)|^2 e^{-\frac{\tau_1}{T_1}}
\\ |a_g(\tau_1)|^2 &\to 1+\left( |a_g(\tau_1)|^2-1 \right)e^{-\frac{\tau_1}{T_1}}
\\ a_e(\tau_1) a_g^*(\tau_1) &\to  a_g(\tau_1)a_e^*(\tau_1)e^{-\frac{\tau_1}{T_2}}.
\end{align}
Here, $T_1$ and $T_2$ are the energy and phase relaxation times of the hybridized MZM island-quantum dot system (not to be mistaken with the coherence times of the topological qubit, which we expect to be much larger).  $T_1$ and $T_2$ depend on the spectral density of the noise~\cite{Schoelkopf02}.

It follows that to lowest order in $t/E_C$,
\begin{equation}
\label{eq:Q}
\begin{split}
&\bra{\psi(\tau_0+\tau_1)} \hat{n}_f\ket{\psi(\tau_0+\tau_1)} \\ 
&=1-\frac{|t|^2}{\varepsilon_+^2+|t|^2}\cos^2\left(B_x\tau_0\right) 
\\ &\quad - \frac{|t|^2E_C}{\varepsilon_+\left(E_C^2+4|t|^2\right)}\cos^2\left(B_x\tau_0\right)e^{-\tau_1/T_1}
\\ &\quad +\frac{2|t|^2}{E_C^2+4|t|^2}\cos^2\left(B_x\tau_0\right)\cos\left(\sqrt{E_C^2+4|t|^2}\tau_1\right)e^{-\tau_1/T_2}
.
\end{split}
\end{equation}
Importantly,  at zero temperature, the charge expectation value has a $B_x$ dependent term that does not decay with $\tau_1$.

 We now comment on how $B_x$ may be extracted from a charge sensing measurement.  We assume that the measurement takes place over a time $\tau_1$ satisfying ${\text{max}[T_1,T_2]\ll \tau_1 \ll B_x^{-1}}$.  The purpose of the lower bound is to ensure that the exponentially decaying terms in Eq.~\eqref{eq:Q} may be neglected.  The upper bound is to justify neglecting the effects of $H_{\text{hyb}}$ when $H_t$ is turned on; violating the upper bound would result in a measurement of $\langle \hat{n}_f \rangle$  averaged over the two MZM parity states. Satisfying the upper bound, instead, ensures that the charge sensing measurement is a strong projective measurement for the MZM parity.  On the other hand, we assume that the time scales of the island-dot charge degrees of freedom are fast, so that the measurement for the quantum dot occupation is weak.  The measurement process therefore is carried out over the following steps. Initially, after a time $\tau_0$ the system will be in particular superposition of the even and odd parity qubit states determined by the phase $B_x \tau_0$ [see \eqref{eq:slow_rabi}]. The measurement collapses this superposition so that the hybridized system is either in the primed or the unprimed sector [see Eq.~\eqref{eq:HCHt}].  For our choice of tunneling amplitudes, the charge expectation value in the primed and unprimed sectors corresponds to $ 1$ or $1 -\frac{|t|^2}{E_C^2}$, respectively.  Repeating the experiment many times for fixed $B_x\tau_0$ gives the charge expectation value Eq.~\eqref{eq:Q} where the primed and unprimed sectors are weighted depending on the amplitudes of the initial superposition.  Finally, by varying $\tau_0$ we can detect the cosine squared dependence on $B_x$, and thus extract the MZM hybridization.  

The above calculation may be modified to consider the case where the tunneling amplitudes are turned off and on in a time scale slow compared $E_C^{-1}$, but fast compared to $B_x^{-1}$.  In this case, the transition from evolution with $H_{\text{hyb}}$ to evolution with $H$ can be made nearly adiabatic.  When the tunneling amplitudes are turned back on, the system will be in a superposition of the two ground states and will have no excited state component.  As such, ${\bra{\psi(\tau_0+\tau_1)}\hat{n}_f\ket{\psi(\tau_0+\tau_1)}}$ will have no decaying terms.

Finally, the results in this appendix survive beyond the point $t_2=t_3=t$. 
Away from this fine-tuned limit, the eigenstates of $H$ in the unprimed sector will be a mixture of ${\ket{0;-1,-1}\otimes\ket{1}}$ and ${\ket{1;1,-1}\otimes\ket{0}}$ and the eigenstates of $H$ in the primed sector will be a mixture fo $\ket{0;1,1}\otimes\ket{1}$ and $\ket{1;-1,1}\otimes\ket{0}$.  
Thus, $\ket{\psi(\tau_0+\tau_1)}$ will be a superposition of all four basis states.  Let $\ket{g}$ denote the ground state for the sector spanned by $\{ \ket{0;-1,-1}\otimes\ket{1}, \ket{1; 1,-1}\otimes\ket{0}\}$, as before, and $\ket{g'}$ denote the ground state for the sector spanned by $\{ \ket{0;1,1}\otimes\ket{1},\ket{1;-1,1}\otimes\ket{0} \}$.  Then, provided that $\bra{g}\hat{n}_f\ket{g}\neq \bra{g'}\hat{n}_f\ket{g'}$, the charge expectation value will still have a $B_x$ dependent term that is not decaying in $\tau_1$.

}


\section{Hexon Details}
\label{app:Hexdetails}

In this appendix, we prove Eq.~(\ref{eq:Xop}) and explain how to shuttle computational MZMs through the qubit. (In the following diagrammatic analysis, we neglect the unimportant overall constants.)

For the two-qubit entangling gate
\begin{equation}
W =\left(
\begin{array}{cccc}
1 & 0 & 0 & 0 \\
0 & i & 0 & 0 \\
0 & 0 & i & 0 \\
0 & 0 & 0 & 1 \\
\end{array}
\right)
,
\end{equation}
we begin by considering the qubit basis states of two hexons in the initial configurations shown below
\begin{widetext}
\begin{equation}
\ket{a,b}=
\pspicture[shift=-1.25](-2,-.75)(2,2)
\psset{arrowscale=1.4,arrowinset=0.15}
  \small
  \psline(-1.75,1.75)(-1.25,1.25)(-.75,1.75)
  \rput(-1.7,1.9){$\gamma_1$}  \rput(-.8,1.9){$\gamma_2$}
  \psline(-.5,1.75)(0,1.25)(.5,1.75)
  \rput(-.45,1.9){$\gamma_3$}  \rput(.45,1.9){$\gamma_4$}
  \psline(.75,1.75)(1.25,1.25)(1.75,1.75)
  \rput(.8,1.9){$\gamma_5$}  \rput(1.7,1.9){$\gamma_6$}
  \psline(-1.25,1.25)(0,0)(1.25,1.25) \rput(0,-.2){$a$}
\endpspicture
\pspicture[shift=-1.25](-2,-.75)(2,2)
\psset{arrowscale=1.4,arrowinset=0.15}
  \small
  \psline(-1.75,1.75)(-1.25,1.25)(-.75,1.75)
  \rput(-1.7,1.9){$\gamma_7$}  \rput(-.8,1.9){$\gamma_8$}
  \psline(-.5,1.75)(0,1.25)(.5,1.75)
  \rput(-.45,1.9){$\gamma_9$}  \rput(.35,1.9){$\gamma_{10}$}
  \psline(.75,1.75)(1.25,1.25)(1.75,1.75)
  \rput(.9,1.9){$\gamma_{11}$}  \rput(1.7,1.9){$\gamma_{12}$}
  \psline(-1.25,1.25)(0,0)(1.25,1.25) \rput(0,-.2){$b$}
\endpspicture.
\end{equation}
As in Fig.~\ref{fig:qubit}, $a,b\in\{0,1\}$ label the fermion parity even or odd states of the outermost pairs of MZMs in a given hexon.

Projecting the fusion channel of MZMs 4 and 5 to vacuum (e.g., using forced measurement) gives
\begin{equation}
\begin{split}
&\Pi_0^{(45)}\ket{a,b} =
\pspicture[shift=-1.25](-1.9,-.75)(1.9,2)
\psset{arrowscale=1.4,arrowinset=0.15}
  \small
  \psline(-1.75,1.75)(-1.25,1.25)(-.75,1.75)
  \rput(-1.7,1.9){$\gamma_1$}  \rput(-.8,1.9){$\gamma_2$}
  \rput(1.2,1.9){$\gamma_5$}  \rput(.05,1.9){$\gamma_4$}
  \psline(0,1.75)(.625,1.125)(1.25,1.75)
  \rput(-.45,1.9){$\gamma_3$}  \rput(1.7,1.9){$\gamma_6$}
  \psline(-1.25,1.25)(0,0)(.625,.625) \rput(0,-.2){$a$}
  \psline(1.75,1.75)(.625,.625)(-.5,1.75)
\endpspicture
\pspicture[shift=-1.25](-1.9,-.75)(1.9,2)
\psset{arrowscale=1.4,arrowinset=0.15}
  \small
  \psline(-1.75,1.75)(-1.25,1.25)(-.75,1.75)
  \rput(-1.7,1.9){$\gamma_7$}  \rput(-.8,1.9){$\gamma_8$}
  \psline(-.5,1.75)(0,1.25)(.5,1.75)
  \rput(-.45,1.9){$\gamma_9$}  \rput(.35,1.9){$\gamma_{10}$}
  \psline(.75,1.75)(1.25,1.25)(1.75,1.75)
  \rput(.9,1.9){$\gamma_{11}$}  \rput(1.7,1.9){$\gamma_{12}$}
  \psline(-1.25,1.25)(0,0)(1.25,1.25) \rput(0,-.2){$b$},
\endpspicture
\\ &= \frac{1}{\sqrt{2}} \left( \pspicture[shift=-1.25](-1.9,-.75)(1.9,2)
\psset{arrowscale=1.4,arrowinset=0.15}
  \small
  \psline(-1.75,1.75)(-1.25,1.25)(-.75,1.75)
  \rput(-1.7,1.9){$\gamma_1$}  \rput(-.8,1.9){$\gamma_2$}
  \psline(-.5,1.75)(0,1.25)(.5,1.75)
  \rput(-.45,1.9){$\gamma_3$}  \rput(.45,1.9){$\gamma_4$}
  \psline(.75,1.75)(1.25,1.25)(1.75,1.75)
  \rput(.8,1.9){$\gamma_5$}  \rput(1.7,1.9){$\gamma_6$}
  \psline(0,1.25)(-.625,.625)(-1.25,1.25)\rput(-.625,.425){$a$}
\endpspicture
\pspicture[shift=-1.25](-1.9,-.75)(1.9,2)
\psset{arrowscale=1.4,arrowinset=0.15}
  \small
  \psline(-1.75,1.75)(-1.25,1.25)(-.75,1.75)
  \rput(-1.7,1.9){$\gamma_7$}  \rput(-.8,1.9){$\gamma_8$}
  \psline(-.5,1.75)(0,1.25)(.5,1.75)
  \rput(-.45,1.9){$\gamma_9$}  \rput(.35,1.9){$\gamma_{10}$}
  \psline(.75,1.75)(1.25,1.25)(1.75,1.75)
  \rput(.9,1.9){$\gamma_{11}$}  \rput(1.7,1.9){$\gamma_{12}$}
  \psline(-1.25,1.25)(0,0)(1.25,1.25) \rput(0,-.2){$b$}
\endpspicture
+\pspicture[shift=-1.25](-1.9,-.75)(1.9,2)
\psset{arrowscale=1.4,arrowinset=0.15}
  \small
  \psline(-1.75,1.75)(-1.25,1.25)(-.75,1.75)
  \rput(-1.7,1.9){$\gamma_1$}  \rput(-.8,1.9){$\gamma_2$}
  \psline(-.5,1.75)(0,1.25)(.5,1.75)
  \rput(-.45,1.9){$\gamma_3$}  \rput(.45,1.9){$\gamma_4$}
  \psline(.75,1.75)(1.25,1.25)(1.75,1.75)
  \rput(.8,1.9){$\gamma_5$}  \rput(1.7,1.9){$\gamma_6$}
  \psline(0,1.25)(-.625,.625)(-1.25,1.25)\rput(-.625,.425){$a$}
  \pszigzag[coilwidth=.2,coilarm=.01,linearc=.05](1.,1.5)(.625,1.125)
  \pszigzag[coilwidth=.2,coilarm=.01,linearc=.05](.625,1.125)(.25,1.5)
\endpspicture
\pspicture[shift=-1.25](-1.9,-.75)(1.9,2)
\psset{arrowscale=1.4,arrowinset=0.15}
  \small
  \psline(-1.75,1.75)(-1.25,1.25)(-.75,1.75)
  \rput(-1.7,1.9){$\gamma_7$}  \rput(-.8,1.9){$\gamma_8$}
  \psline(-.5,1.75)(0,1.25)(.5,1.75)
  \rput(-.45,1.9){$\gamma_9$}  \rput(.35,1.9){$\gamma_{10}$}
  \psline(.75,1.75)(1.25,1.25)(1.75,1.75)
  \rput(.9,1.9){$\gamma_{11}$}  \rput(1.7,1.9){$\gamma_{12}$}
  \psline(-1.25,1.25)(0,0)(1.25,1.25) \rput(0,-.2){$b$}
\endpspicture \right),
\end{split}
\end{equation}
where the wiggly line denotes fusion to a fermion.

Applying the four-MZM projector $\Pi_0^{(5678)}$ to the above superposition yields
\begin{equation}
\Pi_0^{(5678)}\Pi_0^{(45)}\ket{a,b} =
\pspicture[shift=-1.25](-2,-.75)(2,2)
\psset{arrowscale=1.4,arrowinset=0.15}
  \small
  \psline(-1.75,1.75)(-1.25,1.25)(-.75,1.75)
  \rput(-1.7,1.9){$\gamma_1$}  \rput(-.8,1.9){$\gamma_2$}
  \psline(-.5,1.75)(0,1.25)(.5,1.75)
  \rput(-.45,1.9){$\gamma_3$}  \rput(.45,1.9){$\gamma_4$}
  \psline(.75,1.75)(1.25,1.25)(1.75,1.75)
  \rput(.8,1.9){$\gamma_5$}  \rput(1.7,1.9){$\gamma_6$}
  \psline(0,1.25)(-.625,.625)(-1.25,1.25)\rput(-.625,.425){$a$}
  \psline(1.,1.5)(.625,1.125)
  \psline(.625,1.125)(.25,1.5)
  \rput(.625,.925){$b$}
\endpspicture
\pspicture[shift=-1.25](-2,-.75)(2,2)
\psset{arrowscale=1.4,arrowinset=0.15}
  \small
  \psline(-1.75,1.75)(-1.25,1.25)(-.75,1.75)
  \rput(-1.7,1.9){$\gamma_7$}  \rput(-.8,1.9){$\gamma_8$}
  \psline(-.5,1.75)(0,1.25)(.5,1.75)
  \rput(-.45,1.9){$\gamma_9$}  \rput(.35,1.9){$\gamma_{10}$}
  \psline(.75,1.75)(1.25,1.25)(1.75,1.75)
  \rput(.9,1.9){$\gamma_{11}$}  \rput(1.7,1.9){$\gamma_{12}$}
  \psline(-1.25,1.25)(0,0)(1.25,1.25) \rput(0,-.2){$b$}
\endpspicture.
\end{equation}
Then, projecting MZMs 3 and 5 to the vacuum channel gives
\begin{equation}
\begin{split}
\Pi_0^{(35)}\Pi_0^{(5678)}\Pi_0^{(45)}\ket{a,b}&=
\pspicture[shift=-2.25](-2,-.75)(2,3)
\psset{arrowscale=1.4,arrowinset=0.15}
  \small
  \psline(-1.75,1.75)(-1.25,1.25)(-.75,1.75)
  \psline(-1.75,1.75)(-1.75,2.75)
  \psline(-.75,1.75)(-.75,2.75)
  \rput(-1.75,2.9){$\gamma_1$}  \rput(-.75,2.9){$\gamma_2$}
  \psline(.4,2.25)(0,2.25)(-.5,1.75)(0,1.25)(.5,1.75)
  \psline(.5,1.75)(.5,2.75)
  \rput(-.2,2.9){$\gamma_3$}  \rput(.5,2.9){$\gamma_4$}
  \psline(1,2)(1,1.5)(1.25,1.25)(1.75,1.75)(1.75,2.75)
  \rput(.95,2.9){$\gamma_5$}  \rput(1.75,2.9){$\gamma_6$}
  \psline(0,1.25)(-.625,.625)(-1.25,1.25)\rput(-.625,.425){$a$}
  \psline(1.,1.5)(.625,1.125)
  \psline(.625,1.125)(.25,1.5)
  \rput(.625,.925){$b$}
  \psline(1,2)(.75,2.25)(.6,2.25)
  \psline(-.25,2.75)(0,2.5)(.4,2.5)
  \psline(.6,2.5)(.75,2.5)(1,2.75)
\endpspicture
\pspicture[shift=-1.25](-2,-.75)(2,2)
\psset{arrowscale=1.4,arrowinset=0.15}
  \small
  \psline(-1.75,1.75)(-1.25,1.25)(-.75,1.75)
  \rput(-1.7,1.9){$\gamma_7$}  \rput(-.8,1.9){$\gamma_8$}
  \psline(-.5,1.75)(0,1.25)(.5,1.75)
  \rput(-.45,1.9){$\gamma_9$}  \rput(.35,1.9){$\gamma_{10}$}
  \psline(.75,1.75)(1.25,1.25)(1.75,1.75)
  \rput(.9,1.9){$\gamma_{11}$}  \rput(1.7,1.9){$\gamma_{12}$}
  \psline(-1.25,1.25)(0,0)(1.25,1.25) \rput(0,-.2){$b$}
\endpspicture
\\ &= \sum_{a',b'} W^{(5678)}_{ab,a'b'}
\pspicture[shift=-1.25](-2,-.75)(2,2)
\psset{arrowscale=1.4,arrowinset=0.15}
  \small
  \psline(-1.75,1.75)(-1.25,1.25)(-.75,1.75)
  \psline(1.75,1.75)(1.,1)(.25,1.75)
  \psline(-1.25,1.25)(0,0)(1,1)\rput(0,-.2){$a'$}
  \psline(-.25,1.75)(.5,1)(.65,1.15)
  \psline(.85,1.35)(1.25,1.75)
  \rput(-1.7,1.9){$\gamma_1$}  \rput(-.8,1.9){$\gamma_2$}
  \rput(.2,1.9){$\gamma_4$}  \rput(-.2,1.9){$\gamma_3$}
  \rput(1.2,1.9){$\gamma_5$}  \rput(1.7,1.9){$\gamma_6$}
\endpspicture
\pspicture[shift=-1.25](-2,-.75)(2,2)
\psset{arrowscale=1.4,arrowinset=0.15}
  \small
  \psline(-1.75,1.75)(-1.25,1.25)(-.75,1.75)
  \rput(-1.7,1.9){$\gamma_7$}  \rput(-.8,1.9){$\gamma_8$}
  \psline(-.5,1.75)(0,1.25)(.5,1.75)
  \rput(-.45,1.9){$\gamma_9$}  \rput(.35,1.9){$\gamma_{10}$}
  \psline(.75,1.75)(1.25,1.25)(1.75,1.75)
  \rput(.9,1.9){$\gamma_{11}$}  \rput(1.7,1.9){$\gamma_{12}$}
  \psline(-1.25,1.25)(0,0)(1.25,1.25) \rput(0,-.2){$b'$}
\endpspicture.
\end{split}
\end{equation}
This step utilizes the diagrammatic braiding relation of MZMs or Ising anyons:
\begin{equation}
\pspicture[shift=-1.4](-1,-1)(0,1.2)
  \small
  \psline(-.5,.5)(0,1)
  \psline[border=1.5pt](-.25,.25)(0,.5)(-.5,1)
  \psline(-1,1)(0,0)\rput(-1,1.2){$a$}\rput(-.25,0){$b$}
  \rput(-.5,1.2){$\gamma$} \rput(-0,1.2){$\gamma$}
\endpspicture
\quad = \sum_{a',b'=0,1} W_{ab,a'b'}
\pspicture[shift=-1.4](-1,-1)(0,1)
  \small
  \psline(-.5,.5)(0,1)
  \psline[border=1.5pt](-.25,.25)(.5,1)
  \psline(-1,1)(0,0)\rput(-1,1.2){$a'$}\rput(-.25,0){$b'$}
  \rput(.5,1.2){$\gamma$} \rput(-0,1.2){$\gamma$}
\endpspicture \qquad
.
\end{equation}

Finally, projecting MZMs 3 and 4 to  the vacuum channel gives
\begin{equation}
\Pi_0^{(34)}\Pi_0^{(35)}\Pi_0^{(5678)}\Pi_0^{(45)}\ket{a,b} = \sum_{a',b'}W^{(5678)}_{ab,a'b'}
\pspicture[shift=-1.25](-2,-.75)(2,2)
\psset{arrowscale=1.4,arrowinset=0.15}
  \small
  \psline(-1.75,1.75)(-1.25,1.25)(-.75,1.75)
  \rput(-1.7,1.9){$\gamma_1$}  \rput(-.8,1.9){$\gamma_2$}
  \psline(-.5,1.75)(0,1.25)(.5,1.75)
  \rput(-.45,1.9){$\gamma_3$}  \rput(.45,1.9){$\gamma_4$}
  \psline(.75,1.75)(1.25,1.25)(1.75,1.75)
  \rput(.8,1.9){$\gamma_5$}  \rput(1.7,1.9){$\gamma_6$}
  \psline(-1.25,1.25)(0,0)(1.25,1.25) \rput(0,-.2){$a'$}
\endpspicture
\pspicture[shift=-1.25](-2,-.75)(2,2)
\psset{arrowscale=1.4,arrowinset=0.15}
  \small
  \psline(-1.75,1.75)(-1.25,1.25)(-.75,1.75)
  \rput(-1.7,1.9){$\gamma_7$}  \rput(-.8,1.9){$\gamma_8$}
  \psline(-.5,1.75)(0,1.25)(.5,1.75)
  \rput(-.45,1.9){$\gamma_9$}  \rput(.35,1.9){$\gamma_{10}$}
  \psline(.75,1.75)(1.25,1.25)(1.75,1.75)
  \rput(.9,1.9){$\gamma_{11}$}  \rput(1.7,1.9){$\gamma_{12}$}
  \psline(-1.25,1.25)(0,0)(1.25,1.25) \rput(0,-.2){$b'$},
\endpspicture
\end{equation}
which is the desired entangling gate.

An alternative derivation of Eq.~(\ref{eq:Xop}) can be performed by explicitly multiplying the projectors written in terms of Majorana operators, as follows
\begin{align}
&\Pi_0^{(34)}\Pi_0^{(35)}\Pi_0^{(5678)}\Pi_0^{(45)}\Pi_0^{(34)} = \Pi_0^{(34)}\frac{1- i\gamma_3\gamma_5}{2}\frac{1-\gamma_5\gamma_6\gamma_7\gamma_8}{2}\frac{1-i\gamma_4\gamma_5}{2}\Pi_0^{(34)} \nonumber\\
&= 2^{-3} \Pi_0^{(34)} \left( 1-i\gamma_3\gamma_5 -i\gamma_4\gamma_5 +\gamma_3\gamma_4-\gamma_5\gamma_6\gamma_7\gamma_8 +i\gamma_3\gamma_6\gamma_7\gamma_8-i\gamma_4\gamma_6\gamma_7\gamma_8 +\gamma_3\gamma_4\gamma_5\gamma_6\gamma_7\gamma_8 \right) \Pi_0^{(34)} \nonumber \\
&= 2^{-3} \left( 1+\gamma_3\gamma_4 -\gamma_5\gamma_6\gamma_7\gamma_8 +\gamma_3\gamma_4\gamma_5\gamma_6\gamma_7\gamma_8 \right) \Pi_0^{(34)} \nonumber \\
&= 2^{-3} \sqrt{2} \left( \frac{1+i}{\sqrt{2}}\right) \left( 1+i\gamma_5\gamma_6\gamma_7\gamma_8 \right) \Pi_0^{(34)} \nonumber \\
&= \frac{1}{4} e^{i\pi/4}W^{(5678)}\otimes\Pi_0^{(34)} .
\end{align}
Here, we used $\Pi_0^{(34)}i\gamma_{3}\gamma_j\Pi_0^{(34)}=\Pi_0^{(34)}i\gamma_{4}\gamma_j\Pi_0^{(34)}=0$ for $j\neq 3$ or $4$ and the fact that $\Pi_0^{(34)}$ projects $i\gamma_3\gamma_4=-1$.


Finally, the following diagrams illustrate how the computational MZMs may be shuttled through the qubit using anyonic teleportation.
\begin{align}
&\pspicture[shift=-1.25](-2,-.75)(2,2)
\psset{arrowscale=1.4,arrowinset=0.15}
  \small
  \psline(0,0)(-1.25,1.25)\rput(0,-.25){$a$}
  \psline(0,0)(1.25,1.25)
  \psline(-1.25,1.25)(-1.75,1.75)\rput(-1.7,1.9){$\gamma_1$}
  \psline(1.25,1.25)(1.75,1.75)\rput(1.7,1.9){$\gamma_6$}
  \psline(-1.25,1.25)(-.75,1.75)\rput(-.8,1.9){$\gamma_2$}
  \psline(1.25,1.25)(.75,1.75)\rput(.8,1.9){$\gamma_5$}
  \psline(0,1.25)(-.5,1.75)\rput(-.4,1.9){$\gamma_3$}
  \psline(0,1.25)(.5,1.75)\rput(.4,1.9){$\gamma_4$}
\endpspicture
\begin{array}{c}  \Pi_0^{(23)} \\  \rightleftarrows \\  \Pi_0^{(34)} \\ \end{array}
\pspicture[shift=-1.25](-2,-.75)(2,2)
\psset{arrowscale=1.4,arrowinset=0.15}
  \small
  \psline(0,0)(-1.75,1.75)\rput(0,-.25){$a$}
  \psline(0,0)(1.75,1.75)\rput(-1.75,1.9){$\gamma_1$}
  \psline(-.75,1.25)(-1.25,1.75)\rput(-1.2,1.9){$\gamma_2$}
  \psline(-.75,1.25)(-.25,1.75)\rput(-.3,1.9){$\gamma_3$}
  \psline(-.75,.75)(.25,1.75)\rput(.25,1.9){$\gamma_4$}
  \psline(1.25,1.25)(1.75,1.75)\rput(1.7,1.9){$\gamma_6$}
  \psline(1.25,1.25)(.75,1.75)\rput(.8,1.9){$\gamma_5$}
\endpspicture
\begin{array}{c}  \Pi_0^{(12)} \\ \rightleftarrows \\ \Pi_0^{(23)}  \\ \end{array}
\pspicture[shift=-1.25](-2,-.75)(2,2)
\psset{arrowscale=1.4,arrowinset=0.15}
  \small
  \psline(-1.75,1.75)(-1.25,1.25)(-.75,1.75)
  \rput(-1.7,1.9){$\gamma_1$}\rput(-.8,1.9){$\gamma_2$}
  \psline(-.5,1.75)(.625,.625)
  \psline(0,1.25)(.5,1.75)
  \rput(-.45,1.9){$\gamma_3$}\rput(.45,1.9){$\gamma_4$}
  \psline(.625,.625)(1.25,1.25)
  \psline(.75,1.75)(1.25,1.25)(1.75,1.75)
  \rput(.8,1.9){$\gamma_5$} \rput(1.7,1.9){$\gamma_6$}
  \rput(.625,.425){$a$}
\endpspicture
\\&\pspicture[shift=-1.25](-2,-.75)(2,2)
\psset{arrowscale=1.4,arrowinset=0.15}
  \small
  \psline(0,0)(-1.25,1.25)\rput(0,-.25){$a$}
  \psline(0,0)(1.25,1.25)
  \psline(-1.25,1.25)(-1.75,1.75)\rput(-1.7,1.9){$\gamma_1$}
  \psline(1.25,1.25)(1.75,1.75)\rput(1.7,1.9){$\gamma_6$}
  \psline(-1.25,1.25)(-.75,1.75)\rput(-.8,1.9){$\gamma_2$}
  \psline(1.25,1.25)(.75,1.75)\rput(.8,1.9){$\gamma_5$}
  \psline(0,1.25)(-.5,1.75)\rput(-.4,1.9){$\gamma_3$}
  \psline(0,1.25)(.5,1.75)\rput(.4,1.9){$\gamma_4$}
\endpspicture
 \begin{array}{c} \Pi_0^{(45)} \\ \rightleftarrows \\ \Pi_0^{(34)} \end{array}
\pspicture[shift=-1.25](-2,-.75)(2,2)
\psset{arrowscale=1.4,arrowinset=0.15}
  \small
  \psline(0,0)(1.75,1.75)\rput(0,-.25){$a$}
  \psline(0,0)(-1.75,1.75)\rput(1.75,1.9){$\gamma_6$}
  \psline(.75,1.25)(1.25,1.75)\rput(1.2,1.9){$\gamma_5$}
  \psline(.75,1.25)(.25,1.75)\rput(.3,1.9){$\gamma_4$}
  \psline(.75,.75)(-.25,1.75)\rput(-.3,1.9){$\gamma_3$}
  \psline(-1.25,1.25)(-1.75,1.75)\rput(-1.7,1.9){$\gamma_1$}
  \psline(-1.25,1.25)(-.75,1.75)\rput(-.8,1.9){$\gamma_2$}
\endpspicture
\begin{array}{c} \Pi_0^{(56)} \\  \rightleftarrows \\ \Pi_0^{(45)} \\ \end{array}
\pspicture[shift=-1.25](-2,-.75)(2,2)
\psset{arrowscale=1.4,arrowinset=0.15}
  \small
  \psline(-1.75,1.75)(-.625,.625)\rput(-.625,.425){$a$} \rput(-1.7,1.9){$\gamma_1$}
  \psline(-1.25,1.25)(-.75,1.75) \rput(-.8,1.9){$\gamma_2$}
  \psline(-.625,.625)(.5,1.75)\rput(.45,1.9){$\gamma_4$}
  \psline(0,1.25)(-.5,1.75)\rput(-.45,1.9){$\gamma_3$}
  \psline(1.75,1.75)(1.25,1.25)(.75,1.75)
  \rput(1.7,1.9){$\gamma_6$}\rput(.8,1.9){$\gamma_5$}
\endpspicture
\end{align}

\end{widetext}


\bibliography{topo-phases}

\end{document}